\documentclass[graybox]{svmult}
\usepackage{mathptmx}       
\usepackage{helvet}         
\usepackage{courier}        
\usepackage{type1cm}        
\usepackage[utf8]{inputenc}
\usepackage{makeidx}         
\usepackage{graphicx}        
\usepackage[bottom]{footmisc}
\usepackage[compress]{cite}
\usepackage{graphicx}
\usepackage{amssymb}
\usepackage{epstopdf}
\usepackage{amsmath}
\usepackage{color}
\usepackage{mathrsfs}
\usepackage{varioref}
\usepackage[active]{srcltx}
\usepackage{tikz}
\usepackage{lmodern}
\usetikzlibrary{decorations.pathmorphing}
\tikzset{snake it/.style={decorate, decoration=snake}}
\usepackage{fancybox}
\usetikzlibrary{3d,calc}
\usetikzlibrary{decorations.pathmorphing}
\usetikzlibrary{shapes.geometric, arrows}
\tikzstyle{startstop} = [rectangle, rounded corners, minimum width=3.5cm, minimum height=1cm,text centered, text width=3cm, draw=black, fill=red!30]
\tikzstyle{io} = [rectangle, rounded corners, minimum width=3.5cm, minimum height=1cm,text centered, text width=3cm, draw=black, fill=blue!30]
\tikzstyle{process} = [rectangle, rounded corners, minimum width=3.5cm, minimum height=1cm,text centered, text width=3cm, draw=black, fill=orange!30]
\tikzstyle{decision} = [rectangle, rounded corners, minimum width=3.5cm, minimum height=1cm,text centered, text width=3cm, draw=black, fill=green!30]
\tikzstyle{is} = [rectangle, rounded corners, minimum width=3.5cm, minimum height=1cm, text width=3cm, draw=black, fill=yellow!30]
\tikzstyle{arrow} = [thick,->,>=stealth]

\makeindex             
\RequirePackage[colorlinks,citecolor=blue,urlcolor=blue,linkcolor=blue]{hyperref}
\labelformat{section}{Section #1} 
\labelformat{subsection}{Section #1} 
\labelformat{equation}{Eq.~(#1)} 
\labelformat{figure}{Fig.~#1} 
\labelformat{subfigure}{Fig.~\thefigure#1} 
\labelformat{table}{Tab.~#1} 
\newcommand{\be}{\begin{equation}}
\newcommand{\ee}{\end{equation}}
\newcommand{\bea}{\begin{eqnarray}}
\newcommand{\eea}{\end{eqnarray}}

\newcommand{\kets}[1]{\left\vert #1 \right\rangle}
\newcommand{\bras}[1]{\left\langle #1 \right\vert}
\newcommand{\beq}{\begin{equation}}
\newcommand{\eeq}{\end{equation}}
\newcommand{\bqr}{\begin{eqnarray}}
\newcommand{\eqr}{\end{eqnarray}}
\def\CSL{Continuous Spontaneous Localization }
\def\gr{general relativity}
\begin{document}

\title*{Black Holes: Eliminating Information or Illuminating New Physics?}
\author{\bf Sumanta Chakraborty and Kinjalk Lochan}
\institute{ {\bf Sumanta Chakraborty} 
\at IUCAA, Post Bag 4, Ganeshkhind, Pune University Campus, Pune 411 007, India 
\at and
\at Department of Theoretical Physics, Indian Association for the Cultivation of Science, Kolkata 700032, India \\
\email{sumantac.physics@gmail.com}
\and {\bf Kinjalk Lochan} 
\at School of Physics, IISER Thiruvananthapuram, Trivandrum 695016, India
\at and
\at Department of Physical Sciences, IISER Mohali, Manauli 140306, India \\
\email{kinjalk.lochan@gmail.com}}
%
%
\maketitle
\abstract{Black holes, initially thought of as very interesting geometric constructions of nature, over time, have learnt to (often) come up with surprises and challenges. From the era of being described as merely some interesting and exotic solutions of \gr, they have, in modern times, really started to test our confidence in everything else, we thought we know about the nature. They have in this process, also earned a dreadsome reputation in some corners of theoretical physics. The most serious charge on the black holes is that they eat up information, never to release and subsequently erase it. This goes absolutely against the sacred principles of all other branches of fundamental sciences. This realization has shaken the very base of foundational concepts, both in quantum theory and gravity, which we always took for granted. Attempts to exorcise black holes of this charge, have led us to 
crossroads with concepts, hold dearly in quantum theory. The sphere of black hole's tussle with quantum theory has readily and steadily grown, from the advent of the Hawking radiation some four decades back, into domain of quantum information theory in modern times, most aptly, recently put in the form of the firewall puzzle. Do black holes really indicate something sinister about their existence or do they really take the lid off our comfort with ignoring the fundamental issues, our modern theories are seemingly plagued with? In this review, we focus on issues pertaining to black hole evaporation, the development of the information loss paradox, its recent formulation, the leading debates and promising directions in the community.}
\newpage
\section{Being simple is complex}

Twentieth century was a century in theoretical physics when many innovative, visionary  ideas gained firm grounds. Experimental verifications of many remarkable ideas were, deservingly,  celebrated as the decisive steps towards  unraveling the secrets the nature beholds very intimately. From the first decades of the century, sprung out two radically different but conceptually very profound ideas: {\it  Relativity} --- viewing space and time as a geometric fabric acting as stage on which everything else takes place --- and {\it Quantum theory} --- acknowledging the wave character of everything in the universe. These two theories not only came with revolutionary, as well as beautiful insights which kept getting ratified time and again ({\it without fail} till date) by the most sophisticated experiments, but they also did come with some radically counter-intuitive ideas at their cores. With the advent and the subsequent triumph of the quantum theory, we (rather painstakingly) learnt to let go of {\it 
determinism} as an intrinsic property of nature and adopt to probabilistic interpretation for the most fundamental description. This feature of the new realization alone, has caused enough discomfort and unrest with the quantum theory, to raise camps of challengers and sympathizers to it \cite{Adler:1993hm,Bohm:1951xw,Bohm:1951xx,Diosi:1986nu,Everett:1957hd,Bassi:2012bg,Bassi:2003gd,Hawking:1976ra}. Relativity on the other hand, very beautifully describes gravity as merely an artifact of life on a curved fabric named {\it the spacetime}. Describing gravity as the curvature of the spacetime remarkably gives rise to many interesting phenomena. The theory governing the curvature (which we believe to be the famous Einstein's theory of \gr), allows it, in some settings, to blow up during the evolution, causing appearance of singularity and in this process possibly predicting its own demise. This inevitably urges us to think beyond the usual character of gravity as we see or know it. Furthermore, these infinite 
curvature points are ``typically'' {\it expected to be} 
hidden by a causal curtain and can not be seen by us if we happen to be on the other side of the curtain (famously known as the cosmic censorship conjecture \cite{Wald:1997wa,Joshi:2002,Hod:2008zza,Hamid:2014kza}). 
Such an advent of singularity wrapped inside a causal covering is a very beautiful and exotic construct 
of the theory: {\it the black hole} \cite{MTW,Hawking:1973uf,Hawking:2010mca,gravitation,
Mukhanov-Winitzky,Poisson,Wald,Birrell:1982ix,Parker:2009uva,Fabbri:2005mw,Fulling:1989nb}. 

The black hole, at the face value, looks a very elegant and rather innocent construction, the nature seems capable of building up in many dynamical scenarios, which demonstrate the non-trivial, awe-inspiring features of a concept as beautiful as the \gr. The community of relativists was more than delighted when it turned out these black holes, which are formed as end products of many complicated dynamic evolutions, are themselves ``very simple'' in description ({\it all black holes can be totally characterized in terms of very few parameters:} the no hair conjecture \footnote{This is also referred as no-hair theorems after proving the conjecture in some specialized scenarios}) and also seemingly follow some generalized laws of thermodynamics. Thus black holes took the center stage in relativity as they drew concepts from (and stitched together) many branches of physics and mathematics, such as differential geometry, causality and most importantly thermodynamics. These remarkable properties of black 
holes kept us mesmerized so much, that we were totally stumped as we saw this simplicity to grow into one of the most formidable nemesis of the second pillar of our understanding of the nature: the quantum theory. 
\subsection{If you can heat it, it has micro-structure!}

Black holes, by now, are understood to behave really very much like thermodynamic objects \cite{Bekenstein:1974ax,Bardeen:1973gs,Gibbons:1977mu,Hawking:1976de}. They possess entropy \cite{Bekenstein:1973ur,Bekenstein:1972tm,Bekenstein:1974ax} and do radiate thermally \cite{Hawking:1974sw} (in the semiclassical picture). Microscopically speaking, the entropy of any physical system originates from a statistical analysis of the underlying micro-states. Therefore, one expects the black hole as well, to inherit entropy which originates from a micro-state counting, as and when a consistent quantum theory of black holes (which happens to be also the quantum theory of gravity) becomes available. Being on the other side of the causal divide, the singularity does not trouble us or our physics and may lead us to believe that we are safely, far removed from the regime where nitty gritties of quantum gravity is really important. However, there are scenarios which seemingly force us to look towards the quantum description 
of geometry (and black hole), even though the regions we operate from, is very much regular (or classical). 

It turns out that the thermodynamic description has a much deeper root, which transcends the black hole physics. The dynamics of gravity itself, i.e., Einstein's equations have a very nice thermodynamic interpretation which goes both ways --- (a) Starting from the Clausius relation in thermodynamics one can arrive at Einstein's equations using local Rindler observers \cite{Jacobson:1995ab}, (b) Projection of Einstein's equations on a null surface emerges as a thermodynamic identity. Proceeding further, one can also show that the dynamical equations of other modified gravity theories can be cast in a thermodynamic language \cite{Padmanabhan:2003gd,Padmanabhan:2009vy,Padmanabhan:2013nxa,Chakraborty:2015hna,
Chakraborty:2015aja,Chakraborty:2014rga,Chakraborty:2015wma}. However, Einstein's equations have somethings more to reveal, when projected on a two-surface, they become identical to the Navier-Stokes equations of fluid dynamics with a bulk as well as shear viscosity \cite{Chakraborty:2015hna}. All these results indicate that gravity may not be fundamental in nature, but is originating as a thermodynamic limit of some underlying microscopic degrees of freedom, just like elasticity. This viewpoint gets further support from the fact that at every spacetime point one can introduce local Rindler observers, who will observe the spacetime hot. Then following Boltzmann, one can immediately confirm the existence of micro structure of the spacetime. However, we refer our reader to the following reviews \cite{Padmanabhan:2003gd,Padmanabhan:2009vy,Padmanabhan:2016eld,Padmanabhan:2015zmr,Padmanabhan:2015pza,Chakraborty:2016dwb} for more detailed account of the thermodynamic description, which we will be skipping in this 
review.

Classically nothing is supposed to come out of the black hole, which was originally puzzling since thermodynamically, black holes are supposed to be hot and radiating. The work by Hawking \cite{Hawking:1974sw} established that black holes indeed radiate perfectly like a blackbody \emph{if we consider quantum field theory in a classically collapsing background}. Although this result seems to provide a nice  thermodynamic description of black holes, the radiation also causes the black hole to lose energy and shrink. This thermal evaporation of black holes spells doom for unitary quantum theory \cite{Mathur:2009hf}. In fact, thermal radiation and the consequent mass loss of the black hole leads to a tussle between classical gravity and quantum matter in form of the so-called (and now so-infamous) {\it information paradox}. There have been really many versions of it, not all of which are conceptually equivalent. In this review we will be dwelling upon the most popular ones and the attempts to address them. We  
start by the most naive version of the same, which will set the stage for what is to come later. We will be discussing the most prominent and promising ideas being worked 
upon in the field (of course, subject to authors' personal biases, we apologize in advance to the  section of researchers, whose work did not find adequate mention in this review, mostly by the ignorance of the authors or due to limitation of size).  We encourage the reader to look into the references and discussions on the issues discussed here from different perspectives \cite{Harlow:2014yka, Polchinski:2016hrw} for more detailed and technically elaborate descriptions.
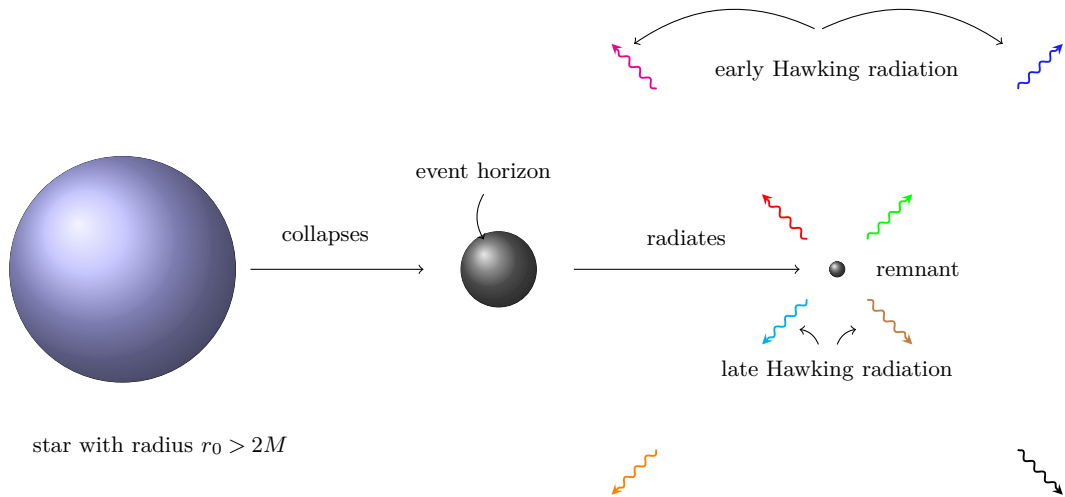
\begin{figure}[t]
\begin{center}
\begin{tikzpicture}
    \draw (-1.5,0) circle (1.5cm);
    \shade[ball color=blue!30!white,opacity=0.80] (-1.5,0) circle (1.5cm);
    \draw[->] (0.2,0) to (2.5,0);  
    \draw (3.5,0) circle (0.5cm);
    \shade[ball color=black!60!white,opacity=0.80] (3.5,0) circle (0.5cm);
    \draw[->] (4.5,0) to (7.5,0);
    \draw (8,0) circle (0.1cm);
    \shade[ball color=black!60!white,opacity=0.80] (8,0) circle (0.1cm);
    \draw[-stealth,decoration={snake,amplitude =0.4mm,segment length = 2mm, post length=0.9mm},decorate,color=red, line width=0.25mm] (7.6,0.4) -- (7.0,1.0);
    \draw[-stealth,decoration={snake,amplitude =0.4mm,segment length = 2mm, post length=0.9mm},decorate,color=green, line width=0.25mm] (8.4,0.4) -- (9.0,1.0);
    \draw[-stealth,decoration={snake,amplitude =0.4mm,segment length = 2mm, post length=0.9mm},decorate,color=brown, line width=0.25mm] (8.4,-0.4) -- (9.0,-1.0);
    \draw[-stealth,decoration={snake,amplitude =0.4mm,segment length = 2mm, post length=0.9mm},decorate,color=cyan, line width=0.25mm] (7.6,-0.4) -- (7.0,-1.0);
    \draw[-stealth,decoration={snake,amplitude =0.4mm,segment length = 2mm, post length=0.9mm},decorate,color=magenta, line width=0.25mm] (5.6,2.4) -- (5.0,3.0);
    \draw[-stealth,decoration={snake,amplitude =0.4mm,segment length = 2mm, post length=0.9mm},decorate,color=blue!90, line width=0.25mm] (10.4,2.4) -- (11.0,3.0);
    \draw[-stealth,decoration={snake,amplitude =0.4mm,segment length = 2mm, post length=0.9mm},decorate,color=orange, line width=0.25mm] (5.6,-2.4) -- (5.0,-3.0);
    \draw[-stealth,decoration={snake,amplitude =0.4mm,segment length = 2mm, post length=0.9mm},decorate,color=black, line width=0.25mm] (10.4,-2.4) -- (11.0,-3.0);
    \node[label=below: \textrm{star with radius} $r_{0}>2M$] at (-1,-2) {};
    \node[label=above: \textrm{collapses}] at (1.2,0.1) {};
    \node[label=above: \textrm{event horizon}] at (3.3,1) {};
    \draw[->,bend right=30] (3.3,1) to (3.3,0.4);
    \node[label=above: \textrm{radiates}] at (6.0,0.1) {};
    \node[label=below: \textrm{early Hawking radiation}] at (8.0,3) {};
    \draw[->,bend right=30] (7.7,3.2) to (5.3,3.0);
    \draw[->,bend left=30] (7.8,3.2) to (10.2,3.0);
    \node[label=below: \textrm{late Hawking radiation}] at (8.0,-1) {};
    \draw[->,bend right=30] (7.75,-1) to (7.5,-0.8);
    \draw[->,bend left=30] (8.0,-1) to (8.25,-0.8);
    \node[label=right: \textrm{remnant}] at (8.3,0) {};
\end{tikzpicture}
\end{center}
\caption{A schematic of the information paradox. A star collapses to form a black hole and forms an event horizon at $r=2M$. Any object falling into the black hole will encode its information inside the event horizon. As a consequence as the black hole gradually evaporates all such information will be washed out leaving a Planck size remnant, with uncorrelated Hawking spectrum. Since it is impossible for the Planck sized remnant to hold the plethora of initial information, apparently the information of what went into the black hole is lost. This results in the information paradox. See text for more discussions.}\label{fig_Paradox_Intro}
\end{figure}
\subsection{Beginner's information Loss}

The story around black hole with its devilish design to destroy information really starts with its birth. Once the black hole forms, the world outside of it becomes oblivious to what formed the black hole at the first place: a pack of pebbles or a bagful of diamonds.

To start with, let us consider a \textit{classical} process in which some matter is thrown into the black hole horizon. Since there are no outgoing timelike/null  world lines from the interior, by the design of black hole,  the matter can not  directly communicate with outside observers. Additionally, since no hair theorems \cite{MTW,Wald} tell us that a classical  black hole  will only be characterized by its mass, charge and angular momentum, the outside geometry  will never be able to tell us about any other intrinsic properties of the matter chunk the black hole has just gobbled up. So, no change in geometry can entirely encode all the information that went inside the hole. This is sometimes (wrongly!) referred to as a problem of loss of information from the point of view of information theory and observers outside. However this issue,  per se, is not a  paradox. The information just gets contained in a region of spacetime, which an outside observer can not visit or see. However, there exists a different 
set of observers (who fall inside the black hole, for instance) for whom all such information is available (unless it is destroyed by the highly curved region around the classical singularity). It is indeed equivalent of locking a encyclopedia inside a room and never revisiting the room again. This is a perfectly classical scenario and does not have any implication for or a tussle with the quantum theory.

Incidentally, this also brings out the notion of observer dependence in the theory. For  observers staying outside the horizon there is really no way  \emph{classically} to get back the information that has gone into the black hole. This inaccessibility of information appears as entropy to these observers\footnote{To such observers, the entropy of the black hole may be accounted for as the entanglement entropy. However, whether the black hole thermodynamics is indeed due to the entanglement entropy remains an open issue, since in many alternative gravity theories, the black hole entropy does not scale as area, while the entanglement entropy always does. Thus, whether entanglement entropy is a mirror to the micro-canonical construction of black hole is not entirely evident.}. On the other hand, an in-falling observer can access all the information by the time (s)he hits the singularity (on a suicide mission, of course). In fact, such observers also do not find any region of spacetime which is causally denied \
cite{
Padmanabhan:2016eld,Padmanabhan:2015zmr,Padmanabhan:2015pza,Chakraborty:2016dwb} to them and they do not see any horizon or associate entropy with a black hole. This notion of observer dependence can seemingly be carried over to other horizons too and turns out to be \emph{essentially} suggesting that gravity better be viewed from a thermodynamic perspective \cite{Jacobson:1995ab,Padmanabhan:2003gd,Padmanabhan:2009vy,Padmanabhan:2013nxa,Chakraborty:2015hna,
Chakraborty:2015aja,Chakraborty:2014rga,Chakraborty:2015wma} and not as a fundamental theory. We will skip the discussion on this topic in this review and move towards more formal discussion of the information paradox.
\section{The way it all began!}

The black holes in \gr\ had pretty much smooth sailing till they came hitting the stumbling blocks put forward by the quantum theory. The semiclassical treatment of matter fields in a curved spacetime harboring a black hole, by Hawking \cite{Hawking:1974sw,Hawking:1974rv}, brought for the first time, the concepts of causalities, face to face with perils of quantum mechanics, a theory  which very delicately and painstakingly handled the intricacies of causality as well as non locality within its framework, in order to come clean with special relativity. The causal and geometric features of black holes seemingly let loose the dreaded monsters of modern physics (in guise of non-causality, non-unitarity), which took enormous efforts to be {\it bottled in} by the proponents of the quantum theory. 

The real problem arises when we start taking the quantum considerations into account. Hawking \cite{Hawking:1974sw,Hawking:1974rv,Gibbons:1976ue}, argued that quantum effects in the semiclassical regime could give rise to pairs of particles of positive and negative energies from the vacuum near the event horizon. While the negative energy particle falls into the event horizon and decreases the black hole mass, the positive energy particle travels to the asymptotic region and appears as the {\it Hawking radiation}. This leads to the interpretation that the mass lost by the black hole in the process, reappears in the form of the energy of this radiation.  In this process, since the particle anti-particle pair popped out of vacuum state together, the out-going modes remain entangled with the in-going modes entering the horizon\footnote{In Quantum Field Theory, we rather talk about mode functions than the particles.}. But if the black hole evaporates completely by this process, there will be nothing left for 
the outgoing modes to remain entangled with! 
The outgoing modes are thermal in description which is described by a mixed state. This process, in which a pure state seemingly evolves into a mixed state, is contrary to the unitary quantum evolution \cite{Mann:2014yxa}. The thermal nature turns out to be particularly problematic. Information theoretically a thermal radiation is information free \cite{Mathur:2009hf} and its correlations are also very simple. Thus a thermal radiation can not hide information in its correlation. There are suggestions that the state actually remains pure during evolution such that the outgoing radiation has {\it effectively} thermal profile with information residing in the correlations \cite{Kiefer:2001wn, Demers:1995tr, Maldacena:2013xja}. It must be noted that there exist multiple sources of possible distortions to the thermal Hawking radiation \cite{Visser:2014ypa}, none of these turn out to be strong enough to make the theory unitary \cite{Mathur:2009hf}. Further, the radiation modified by these suggestion is still 
expected to reveal 
only the mass, charge and angular momentum of the hole. No other information of the field contents gets visible to the asymptotic observers. So once the black hole disappears radiating, the resulting radiation is found wanting in terms of restoration of information which the hole ate up. One must contrast this situation with that of burning of a coal, where also the final radiation is nearly thermal, but the correlations between the early radiation quanta and late radiation quanta lead to a complete recovery of the initial state of the coal \cite{Alonso-Serrano:2015trn,Alonso-Serrano:2015bcr}. However, in black holes, even the correlations are devoid of any information, courtesy the no-hair properties of black holes. We will discuss more on this issue in later sections.

There have been many different attempts in the literature to tackle this conflict. There are suggestions that the black hole actually never evaporates completely and leaves behind a Planck size remnant at the end of the process (when the semi-classical description breaks down), which hides the missing information. Still, how such a Planck size remnant could  accommodate the landscape of varying initial configurations remains a mystery. There are interesting suggestions dealing with the horizon structure of spacetime \cite{Mathur:2008kg} which could, in principle, restore faith in essential tenets of both classical gravity and the quantum theory. There are also radical proposals which demand drastic modification of structures of quantum theory to accommodate such a process \cite{Papadodimas:2013wnh, Modak:2014vya}. However, such modifications may have implications for other physical scenarios where the triumph of the erstwhile concepts have been vividly demonstrated. 

So the crux of the information paradox can be summarized as follows: when the black hole evaporates completely without leaving any remnant, one is justified in assuming that the entire information content of the collapsing body must be encoded in the resulting radiation. However, remnant radiation in this process is \emph{dominantly} thermal, which is thermodynamically prohibited to contain much of information. Therefore, \emph{most of} the information  content of the matter which made the black hole in the first place seems inexplicably lost (see \ref{fig_Paradox_Intro}). 

Thus, first encounter of quantum theory with the {\it freak} of \gr\ turned out to be so mammoth in proportions that the grappling is still going on after four decades, with none turning out to be a clear potential winner. The tussle has taken many  twists and turns in the meanwhile with people anticipating for (and against) every possible outcome: submission of black holes, submission of \gr, submission of quantum theory, submission of quantum gravity. The saga of the (one of the, if not absolutely the) most profound, fascinating and  conceptually inviting paradox of modern theoretical physics is compelling and interesting enough to review the advent and subsequent evolution into one of the most formidable problems, which we try to do in the subsequent sections; for now starting with the first thing first: The arrival of the information loss!
\subsection{Semi-classical information loss}

Hawking, in his seminal work deduced that not only the classical intuition that nothing comes out of the black hole was wrong at the semiclassical level, but the black holes indeed radiate as a thermal body whose temperature is obtained from the surface gravity, exactly the way in which laws of black hole thermodynamics are prescribed. We will very briefly discuss this foundational result which is the genesis  of all the modern day versions of information paradox.

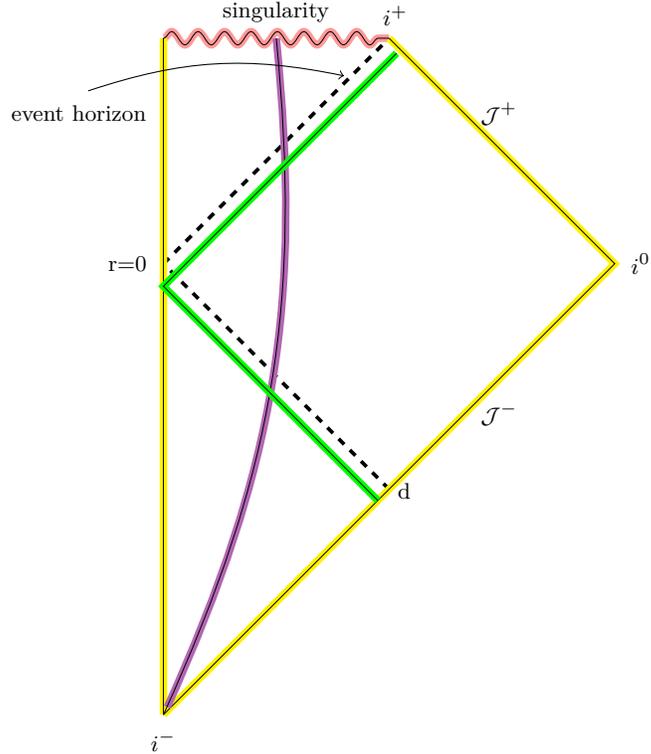
\begin{figure}[t]
\sidecaption[t]
\begin{tikzpicture}
    \draw[dashed,black,line width=0.5mm] (0,0) -- (3,3);
    \draw[black,dashed,line width=0.5mm] (3,-3) -- (0,0);
    \draw[red!40,line width=1mm, snake it] (0,3) -- (3,3);
    \draw[yellow,line width=1mm] (0,-6) -- (0,3);
    \draw[yellow,line width=1mm] (3,3) -- (6,0) -- (0,-6);
    \draw[black, line width=0.1mm] (0,-6) -- (0,3);
    \draw[black,line width=0.1mm, snake it] (0,3) -- (3,3);
    \draw[black, line width=0.1mm] (0,-6) -- (6,0) -- (3,3);
    \draw[violet!60,line width=1mm,bend right=15] (0.05,-5.9) to (1.5,3);
    \draw[black,line width=0.1mm,bend right=15] (0,-6) to (1.5,3); 
    \draw[green,line width=1mm] (3.10,2.80) -- (0,-0.3) -- (2.85,-3.15);
    \draw[black,line width=0.1mm] (3.10,2.80) -- (0,-0.3) -- (2.85,-3.15);
    \node[label=right:$i^+$] at (2.7,3.3) {};
    \node[label=below:$i^-$] at (0,-6) {};
    \node[label=right:$i^0$] at (6,0) {};
    \node[label=below: d] at (3.2,-2.7) {};         
    \node[label=right:$\mathcal{J}^+$] at (4,2) {};
    \node[label=right:$\mathcal{J}^-$] at (4.0,-2.0) {};
    \node[label=left:\textrm{r=0}] at (0,0) {};
    \node[label=left:\textrm{event horizon}] at (0,2) {};
    \node[label=above:\textrm{singularity}] at (1.5,3) {};
    \draw[->,bend left=15] (-1.0,2.3) to (2.4,2.5);
\end{tikzpicture}
\caption{A schematic illustration of the Penrose diagram for Schwarzschild spacetime, which is being formed by collapse of a star (indicated by thick violet line). The point $i^{-}$ stands for past timelike infinity (i.e., $t\rightarrow -\infty$), the point $i^{+}$ is the future timelike infinity (i.e., $t\rightarrow \infty$) and $i^{0}$ is the spatial null infinity (i.e., $r\rightarrow \infty$). The two null surfaces $\mathcal{J}^{+}$ and $\mathcal{J}^{-}$ represents future and past null infinity respectively. The black dotted line represents the event horizon, which is marked by `d' on $\mathcal{J}^{-}$. Any null ray to the left of it will end up in $\mathcal{J}^{+}$, while those on the right ends up in the singularity.}\label{fig_Penrose_Intro}
\end{figure}

The case we discuss first is that of a  real scalar quantum field living in a collapsing spacetime, which eventually would harbor a black hole (see for example \ref{fig_Penrose_Intro}). The initial state of the field is specified at the past null infinity (${\cal J}^{-}$). The geometry at ${\cal J}^{-}$ is Minkowski-like and the corresponding modes describing the quantum field will be the flat spacetime modes. For the flat spacetime free-field theory, the mode decomposition of the in-falling field (which falls in and forms/becomes a part of the black hole) is given as
\bea
\hat{\phi}({x}) = \int \frac{d^3 {\bf k}}{\sqrt{2 \omega_{\bf k}}}\left\lbrace \hat{a}_{{\bf k}}u_k(x)+\hat{a}_{{\bf k}}^{\dagger}u_k^*(x)\right\rbrace
\eea
with $u_k(x)$ being the mode functions satisfying the field equations for the field under consideration.

We will study a field which undergoes spherically symmetric (s-wave) collapse, slowly \cite{Chakraborty:2015nwa,Singh:2014paa} to form a large mass black hole. We consider the Klein Gordon equation for the scalar field sitting atop this geometry. From the solution of the Klein-Gordon equation we obtain the relevant  positive frequency modes describing the state of the field on ${\cal J}^{-}$ (which is a Cauchy surface) as
\bea
u_{\omega}(t,r,\theta,\phi) \sim \frac{1}{r\sqrt{\omega}}e^{-i\omega(t+r)}S(\theta,\phi)
\eea
where $S(\theta,\phi)$ gives a combination of spherical harmonics $Y_{lm}(\theta,\phi)$. For this collapsing case we take the initial state to be in-moving  at ${\cal J}^{-}$ which is totally spherically symmetric, i.e., $l=0$. Once an event horizon is formed through the collapse process, the full state can again be described using a combined description at the event horizon ${\cal H}$ and on the future null infinity (${\cal J}^{+}$) \cite{Padmanabhan:2009vy,Parker:2009uva}, which together form a Cauchy surface, i.e., the field configuration of spacetime can also be described using positive and negative frequency modes compatible to ${\cal J}^{+}$ as well as on the horizon ${\cal H}$. For an asymptotic observer in the remote future, the end state configuration of the field will be called the out-state, described using modes at ${\cal J}^{+}$ (only), which are again flat spacetime modes owing to the asymptotic flatness of the geometry at hand. The only difference between the modes on ${\cal J}^{-}$ and ${\cal 
J}^{+}$ (of course apart from their frequencies, which change due to change of geometry) 
is the mode which moves inwards from ${\cal J}^{-}$ appears as an outgoing on ${\cal J}^{+}.$ Due to linear property of Klein-Gordon equations any set of solutions can be mapped into each other by linear transformation  known as the {\it Bogoliubov transformations} \cite{Birrell:1982ix,Parker:2009uva,Fulling:1989nb,Mukhanov-Winitzky,Fabbri:2005mw}. The Bogoliubov coefficients of transformations are obtained by tracing an outgoing null ray on ${\cal J}^{+}$ back to ${\cal J}^{-}$. This tracing back leads to the fact that all the modes reaching ${\cal J}^{+}$ originate from a portion of ${\cal J}^{-}$. Therefore the full coordinate range on ${\cal J}^{+}$ is logarithmically related to the coordinates in a portion of ${\cal J}^{-}$. This logarithmic dependence of coordinates relate the frequencies on these surfaces exponentially. It must be kept in mind that when one traces a ray from ${\cal J}^{+}$ backwards one needs to be aware of the geometry of the spacetime in order to calculate the gravitational effects (
scattering, lensing, spreading of geodesics etc.) which makes the calculations intractable. However, when we 
are interested in the rays which reach ${\cal J}^{+}$ very late in time (or, equivalently stay very close to the horizon in its outward journey), these modes are very high frequency rays which do not care about any other scale in the theory and the optical ray tracing  can be done effectively. Therefore, we apply optical ray tracing to all the null rays expecting that we will not be making much of error while doing so for low frequency modes. In any case, for the exact or the approximated ray tracing the asymptotic form of these Bogoliubov coefficients are obtained as \cite{Hawking:1974sw}
\bea 
\alpha_{\Omega \omega} &=&\frac{1}{2\pi \kappa} \sqrt{\frac{\Omega}{\omega}}\exp{\left[\frac{\pi\Omega}{2\kappa}\right]}\exp{[i(\Omega-\omega)d]}\exp{\left[\frac{i\Omega }{\kappa} \log{\frac{\omega}{C}} \right]}\Gamma\left[-\frac{i\Omega }{\kappa}\right],\nonumber\\
\beta_{\Omega \omega} &=&-\frac{1}{2\pi \kappa} \sqrt{\frac{\Omega}{\omega}}\exp{\left[-\frac{\pi\Omega}{2\kappa}\right]}\exp{[i(\Omega+\omega)d]}\exp{\left[\frac{i\Omega }{\kappa} \log{\frac{\omega}{C}} \right]}\Gamma\left[-\frac{i\Omega }{\kappa}\right], \label{BT}
\eea
where $\Omega$ is the frequency of the out-modes at ${\cal J}^{+}$, the parameter $\kappa$ is the surface gravity of the black hole, while $C$ is a product of affine parametrization of incoming and outgoing null rays \cite{Hawking:1974sw,Padmanabhan:2009vy,Parker:2009uva} and $d$ is an arbitrary constant marking the last null ray reaching ${\cal J}^{+}$. The field content of the out-state can be obtained using the Bogoliubov coefficients between the modes at ${\cal J}^{-}$ and ${\cal J}^{+}$ \cite{Hawking:1974sw,Padmanabhan:2009vy,Parker:2009uva}.
We can set $d=0$ through coordinate transformations on ${\cal J}^{-}$. These Bogoliubov coefficients are accurate for large values of $\omega$. At small $\omega$ values, the expressions in \ref{BT} will receive corrections as we discussed above, the ray tracing will become less trustworthy. However, when we are interested in  the late-time radiation at future null infinity (${\cal J}^{+}$), one can show that the dominating spectra will come from those modes which have just narrowly escaped the black hole, i.e. which were scattered just before the formation of the event horizon. Such modes are the ones with high frequencies at the past null infinity. So the calculations done with \ref{BT} will be accurate to the leading order. If a field has a configuration which is described as vacuum on ${\cal J}^{-}$, then the state of the field will be obtained on ${\cal J}^{+}$ as populated, primarily because all the quantum correlations on ${\cal J}^{-}$ which contain the information that the field is in vacuum 
configurations in a flat region of the spacetime are unable to land up 
completely on ${\cal J}^{+}$. The population on ${\cal J}^{+}$ is obtained from \cite{Hawking:1974sw,Birrell:1982ix,Padmanabhan:2009vy,Parker:2009uva}
\bea
N_{\Omega} = \int d \omega | \beta_{\Omega \omega}|^2 = \frac{1}{e^{4M \omega} -1} \delta (0), \label{BHR}
\eea
which predicts a thermal profile with a temperature $\sim 1/4M$. The above \ref{BHR} also reflects the change in the notion of vacuum once the geometry evolves dynamically. The {\it notion of the vacuum} before and after the formation of the black hole becomes different (remember, the state remains the same, as in Heisenberg picture, which we use here, it does not evolve), and the state which was declared as a vacuum before the collapse, appears as an occupied state when the collapse is completed and the {\it notion of vacuum is awarded to some other state}. Quantum field theoretically, in-vacuum is different from the out-vacuum.

Thus, presence of horizon (which masks a portion of initial profile) together with the optical ray tracing struck gold : a black hole indeed radiates thermally (see \ref{BHR}). This was huge relief for the laws of black hole thermodynamics which were predicting the black holes to radiate!
\begin{itemize}
\item {\bf It's not  only about horizon: Hiding is not destroying}\vspace{0.5em}

Till now we have obtained that at the future times an observer will feel to be set up in thermal setting had she started with a vacuum state. But this really does not amount to saying the black holes do radiate. Masking off a portion of initial vacuum data can be brought about by any horizon, even by the Rindler horizon at the least (or the de Sitter horizon).  Therefore, even in Minkowski space time, such horizons can exist which mask off a portion of the vacuum field configuration from a specified family of observers. This does not amount to say that even Minkowski spacetimes radiate. They do not! There is no real flux of radiation in Minkowski spacetime, precisely because it is homogeneous, isotropic and time translational symmetric.\\

What the horizons are capable of, is to give rise to thermal environment for a special set of observers measuring real flux of radiation. The genesis of radiation is brought about by breaking some of the symmetries the spacetime possess, and this is the story of a black hole.\\
\item {\bf It's also about geometry non-trivia:}\vspace{0.5em}

The main difference a black hole has from the other horizon settings we commented above is that the black hole not only breaks the homogeneity due to curvature, it also breaks time reversal invariance in order to give rise to time irreversible phenomenon like radiation. In fact the $\delta(0)$ sitting in \ref{BHR} is understood to be total time, such that particle creation rate per unit time becomes exactly that of a radiating body. Thus, it is the breaking of these symmetries which allow the flux of energy radiated away to be non zero \cite{Chakraborty:2015nwa,Singh:2014paa} when calculated from $\langle T_{\mu \nu} \rangle$. Thus, non trivial geometry of the spacetime added with the presence of horizon makes the black hole radiate thermally (predominantly).\\ 

This is the point exactly where we make true contact with the thermodynamics of the black hole we kept believing in (even when a concrete proof was not available), and  this, unfortunately, is also exactly where the problem at hand starts unfolding.\\
\item {\bf ... and about finiteness of size:}\vspace{0.5em}

Now that we have credible theoretical understanding that the black holes leak out energy in thermal flux, we can think of the black hole losing mass in this way and evaporating away in time. We did not require anything extra than our belief in the quantum field theory on a regular manifold (however curved it may be). Therefore, unless the black hole becomes so small where the differential structure of spacetime is no longer applicable, we do not see any other reason to discredit or mistrust our calculations (for the case of primordial black holes, see \cite{Khlopov:2008qy,Khlopov:2004tn,Khlopov:1985jw}). This, then immediately leads us to expect that being compact and hence finite sized, the black holes {\it evaporate} away {\it practically completely}. Then there is a crisis at our hands!
\end{itemize}
\subsection{The paradox}

The black hole radiates thermally with the temperature decided by the mass it has. Therefore again the semiclassical radiation cares only about the classical charge a black hole carries. If we collect the total radiation the black hole emitted we will be able to track how the black hole lost its mass. But once the black hole is no more, apart from this information there is no way to get any other kind information out. Such as what were the different types of matter inside the black hole? How many particles were there? Every other detail should have been converted as the energy of the black hole got converted into radiation. But thermal character of resulting radiation forbids it to carry any such information (a thermal radiation is just characterized by its temperature, we do not learn anything more from its correlations). Thus the thermal character of the black hole while rescuing the laws of thermodynamics seemingly destroys something very sacred: the conservation of information.

An easy way out is to think that if thermality is the problem let us do something to the radiation such that it does not remain exactly thermal. Indeed there are many sources which can bring about departure of the resulting radiation from being exactly thermal \cite{Visser:2014ypa} (however also see \cite{Gray:2015pma,Hod:2000kb,Hod:2000it,Chakraborty:2016fye,Lochan:2015bha,Saini:2015dea}). Unfortunately though, most of such corrections again only do care about the classical parameters a black hole is parameterized with, and there are only very few of them. Thus resulting radiation which may well be non-thermal, solely depends on the classical charges of the hole and the story of information loss follows exactly in the same way as outlined above. In fact, with more elaborate understanding, it increasingly looks like that the issue is probably not about thermality (or the lack of it) of the outgoing radiation, but more about how we implement quantum theory in this setting. We need not only make the radiation 
non thermal, 
we need to make it non thermal in very sophisticated manner.

In quantum theoretical terms if the outgoing radiation in the exterior really does not care about the microscopic details of the interior then we can think the full Hilbert space to be separable into those supported on exterior and interior. Thus, if we have access to exterior radiation only, then effective description becomes that of a mixed state. This can be viewed in the following manner as well. One of the basic tenet of \gr\ is the Equivalence principle, which states that the spacetime near the horizon can be viewed locally as Minkowski spacetime. Therefore very close to horizon the Rindler description of the spacetime can be adopted. Therefore the horizon hides correlations in the same fashion as a Rindler horizon does. Unless the correlations across the horizon are turned off, the description on the outside will remain mixed. {\it Thus, the Equivalence Principle also necessitates a mixed state description for the exterior.}

As long as the black hole is present, the interior of the black hole has the part of correlations, which added together with the mixed state description outside, lets us know of the full information. Thus the complete state in the full Hilbert space is a pure state. However, we keep trusting this structure of information distribution and suddenly, late in time, we realize that the black hole is no more there, to support the missing piece of information from the outside. It has evaporated away and we are left with a mixed state description only. This is vehemently against the principles of unitarity. {\it A pure state can not evolve into a mixed state.} Now this description seems true for any pure state. So where were we wronged? Was the effective description in the outside not mixed? Has believing in the Equivalence Principle tricked us into confrontation with unitarity of quantum theory? Or, the full Hilbert space was really not as nicely separable into exterior and interior as we thought? Or, the 
unitarity which we have always sworn by, is not really a true law of nature to be respected, particularly in these settings? More sophisticated treatments of the problem have led us to explore in all possible directions mentioned above and many more. Rigorous research in these arenas has really moved the sphere of debate literally from thermality (or the lack of it) of the Hawking radiation, to scrutiny of the concepts in quantum information theory as well as to the notion of geometry at micro scales. Existence of equivalence principle, quantum unitarity and even black holes have all been challenged, in the course. We will review some of the leading discussions of the current time and the insights they bring about the structure of theories involved, if existence of black holes is accepted.   
\section{Hiding information in correlations}

As we discussed above, the thermal radiation the black hole seems emitting mostly throughout its life, a very satisfying feature from the black hole thermodynamics point of view, turns out to be rather unsettling from the information theory point of view. If black hole dies, emitting thermal radiation, where did all the information go? And if it radiated non thermally, was that non thermality sufficient to know all about what was on the other side of the horizon?

It was proposed right after, that all such information basically reside in the correlation between the black hole interior and the Hawking radiation. So we should not freak out witnessing a thermal radiation from the black hole. {\it Black hole, of course was not the first thing which we found radiates thermally. If we did not worry for a thermal blackbody why should we be alarmed by the black hole?} Like any other blackbody, there will be a time after which the correlations start leaking out into the exterior. We just have to wait long enough to catch them. In this section we will elaborate upon these schemes. We soon realize that unlike any other blackbody, the black hole comes dressed with an event horizon, a causal separation surface, and this is the reason of undoing the applicability of the concepts, which govern the black bodies, on black holes.
\subsection{The black hole history} \label{PageCurve}

The thermal character of the Hawking radiation was a problem in itself and some further calculations showed that there are no appreciable correlations which can support the full information. In his seminal work \cite{Page:1993wv} Page argued that absence of any appreciable correlations is also there for a perfectly unitary evolution (e.g. a ``pure state'' coal burning) for initial times. He went on to estimate the time before the information about the black hole interior starts leaking out to the exterior. 

The black hole and the radiation in the exterior is thought of as subsystems of the full system. The state which describes the black hole and the radiation, together is described by a pure state in the Hilbert space. The information in the individual subsystems (i.e., the black hole interior and/or the radiation in the exterior) is quantified by the deviation of the {\it averaged entanglement entropy} from being the maximal entropy as allowed by the size of the Hilbert space of the subsystem. For example, if the black hole interior (A), with $\textrm{dim}(A) =m$, and black hole exterior (B), with $\textrm{dim}(B) =n$, are subsystems of a composite system (AB) with Hilbert space dimension $\textrm{dim}(AB)=mn$, the information content in the either subsystem is defined as
\begin{eqnarray}
I_A = \log{m} -\langle S_A^{\text{ent}}\rangle_{\Psi} \\
I_B = \log{n} -\langle S_B^{\text{ent}}\rangle_{\Psi},  \label{InfoinRad}
\end{eqnarray}
where $\log{d}$ is the maximal entropy $S_{\text{max}}^d$ of a system residing in the Hilbert space of dimension $d$ (when all the states in the Hilbert space are equally likely to occur for a given macroscopic description) and the entanglement entropy for a subsystem (e.g. $A$) is obtained from the full system description $\rho$ as
\begin{equation}
  S_A^{\text{ent}} = \textrm{Tr}_{B}\left(\rho\right).
\end{equation}
Since the combined system (AB) is in a pure state $S_{AB}^{\text{ent}}=0$. Also for the pure states $S_A^{\text{ent}}=S_B^{\text{ent}}$. The average $\langle...\rangle_{\Psi}$ is over the {\it full} Hilbert space, which is obtained by using a unitarily invariant Haar measure on the set of normalized state vectors $|\Psi \rangle$. Therefore, what we obtain is the {\it average information}\footnote{ In fact, for a large dimensional system such as a macroscopic black hole, behavior of a specific unitary process is very close to what an average over various unitaries will suggest : Levy's lemma.} in the exterior (or conversely in the interior) once the dimensionality of Hilbert spaces corresponding to $A$ and $B$ (i.e. $m$ and $n$) are specified. We must also note here that assuming the dimensionality of the combined system to be $mn$ we have intrinsically assumed that the interior and exterior Hilbert spaces are totally independent. That is to say, the full system's Hilbert space is factorizable in interior and 
exterior Hilbert spaces
\begin{equation}
\mathcal{H}_{AB} = \mathcal{H}_{A} \otimes \mathcal{H}_{B}.
\end{equation}
In the evaporating black hole scenario at the earlier phase of evaporation most of the degrees of freedom will lie in the interior of the black holes and very few degrees of freedom will be in the exterior. This will reflect into the their respective dimensionality of the Hilbert spaces as $m\gg n$. As the black holes progresses in the course of evaporation $m$ and $n$ will come close to each other and eventually the dimensionality of the exterior Hilbert space will grow larger than the interior Hilbert space.  Page estimated the information content for these scenarios and obtained for $m \geq n$ \cite{Page:1993wv}
\begin{eqnarray}
I_B = \log{n} + \frac{n-1}{2m} - \sum_{k=m+1}^{mn}\frac{1}{k} \label{InfoinRad1}
\end{eqnarray}
and specifically, when both the subsystems are macroscopic (which is the case for most of the time in the course of black hole evaporation) $m\geq n\gg1$,
\begin{equation}
I_B = \frac{n}{2m} \sim e^{ S_{\text{max}}^B-S_{\text{max}}^A}, \label{InfoinRad2}
\end{equation}
While the interior will have the information
\begin{eqnarray}
I_A = \log{m} + \frac{n-1}{2m} - \sum_{k=m+1}^{mn}\frac{1}{k}, \label{InfoinRem1}
\end{eqnarray}
which can be approximated in the regime $m\geq n\gg1 $, by
\begin{eqnarray}
I_A \sim \log{\frac{m}{n}} + \frac{n}{2m}.\label{InfoinRem2}
\end{eqnarray}
The above set of expressions provide remarkable insights about the average information content of the radiation or the surviving remnant of the black hole. \ref{InfoinRad2} tells that unless the radiation in the exterior gathers enough particles, the information content in their correlations, considered separately, is exponentially suppressed. Therefore, {\it until the time the radiation collects number of degrees of freedom such that the dimension of the Hilbert spaces of the radiation and the remaining black hole are of the same order (such a time is known as the {\it Page time}), there is virtually no information in the radiation.} Secondly, till the time, the radiation has enough quanta, the information in it is practically zero, suggesting (see \ref{InfoinRad}) that the entanglement entropy of radiation trades very close to its maximal value.

Once the radiation takes over the remaining black hole, the correlations in the exterior start becoming rich in information, with the expression given by \ref{InfoinRem1} adapted to $B$ with $m\leftrightarrow n$. That is to say it is only  after the {\it Page time} the black hole starts leaking out information in the radiation. Also an important insight one can gain is that 
under such a factorization of the Hilbert space the information really does not get factorized. If we add \ref{InfoinRad2} and \ref{InfoinRem2}, we  obtain the sum of the information in the subsystems' individual correlations to be $\log{\frac{m}{n}}+\frac{n}{m}$ while the full system had an information content $mn\log{mn}$ (since $S_{AB}^{\text{ent}}=0$). Since we have already invested our trust into unitary physics at the moment (in this section), we better expect that the information must not be destroyed in this factorization. That promptly suggests that the remaining $\log{n^2}-\frac{n}{m}$ units of information should be available with the correlations between subsystems $A$ and $B$, i.e., the black hole interior and the exterior radiation. Now, this is the point on which the causal divide of event horizon, added with the celebrated no-hair behavior of black holes, hit upon and create a problem with this justifiable expectation.
\subsection{No hiding theorem --- No information in correlations between separable Hilbert spaces}

Braunstein {\it et. al. }  \cite{Braunstein:2006sj} proved an important result that if we treat the states of the in-fallen matter quantum mechanically then the black-hole information paradox becomes more severe. If the Hawking radiation is described by only the classical charges (aka the local geometry) and is totally insensitive to other information which get deposited inside the hole, the radiation can be described by a quantum state in a Hilbert space which is totally independent of the interior states, then it is possible to rule out that the information of the in-falling matter might be hidden in correlations between semi-classical Hawking radiation and the internal states of the black hole. For such a coding to happen, either unitarity or Hawking's semi-classical predictions must break down \cite{Braunstein:2005zz,Braunstein:2006sj,Braunstein:2009my,Bose:1996pi}. This becomes apparent with the following line of arguments they proposed.

Let us consider a process which takes an arbitrary input state $\rho_I$ from subspace $I$ and maps it into a larger Hilbert space. This operation will correspond to a  hiding process if there exists some (say, $K$-dimensional) subspace $O$ (the output) in the larger Hilbert space, whose state $\sigma_O$ has no dependence on the input state. Thus, any observer concentrating (or living)  only  on $O$ will not have any clue of the input state and the mapping process would have successfully hidden the information about the state being mapped, from a subspace of the full Hilbert space \footnote{We think of the Hawking radiation as such a mapping}. In other words, this process maps $\rho_I \rightarrow \sigma_O$ with $\sigma_O$ fixed for all $\rho$. The remainder of the encoded Hilbert space may be regarded as an ancilla $A$ \footnote{Ancilla : A (foreknown) quantum state which tags along the main state of study}. Thus, the entire system may be represented in terms of two subsystems $O$ and $A$. The expectation 
from a standard unitary 
quantum theory 
of black hole evaporation demands this process to be governed by a linear and unitary transformation. By linearity we expect that it is sufficient to study the action on an arbitrary pure state $\rho_I = |\psi\rangle_I  {}_I\langle \psi |$. Unitarity tells us that the ancilla space must be large enough so that the hiding process can be represented as a mapping from pure states to pure states, 
\begin{equation}
|\psi\rangle_I \rightarrow \sum_k p_k |k \rangle_O \otimes |A_k(\psi)\rangle_A.\label{FactoredAncill1}
\end{equation}
where $|k \rangle_O$ and $|A_k(\psi)\rangle_A $ are the orthogonal set of states belonging to the system (Hawking radiation) and the Ancilla (interior). For $d-$ orthogonal states $|\psi_j\rangle$ (corresponding to the $d-$ dimensional input state space) , $ |A_k(\psi_j)\rangle$ should provide an orthonormal basis in Ancilla space as well \cite{Braunstein:2006sj}, thereby making it a vector in $Kd-$ dimensional Hilbert space. However,  since $(|q_k\rangle \otimes |\psi_j\rangle \oplus 0)_A $ also provides a good orthonormal basis in the Ancilla space, if $|q_k\rangle$ provide an orthonormal basis in the $K-$ dimensional subspace of the Ancilla space, $|A_k(\psi)\rangle_A$ must be related to this basis by unitary transformation. Using arguments of linearity and  equivalence of basis states under unitary transformation\footnote{In a similar spirit in which, position basis $|x \rangle$ and momenta basis $|p\rangle$ are unitarily related and equally justified for decomposing a state into.}, it is evident that 
the 
above equation can always be written in the new basis as
\begin{equation}
 |\psi\rangle_I \rightarrow \sum_k p_k |k \rangle_O \otimes (|q_k\rangle \otimes |\psi\rangle \oplus 0)_A, \label{FactoredAncill2}
\end{equation}
with $\oplus 0$ depicting the other unused dimensions of the Ancilla, where the linearity (in $|\psi_j\rangle$) of the new basis is implemented .
Evidently, in \ref{FactoredAncill2} the Ancilla basis part itself is factorized into two parts: a part $|q_k\rangle$ which depends on $k$ and cares about the system through $p_k$ and a part which does not. The complete wave function itself becomes factorizable into $k-$ dependent and independent parts. Therefore, we see the state dependence has factored out in the Ancilla section and the system dependence is separable from the state           $|\psi\rangle _{I}$. The state does not really get entangled with what comes outside, in such a mapping. Hence any system-system correlation or system-Ancilla cross correlation will not contain any information about $|\psi\rangle _{I}$. 

So, this shows that it is too naive to expect both the no-hair theorem (which essentially suggests above-mentioned factorizability of the state in the Hilbert space) and the information rich correlations to be at peace. The road for a bipartite consideration seems very well to be over in unitary quantum framework. So, is black hole physics officially at war  with quantum concepts now on?

However, with more reflection of the lessons learnt it turns out that {\it it  may probably still be possible to hide information in the correlations of a tripartite system.} So, we should probably view the configuration of  a radiating black hole as a tripartite system. This idea was indeed quickly sold. After all, Page suggested that the radiation will become information rich {\it after the Page time} and not before that.
So the radiation itself gets divided into two categories: early radiation (which is thermal and information free) and late time radiation (which is non-thermal and rich in information, see \ref{fig_Paradox_Intro}). So the interior along with the exterior early time and exterior late time radiation as a tripartite system can hold information in correlation.

Unfortunately, it turns out this is the point where the event horizon delivers a causative blow to this idea, making the problem more sharply defined. If early time and late time radiation, together have the full information, what happens to the information which was in the interior (and nothing from the inside ever came outside, unlike a burning coal)? Did we just prepare two copies of information of the matter which fell into the hole, ignoring the red flags of unitary quantum theory?

This aspect is more clearly understood and fought with, in the concept of black hole complementarity, which fundamentally asks the question, how much and how long to trust the no-cloning? The same problem can be posed in guise of principles of sub-additivity of a tripartite system. The peculiar casual structure of black holes seemingly enables us to violate the rules of sub-additivity. The tripartite system may also involve exotic possibilities such as a stable remnant, a late time flash etc. We discuss these ideas and those of black hole complementarity next.
\section{Black Hole Complementarity}

One of the most fascinating and oldest proposal towards resolving this version of information paradox, discussed above, within the context of unitary quantum mechanics, was by Susskind {\it et. al. } when they stumbled upon the idea of black hole complementarity \cite{Susskind:1993if}. They proposed  a very significant departure from the traditional notion of unitarity of quantum theory. The proposal, rightfully, has fueled a huge debate in the physics community, recently leading to a more information theoretic formulation of the paradox, known as the firewall puzzle \cite{Almheiri:2012rt}, which not only has sharpened the information paradox in a more modern language but has also acted as a catalyst for the birth of many new and promising directions  towards the resolution of the paradox. In this section, we will discuss the original proposal by Susskind as well as the firewall puzzle in detail. To start with consider the information paradox in a different setting, for which it is important to consider the 
Penrose diagram of an evaporating black hole as presented in \ref{fig_Complimentarity}. One can foliate the spacetime with space-like hypersurfaces, known as Cauchy surfaces. Before the black hole forms one such Cauchy surface would be $\Sigma$. After the formation of the event horizon, any Cauchy surface (e.g., $\Sigma _{P}$), can be subdivided into two disjoint surfaces: inside and outside, i.e., $\Sigma _{P}=\Sigma _{\rm out}\times \Sigma _{\rm bh}$. Similarly, after the black hole has evaporated away Cauchy surfaces will be represented as $\Sigma '$ (see \ref{fig_Complimentarity}).
\subsection{Posing The Problem}

\begin{figure}[t]
\sidecaption[t]
        \begin{tikzpicture}[scale=1]   
            \draw[red!60,line width=1mm] (0,-6) -- (6,0) -- (2,4);
            \draw[red!60,line width=1mm] (2,4) -- (2,-0.5);
            \draw[red!60,line width=1mm] (0,-6) -- (0,-0.5);
            \draw[black,line width=0.01mm] (0,-6) -- (6,0) -- (2,4);
            \draw[black,line width=0.01mm] (2,4) -- (2,-0.5);
            \draw[black,line width=0.01mm] (0,-6) -- (0,-0.5);
            \draw[green!80,line width=0.8mm,snake it] (0,-0.5) -- (2,-0.5); 
            \draw[black!80,line width=0.08mm,snake it] (0,-0.5) -- (2,-0.5);            
            \node[label=right:$\mathcal{J}^{-}$] (4) at (3.5,-2.5) {};
            \node[label=left:\textrm{r=0}] (5) at (0,-2.5) {};
            \node[label=left:\textrm{singularity}] (7) at (2,0) {};
            \node[label=left:\textrm{r=0}] (8) at (2,1.5) {};
            \node[label=right:$\mathcal{J}^{+}$] (9) at (4,2) {};
            \node[label=above:$i^{+}$] at (2,4) {};
            \node[label=below:$i^{-}$] at (0,-6) {};
	    \node[label=below:$i_{0}$] at (6.2,0.2) {};
            \draw[blue!90,dashed,line width=0.5mm,bend right=15] (6,0) to (2,1.5);
            \draw[blue!90,dashed,line width=0.5mm,bend right=15] (6,0) to (0,-1.4);
            \draw[blue!90,dashed,line width=0.5mm,bend right=15] (6,0) to (0,-3.8);
            \draw[black!90,dashdotted,line width=0.5mm,] (2,-0.5) to (0,-2.2);
            \draw[->,bend right=35] (-2,-1) to (0.8,-1.4);        
            \node[label=above: $\Sigma'$] at (3,1.2) {};
            \node[label=below: $\Sigma_{\rm out}$] at (3,0.5) {};
            \node[label=below: $\Sigma$] at (3,-0.6) {};
            \node[label=below: $\Sigma_{\rm bh}$] at (0.4,-0.4) {};
            \node[label=below: \textrm{event horizon}] at (-2,-0.4) {};
            \node[label=below: \textrm{P}] at (2.1,-0.5) {};        
        \end{tikzpicture}
    \caption{Penrose Diagram for black hole evaporation. The singularity is depicted by green curve. On the other hand the black dashed dotted curved represents the event horizon and the violet dotted curves stands for the Cauchy surfaces in the space time. Among others $\mathcal{J}^{-}$ is the past and $\mathcal{J}^{+}$ is the future null infinity, while $i^{\pm}$ are past and future timelike infinity and $i^{0}$ is the spatial infinity. Further discussions can be found in the main text.}
    \label{fig_Complimentarity}
\end{figure}
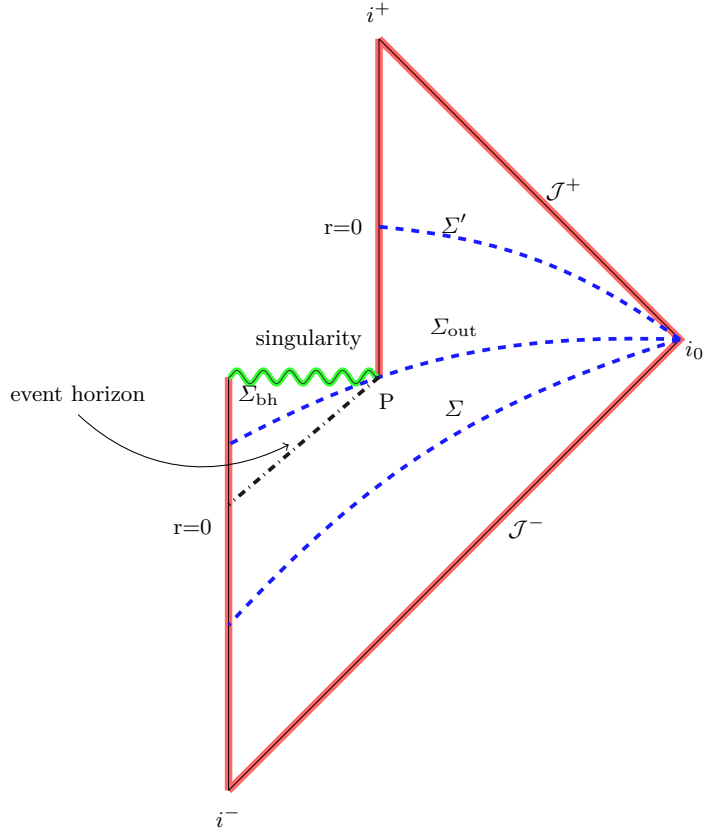

For this discussion we again assume unitarity and linearity of quantum theory, i.e., a linear Schr\"{o}dinger equation derived from local quantum mechanics, evolves a quantum state on one Cauchy surface to another. Let such a quantum state on $\Sigma$ be given by, $|\psi (\Sigma)\rangle$, whose evolution will hit the singularity as the surface $\Sigma _{P}$ is reached  (see \ref{fig_Complimentarity}). Since $\Sigma _{\rm bh}$ and $\Sigma _{\rm out}$ are causally disconnected (from the exterior observer's point of view the future domain of dependence of $\Sigma _{\rm out}$ does not care for $\Sigma _{\rm bh}$), one can write the Hilbert space $\mathcal{H}$ on $\Sigma _{P}$ as a direct product of Hilbert spaces $\mathcal{H}_{\rm bh}$ and $\mathcal{H}_{\rm out}$ corresponding to $\Sigma _{\rm bh}$ and $\Sigma _{\rm out}$ respectively, i.e., $\mathcal{H}=\mathcal{H}_{\rm bh}\otimes \mathcal{H}_{\rm out}$.

On the other hand, after the black hole has evaporated, on the Cauchy surface $\Sigma'$ the quantum state would be $|\psi (\Sigma ')\rangle$. If unitarity exists, $|\psi (\Sigma ')\rangle$ must be a pure state, such that,
\begin{equation}\label{Eq01_Unitary}
|\psi (\Sigma ')\rangle =U|\psi (\Sigma)\rangle
\end{equation}
where $U$ is an unitary operator enforcing evolution. However, since $\Sigma '$ is causally connected to $\Sigma _{\rm out}$ only, unitary evolution suggests that, 
\begin{equation}\label{Eq02_Unitary}
|\psi (\Sigma ')\rangle =\tilde{U}|\chi (\Sigma _{\rm out})\rangle
\end{equation}
where, $\tilde{U}$ corresponds to the relevant unitary evolution operator from $\Sigma _{\rm out}$ to $\Sigma'$ and $|\chi (\Sigma _{\rm out})\rangle \in \mathcal{H}_{\rm out}$ must be a pure state. Hence, $|\psi (\Sigma _{P})\rangle$ must be a product state,
\begin{equation}
|\psi (\Sigma _{P})\rangle=|\phi(\Sigma _{\rm bh})\rangle \otimes |\chi (\Sigma _{\rm out})\rangle
\end{equation}
where $|\phi(\Sigma _{\rm bh})\rangle$ is a state in $\Sigma _{\rm bh}$ and the state $|\psi (\Sigma _{P})\rangle$ must originate due to linear evolution from $|\psi (\Sigma)\rangle$. However, from \ref{Eq01_Unitary} and \ref{Eq02_Unitary} it is evident that $|\chi (\Sigma _{\rm out})\rangle$ also depends linearly on $|\psi (\Sigma)\rangle$. Now, that leaves the state $|\phi(\Sigma _{\rm bh})\rangle$ to have no dependence on $|\psi (\Sigma)\rangle$ at all (otherwise $|\psi (\Sigma _{P})\rangle$ will not depend linearly on $|\psi (\Sigma)\rangle$). This is a drastic conclusion. The states inside the horizon somehow get rid of all dependence of what forms the black hole! This suggests something peculiar happens at the event horizon, violating the equivalence principle. {\it If we wish to save the equivalence principle, then we can not have a pure state $|\chi (\Sigma _{\rm out})\rangle$ on $\Sigma _{\rm out}$.} Thus the state on $\Sigma _{\rm out}$ has to be mixed, for the sake of equivalence 
(and hence so is the state on $\Sigma'$, violating unitarity!).
\subsection{A Possible Resolution}

There possibly exists one way to evade this conclusion, which corresponds to black hole complementarity \cite{Susskind:1993if}. The assumption that $|\psi (\Sigma _{P})\rangle$ describes both the interior and exterior of black hole spacetime requires correlations between inside of event horizon and outside. However since there is no causal connection between the two, these correlations have no operational meaning, i.e., no observations can detect these correlations as far as the exterior observers are concerned. As an observer gets into the horizon, (s)he can not communicate the findings to the outside colleague. Thus there exists no observer in the spacetime who can measure, the full $|\psi (\Sigma _{P})\rangle$, they can measure either $|\chi (\Sigma _{\rm out})\rangle$ or $|\phi(\Sigma _{\rm bh})\rangle$ but not both. 

\begin{figure}[t]
\begin{center}
\begin{tikzpicture}
    \draw (-1.5,0) circle (1.5cm);
    \shade[ball color=blue!30!white,opacity=0.80] (-1.5,0) circle (1.5cm);
    \draw[->] (0.2,0) to (2.5,0);  
    \draw (3.5,0) circle (0.5cm);
    \draw (3.5,0) circle (0.6cm);
    \shade[ball color=blue!10!white,opacity=0.80] (3.5,0) circle (0.6cm);
    \shade[ball color=black!60!white,opacity=0.80] (3.5,0) circle (0.5cm);
    \draw[->] (4.5,0) to (7.5,0);
    \draw (8,0) circle (0.2cm);
    \draw (8,0) circle (0.3cm);
    \shade[ball color=blue!30!white,opacity=0.60] (8,0) circle (0.3cm);
    \shade[ball color=black!60!white,opacity=0.80] (8,0) circle (0.2cm);
    \draw[-stealth,decoration={snake,amplitude =0.4mm,segment length = 2mm, post length=0.9mm},decorate,color=red, line width=0.25mm] (7.6,0.4) -- (7.0,1.0);
    \draw[-stealth,decoration={snake,amplitude =0.4mm,segment length = 2mm, post length=0.9mm},decorate,color=green, line width=0.25mm] (8.4,0.4) -- (9.0,1.0);
    \draw[-stealth,decoration={snake,amplitude =0.4mm,segment length = 2mm, post length=0.9mm},decorate,color=brown, line width=0.25mm] (8.4,-0.4) -- (9.0,-1.0);
    \draw[-stealth,decoration={snake,amplitude =0.4mm,segment length = 2mm, post length=0.9mm},decorate,color=cyan, line width=0.25mm] (7.6,-0.4) -- (7.0,-1.0);
    \draw[-stealth,decoration={snake,amplitude =0.4mm,segment length = 2mm, post length=0.9mm},decorate,color=cyan, line width=0.25mm] (5.6,2.4) -- (5.0,3.0);
    \draw[-stealth,decoration={snake,amplitude =0.4mm,segment length = 2mm, post length=0.9mm},decorate,color=red, line width=0.25mm] (10.4,2.4) -- (11.0,3.0);
    \draw[-stealth,decoration={snake,amplitude =0.4mm,segment length = 2mm, post length=0.9mm},decorate,color=green, line width=0.25mm] (5.6,-2.4) -- (5.0,-3.0);
    \draw[-stealth,decoration={snake,amplitude =0.4mm,segment length = 2mm, post length=0.9mm},decorate,color=brown, line width=0.25mm] (10.4,-2.4) -- (11.0,-3.0);
    \node[label=below: \textrm{star with radius} $r_{0}>2M$] at (-1,-2) {};
    \node[label=above: \textrm{collapses}] at (1.2,0.1) {};
    \node[label=above: \textrm{event horizon}] at (3.3,1) {};
    \draw[->,bend right=30] (3.3,1) to (3.3,0.4);
    \node[label=below: \textrm{stretched horizon}] at (3.3,-1) {};
    \draw[->,bend left=30] (3.3,-1) to (3.3,-0.6);
    \node[label=above: \textrm{radiates}] at (6.0,0.1) {};
    \node[label=below: \textrm{early Hawking radiation}] at (8.0,3) {};
    \draw[->,bend right=30] (7.7,3.2) to (5.3,3.0);
    \draw[->,bend left=30] (7.8,3.2) to (10.2,3.0);
    \node[label=below: \textrm{late Hawking radiation}] at (8.0,-1) {};
    \draw[->,bend right=30] (7.75,-1) to (7.5,-0.8);
    \draw[->,bend left=30] (8.0,-1) to (8.25,-0.8);
\end{tikzpicture}
\end{center}
\caption{A schematic of the complementarity hypothesis. When a star collapses to form a black hole, a stretched horizon very near to the actual event horizon is formed as far the outside observers are concerned and encapsulates all the information that has fallen in the black hole. This information emerges in the Hawking radiation by non-trivial correlations between early and late Hawking radiation (as depicted in the figure, the Hawking quanta with same color are correlated.) On the other hand, for an infalling observer, there is no stretched horizon, the observer can enter the black hole event horizon and in principle recover all the information. But she has no way to communicate the same with her outside colleague and hence these two descriptions are disjoint and does not lead to cloning. More discussions can be found in the main text.}\label{fig_03}
\end{figure}
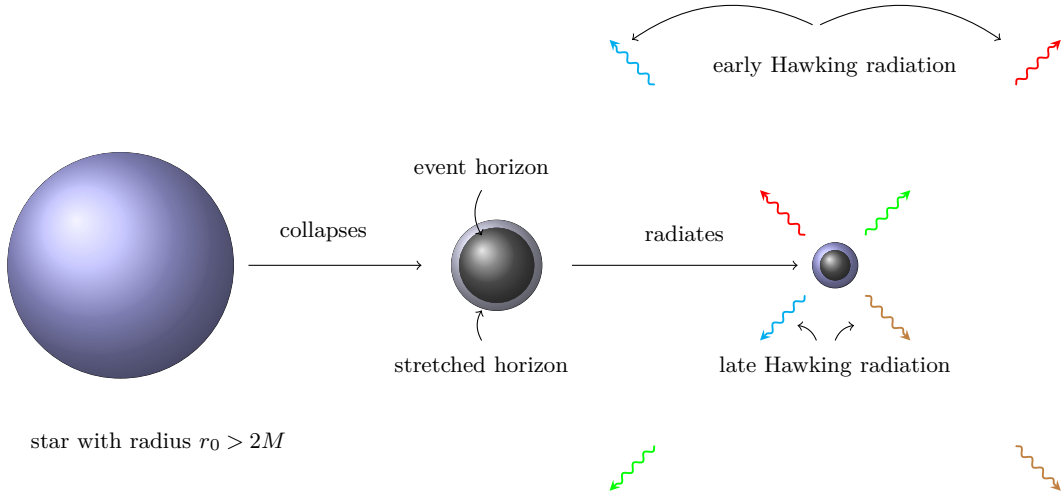

Thus one postulates that to an outside, distant observer, the black hole horizon will behave as a stretched horizon, a horizon with an infinitesimal thickness, with thermodynamic properties. For example, it will have temperature, will have Navier-Stokes equation valid near it having negative bulk viscosity. Note that the idea of stretched horizon has been introduced much earlier in the context of membrane paradigm for black holes \cite{Price:1986yy,Parikh:1998mg}. This stretched horizon encodes all the information that has fallen in the black hole. On the other hand, the infalling observer will not encounter any stretched horizon, but there is no way to communicate these results to the outside world. Thus the static and infalling observers are complementary to one another (see \ref{fig_03}). 

In this picture, there is no information loss, the information is returned to the outside observer through long-time non-thermal correlations between quanta emitted at very different times. The following three postulates are being made:\\
\begin{itemize}

\item Formation and evaporation of black hole, viewed by a distant observer can be described entirely using unitary quantum mechanics.\\

\item Outside stretched horizon, semi-classical physics holds good.\\

\item To a distant observer, the black hole is a quantum system with discrete energy levels and number of states being exponential of Bekenstein entropy. 

\end{itemize}

\subsection{Problem with these postulates: Firewall?}

Everything about the proposal of complementarity rings sound and healthy. However, there also exists a particular problem with these three postulates, a realization that provoked an idea of firewall \cite{Braunstein:2009my, Almheiri:2012rt} at the horizon. Let us suppose a black hole has been formed from a pure state and has evaporated partially, till a certain time which we denote by $t=0$. Let $R$ be the early Hawking radiation, i.e., emitted during $t<0$ and let $R'$ be the final Hawking radiation, emitted afterwards. Let $|\psi\rangle$ be the initial wave function that collapses and forms a black hole, while the final state being $|R\rangle \otimes |R'\rangle$. By unitarity one must have,
\begin{equation}
|\psi'\rangle \equiv|R\rangle \otimes |R'\rangle=U|\psi \rangle
\end{equation}
where, $U$ is an unitary matrix. It is assumed that: $R$ is a larger subsystem, i.e., the dimension of the Hilbert space associated with $R$, $\textrm{dim}\mathcal{H}_{R}$ is larger compared to that of $R'$, $\textrm{dim}\mathcal{H}_{R'}$. As we saw previously in \ref{PageCurve} that in a smaller subsystem of a pure state, the entanglement entropy is (almost) maximal. Thus,
\begin{equation}
\text{If } S_{RR'}=0;\qquad S_{R'}= -\textrm{Tr} \rho_{R'}\ln \rho_{R'}\sim\ln \textrm{dim}\mathcal{H}_{R'}
\end{equation}
where $\rho _{R'}$ is obtained by a partial tracing over $\mathcal{H}_{R}$. Therefore, at $t=0$, one has the early Hawking radiation $R$ and remnant of the hole, maximally entangled with each other. 

Complementarity distinguishes an outside static observer `Bob' with an infalling observer `Alice'. Under this proposal, both of these observers have a consistent description of physics, without violating unitarity or equivalence principle, until they can exchange information. According to Bob, the late time Hawking radiation originates from regions near the horizon which, consists of two subsystems --- (a) the stretched horizon `H' and (b) the near horizon zone `B', such that,
\begin{equation}
|R'\rangle =|B\rangle \otimes |H\rangle
\end{equation}
Note that Bob can probe `B' without experiencing acceleration of the Planck scale order, while `H' can not be probed by him, without going to Planck scales. On the other hand, local physics being inertial, Alice will observe the near horizon regime\footnote{By near horizon regime we imply a spacetime region around every point on the event horizon, where one may neglect the effect of curvature and treat the spacetime to be flat.} to be Minkowski vacuum and the horizon as a Rindler horizon. Then `B' will lie in the right Rindler wedge for Alice\footnote{Given a spacetime point, one can always expand any metric around the Minkowski background with additional curvature corrections introducing an associated length scale into the problem. If one is interested in physics beyond this length scale, the notion of local inertial physics is no longer applicable. The region $B$ is thought to be living within this length scale.}. Since fields having support in the local 
Rindler wedges are  maximally entangled if the quantum state coincides with the local Minkowski vacuum, Alice will find the black hole interior `A' to be maximally entangled with `B'. Moreover, since the Minkowski vacuum is a pure state, one further has,
\begin{equation}\label{Eq01_Comp}
S_{AB}=0;\qquad S_{B}\sim \ln ~\textrm{dim} \mathcal{H}_{B}
\end{equation}
Now $B\subset R'$, as observed by Bob, i.e., subset of late Hawking radiation. There is also a subsystem $R_{BH}\subset R$  of the early Hawking radiation, that purifies $R'$. Moreover, there exists another system $R_{B}\subset R_{BH}$, which will purify the near horizon region $B$, implying
\begin{equation}\label{Eq02_Comp}
S_{R_{B}B}=0
\end{equation}
Strong sub-additivity of entropy ensures that a system can not be maximally entangled with two distinct other systems. Thus \ref{Eq01_Comp} and \ref{Eq02_Comp} are incompatible with one another. Hence follows the paradox. One need to abandon equivalence principle. Note that this is a much revised and sharpened version of the original firewall proposal \cite{Almheiri:2012rt,Bousso:2012as,Avery:2012tf,Chowdhury:2012vd,Mathur:2013gua,Almheiri:2013hfa,Lee:2013vga,
Harlow:2013tf,Bousso:2013wia,Papadodimas:2012aq,Nomura:2012cx,Susskind:2012uw,Brustein:2012jn,Giveon:2012kp,Shenker:2013pqa} (however also see \cite{Hooft:2016vug}).

The above demonstrates a somewhat mathematical exposition of the problem. It always benefits to understand the same from a more physical viewpoint, on which we elaborate below. According to complementarity, theories of Bob and Alice will both be based on unitary quantum mechanics as well as semi-classical gravity and they only have to match on observations that are causally connected. For example, when Alice enters the stretched horizon regime `H' there is no option left for her \footnote{one requires an Planck scale energy to come out of `H'} other than to go inside the horizon. Hence experiences of Alice for the region $A\cup H$ can not be communicated to Bob. Hence the theory of Alice may be inconsistent with that of Bob and a combination of both to a global picture can lead to contradictions. What is crucial for the unitarity to hold, is that Bob must see the Hawking radiation as pure, while Alice will see the pure state within event horizon. But this does not violate no-cloning theorem as, none of the 
observers can see both the copies. 

It is instructive to see whether the same applies to Firewall paradox as well. Both Bob and Alice have equal access to early Hawking radiation $R$. Thus it might happen that only after detecting the early Hawking radiation, Alice decides to jump into the black hole. Since Bob has access to $R'$ he can find out that a subsystem of $R'$ ($B$ in particular) is being purified by a subsystem of $R$ ($R_{B}$ to be precise) and hence can verify \ref{Eq02_Comp}. Since he has no hint about black hole interior $A$ he can not verify \ref{Eq01_Unitary} and hence no contradiction. 

While Alice after measuring $R$ jumps into the black hole. Thus she has no access to $R'$ and can not verify $S_{RR'}=0$. However she can verify \ref{Eq01_Comp}, since she can measure the Minkowski vacuum and hence infer that Rindler wedges $A$ and $B$ are maximally entangled. But she has no way to verify \ref{Eq02_Comp} and hence no contradiction. For this to be valid one has to ensure that it is impossible for Alice to measure `B' before crossing `H'. If she can measure `B' before entering `H' she will know about a subsystem of $R'$ and then fire rockets and her observation must match with that of Bob. But then Alice has came to know both \ref{Eq01_Comp} as well as \ref{Eq02_Comp} leading to contradiction. Thus if Alice can have information about `B' before entering `H' there is no way to save the complementarity postulates. 

It further turns out that the arguments coined above are reasonably robust and do not require the systems to really be maximally entangled, strong entanglement between the systems suffice \cite{Bousso:2012as,Bousso:2013wia}. Let $X$ and $Y$ be two subsystems of a pure system $XY$. Then purity ensures that,
\begin{equation}\label{Eq_Ent_01}
S_{XY}=0;\qquad S_{Y}>0
\end{equation}
where the last condition ensures that the systems are \emph{not} maximally entangled. Then the fact that a third system $Z$, completely different from $X$, i.e., $X\cap Z=\phi$ is also entangled with $Y$ can not happen. Since in that case $S_{YZ}=0$ and from the strong sub-additivity of entropy 
\begin{equation}
S_{XYZ}+S_{Y}\leq S_{XY}+S_{YZ}
\end{equation}
it follows that, $S_{XYZ}<0$, which is against the positivity of entropy. Now if one identifies $X=$ inside regime of event horizon, $Y$ as the near horizon region including the stretched horizon, such that $XY$ together is in a vacuum configuration, then $Z=$ early Hawking radiation can not form a pure state with $Y$ without violating unitarity. On the other hand, if $YZ$ forms a pure system, unitarity would be restored but then $XY$ pair has to break the correlations accordingly and that can not be done with a vacuum state and hence equivalence principle has to be abandoned. {\it The cost of breaking such correlations is a highly excited configuration at the horizon} (see the supplement of \cite{Braunstein:2009my} for a more clear elaboration). This is the advent of the {\it proposal of firewall} from a mathematical side. Note that this does not require the systems to be maximally entangled, for maximal entanglement would even destroy classical correlations (see \cite{Bousso:2012as,Bousso:2013wia} for a 
detailed discussion). 
\subsection{The Role of In-falling Vacuum}

One can also try to solve the problem by postulating that the system $X$ and the system $Z$ are equivalent, i.e., the Hilbert space of the early Hawking radiation is same as the Hilbert space belonging to the interior to the hole. Then one would obtain,
\begin{equation}\label{Eq_Infalling}
X\otimes Y=Z\otimes Y\neq X\otimes Y\otimes Z
\end{equation}
The xeroxing problem does not hit us, since the system $Z$ is visible to Bob (i.e., the static observers outside the hole) but $X$ is not visible to him. On the other hand $X$ is visible to Alice (the in-falling observer in a suicide mission) but she has no way to communicate with $Z$ after realizing the full Hilbert space of $X$. Hence no two observers observe both the patches. 

However by equivalence principle one demands that the Hilbert space $X\otimes Y$ consists of a single state, the Minkowski vacuum $|0\rangle _{M}$, which is unique. On the other hand, the Hilbert space $Z\otimes Y$ contains the out state $|\psi \rangle$ but it can accommodate many different states depending on various initial configurations of matter that collapsed and formed the black hole. But then how is the equality in \ref{Eq_Infalling} maintained? This being another version of the firewall puzzle.
\subsection{The Problem with Projection Operator: Can firewall exist?}

Having discussed a somewhat different and sharpened version of the firewall paradox we will now discuss what the original firewall proposal due to \cite{Almheiri:2012rt} actually argued for. After the black hole has emitted half of its Bekenstein-Hawking entropy, the system of old black hole with Hawking radiation can be written as,
\begin{equation}
|\Psi \rangle =\sum _{i}c_{i}|\psi _{i}\rangle \otimes |i\rangle_L
\end{equation}
where $|\psi _{i}\rangle$ is a state describing the early Hawking radiation, whose Hilbert space is $\mathcal{H}_{\rm rad}$. Similarly, $|i\rangle$ is a state describing the remnant black hole with Hilbert space $\mathcal{H}_{\rm bh}$. This remnant ultimately gets converted to late time radiation $|i\rangle_L$. Since the black hole has crossed the Page time, one obtains $\textrm{dim}\mathcal{H}_{\rm rad}\gg \textrm{dim}\mathcal{H}_{\rm bh}$. Note that this ensures the entanglement, when the black hole interior is $|i\rangle$ then early Hawking radiation must be $|\psi _{i} \rangle$. Signaling the presence of maximal entanglement. Then one can expect the states $|\psi _{i}\rangle$ to almost span the full Hilbert space of early radiation and provide an orthonormal basis, i.e., $\langle \psi _{i}|\psi _{j}\rangle =\delta _{ij}$.

We can also introduce a projection operator $P_{i}=|\psi _{i}\rangle \langle \psi _{i}|$, which projects state vectors on the Hilbert space of early radiation $\mathcal{H}_{\rm rad}$. Thus application of the same on $|\Psi\rangle$ yields,
\begin{equation}\label{Eq_Class_01}
P_{i}|\Psi\rangle =\sum _{j}c_{j}|\psi _{i}\rangle \langle \psi _{i}|\psi _{j}\rangle \otimes |j\rangle _L
=c_{i}|\psi _{i}\rangle \otimes |i\rangle_L.
\end{equation}
Both Bob (the distant observer) and Alice (the infalling observer) have causal access to the early Hawking radiation and hence they can access $|\psi _{i}\rangle$ and do measurements on that Hilbert subspace. For Bob, the remaining degrees of freedom are located near the stretched horizon which are emitted as the late time Hawking radiation and become accessible, whereas Alice will be able to measure the states in $\mathcal{H}_{\rm bh}$ exactly while traveling through them. Since Alice has access to the Hilbert space spanned by $|\psi_i\rangle$, she could do a projective measurement as well, which will make the the late time radiation state to land in a state $|i\rangle$.

Let us consider an instant when $|i\rangle$ is an eigenstate of the number operator $b^{\dagger}_{\Omega}b_{\Omega}$ associated with late Hawking radiation. Since the vacuum in the near horizon region for a freely falling observer is not a vacuum for the late Hawking radiation, one obtains a Bogoliubov transformation between operators of recipient of late time radiation and the freely falling ones,
\begin{equation}
b_{\Omega}=\int _{0}^{\infty}d\omega~\left[B(\omega,\Omega)a_{\omega}+C(\omega,\Omega)a^{\dagger}_{\omega}\right]
\end{equation}
Now if the radially infalling observer, Alice  does some measurement such that it amounts to projection of $|\Psi \rangle$ using $P_{i}$, where, 
\begin{equation}
b^{\dagger}_{\Omega}b_{\Omega}|i\rangle \propto |i\rangle,
\end{equation}
as discussed above, then as Alice tries to measure the near horizon region (once she becomes freely falling), $a_{\omega}|i\rangle$ will be non zero for large $\omega$ due to the Bogoliubov transformation. The state has been cast in a number eigenstate of the late time radiation modes, which is not a vacuum for freely falling modes. The state she witnesses is an excited one, which receives contribution form all frequencies (even large ones). As a result, an in-falling observer having information about the early radiation encounters a firewall at the horizon.

In other words, the mode $|i\rangle$ connected with $b_{\Omega}$ will be a superposed one for observables constructed using $a_{\omega}$, including $\omega \rightarrow \infty$ contributions. Hence Alice will witness the horizon as a highly energetic place, in contradiction with the equivalence principle (as she measures the number expectation constructed using $a_{\omega}$).

The firewall outcome looks pretty much inescapable, isn't it ? However, there is still some room for doubts perhaps. A possible problem with the above argument can be that the existence of such a projection operator can not imply that such a measurement will lead to a classical world with \ref{Eq_Class_01} as the quantum state. In other words, the state in \ref{Eq_Class_01} can not lead to a classical world. The measurement of early Hawking radiation will 
select a state, e.g., $|\psi _{I}\rangle$ in $\mathcal{H}_{\rm rad}$, having well defined classical configuration compatible with it. This state need not be maximally entangled with the eigenstate of 
$b^{\dagger}_{\Omega}b_{\Omega}$. The dimension of the Hilbert space spanned by $b^{\dagger}_{\Omega}b_{\Omega}$ is very small and therefore, one can expect that $|\psi _{I}\rangle$ would practically never 
coincide with $|\psi _{i}\rangle$, if $|i\rangle$ is an eigenstate of $b^{\dagger}_{\Omega}b_{\Omega}$. To put it in another form, the eigenbasis of $b^{\dagger}_{\Omega}b_{\Omega}$ being maximally entangled 
with states in $\mathcal{H}_{\rm rad}$ is different from the one that is entangled with in-falling observer and is picked by  measurement leading to classical description. Thus the projection 
operator $P_{i}$, with $|i\rangle$ being an eigenstate of $b^{\dagger}_{\Omega}b_{\Omega}$, leads to a superposition of macroscopically different world not about a single semi-classical world. Thus there is no contradiction between equivalence principle and complementarity. This suggests that a state of the early Hawking radiation 
describing the classical world reasonably well is a superposition of $|\psi _{i}\rangle$'s which are maximally entangled with the eigenstates of $b^{\dagger}_{\Omega}b_{\Omega}$. How likely is that the measurement of the early radiation will cast the late time radiation in a number eigenstate? Let us calculate the probability of such a process.

To see this in an explicit manner consider, the eigenbasis of $b^{\dagger}_{\Omega}b_{\Omega}$, which we denote as $|i\rangle$. Further, $N_{\rm rad}=\textrm{dim}\mathcal{H}_{\rm rad}$ and  $N_{\rm bh}=\textrm{dim}\mathcal{H}_{\rm bh}$. Then expansion of a pure state in the full Hilbert space $\mathcal{H}_{\rm rad}\otimes \mathcal{H}_{\rm bh}$ can be performed as,
\begin{equation}
|\Psi\rangle =\sum _{i=1}^{N_{\rm bh}}c_{i}|\psi _{i}\rangle \otimes |i\rangle
\end{equation}
Further $\mathcal{H}_{\rm rad}$ can be expanded in a classical state basis $\lbrace |e_{n}\rangle \rbrace$ (i.e., the basis in which the state $|\Psi _{I}\rangle$ is expanded and the eigenvalues represent the classical world), where $n=1,2,\ldots N_{\rm rad}$. We would like to determine the overlap between $|\psi _{i}\rangle$ and $|e_{n}\rangle$. In other words we are interested in determining the probability that such a state $|i\rangle$ exists, such that the entangled state $|\psi _{i}\rangle$, satisfies the relation $|\langle e_{n}|\psi _{i}\rangle|>1-\epsilon$, for some small positive and real number $\epsilon$. 

Choosing the states $|\psi _{j}\rangle$ and $|e_{m}\rangle$ in the $2N_{\rm rad}$ dimensional \emph{real} vector space, such that $|e_{m}\rangle=(1,0,0,\ldots)$. Then $|\psi _{j}\rangle$ is a point on the $2N_{\rm rad}-1$ dimensional unit sphere with $\psi _{j}^{1}>1-\epsilon$. This denotes two small circle around the north and south pole. Since $\cos \theta =1-\epsilon$, the radius $\theta \sim \sqrt{2\epsilon}$, and hence the probability becomes,
\begin{align}
\textrm{Probability}&=\frac{\rm the ~area~ of~ the~ small~ circular~ regions ~near~ the~two~ poles}{\rm total ~ area~ of ~the~ sphere}
\nonumber
\\
&\sim \frac{(\sqrt{2\epsilon})^{2N_{\rm rad}-1}}{\sqrt{N_{\rm rad}}}
\end{align}
Total probability would become multiplication of the above by possible $|\psi _{j}\rangle$ states, which is $N_{\rm bh}$ and by possible $|e_{m}\rangle$ states, being $N_{\rm rad}$. Thus finally,
\begin{equation}
P\sim N_{\rm bh}\sqrt{N_{\rm rad}}~(2\epsilon)^{N_{\rm rad}}
\end{equation}
Thus for small $\epsilon$ the probability is hugely suppressed. 

Another objection against the firewall hypothesis can be put forward from the perspective of quantum computation \cite{Harlow:2013tf}. In particular, the firewall proposal depends crucially on the fact that the infalling observer must be able to confirm the entanglement of the early Hawking radiation with modes in the near horizon regime. But for that to happen Alice must have sufficient time to process the early Hawking quanta in her quantum computer and obtain the corresponding correlations with late Hawking quanta. Following this argument it has been observed in \cite{Harlow:2013tf} that no such measurement can be performed faster than the time scale of evaporation of the black hole and hence the correlations between early and late time Hawking quanta can never be realized by an radially infalling observer. This certainly speaks against the firewall puzzle \cite{Harlow:2013tf}(see also \cite{Ong:2014maa}).

Further, one of the main backbone for the firewall claim being that, a single mode of the late time Hawking radiation located near the horizon is maximally entangled with the modes residing inside the event horizon and simultaneously with the modes of early Hawking radiation. It is also possible that this situation never happens during the evolution of a black hole, due to the unitary evolution itself, which maximally entangles a late time Hawking radiation mode located just outside the horizon with a combination of the early Hawking radiation and black hole interior states, instead of either of them separately \cite{Hutchinson:2013kka}. 

The above sections demonstrates the current state of affairs with the firewall puzzle --- there are proposals evading the firewall claim (two more of such proposals will be discussed in \ref{Nonlocality}) as well as arguments strengthening the firewall puzzle. Till date, no definite conclusions have been reached to either favor or disfavor the existence of firewall at the horizon.
\section{The overreach of quantum gravity}

All this while,  we have been content with the idea that the quantum effects of gravity are really not important at the horizon of a macroscopic black hole (since the curvature there is too small). Is curvature the only quantity which paves the way for the onset of quantum gravity?  Or there are other quantities to a black hole, more than its horizon curvature, which can cause an early arrival of quantum gravity?  Truly speaking we do not even know how to identify the event horizon at the level of quantum gravity. The idea of causality itself may be compromised at that level\footnote{One can think of a situation where a null shell is collapsing to form a black hole, resulting in formation of an event horizon, say at earth right now, due to its teleological property. By this we mean that in a null shell collapse forming a black hole, the null shell crosses the horizon to form a black hole, i.e. an event horizon is formed in the region of the causal past  of the shell as well. However one should \emph{not} 
expect some quantum gravity effects to become prominent at earth even long before the black hole actually forms.}. So, if we are willing to let go of the character of differential geometry when quantum gravity arrives,
 why not use those parameters which still keep making sense (unlike curvature) when quantum gravity is in full force. In this section, we discuss proposals which call for a premature intervention of quantum gravity on account of features a black hole possesses other than the mundane curvature at the horizon.
\subsection{The fuzzball paradigm}\label{Fuzzball}

One of the most straightforward resolution to the information paradox could have been achieved if the event horizon of the black hole were not there, leading to order unity corrections to the thermal radiation, due to some ultraviolet behavior of gravity theory. This looks like a very radical thought at first sight and a plethora of questions arise at first glance, such as: (a) Event horizons appear in classical general relativity. Any modifications of the short distance behavior of gravity should, of course, modify the singularity structure in \gr, but how come the event horizons, which are free from singularities and are seemingly regular place get affected? (b) What happens to radially infalling observers, who were, otherwise supposed to cross the horizon? In particular, what will they observe as they approach the point, where the horizon was supposed to be present, classically? (c) This proposal must violate no-hair theorems, as something else is going to replace the horizon of a black hole, 
supposedly adding new features to it, apart from the classical charges. So which of the assumptions of the no-hair theorems are being tampered with? Interestingly, {\it the fuzzball proposal} \cite{Mathur:2016ffb,Mathur:2014roa,Mathur:2008nj,Mathur:2005zp} attempts to give consistent answer to each of these questions which we will describe next. However due to mathematical complexity of the proposal we will content ourselves with a broad and somewhat heuristic overview of the physics at work (for more accurate accounting, see \cite{Mathur:2016ffb,Mathur:2014roa,Mathur:2008nj,Mathur:2005zp}). 
\begin{figure}[t]
\begin{center}
\begin{tikzpicture}
    \draw (0,0) circle (1.5cm);
    \shade[ball color=blue!30!white,opacity=0.80] (0,0) circle (1.5cm);
    \draw[->] (2,0) to (4.5,0);  
    \shade[ball color=blue!10!white,opacity=0.40] (7.0,0) circle (0.5cm);
    \draw[color=red, line width=0.25mm] (6.9,2.1) -- (6.9,0.6)
        arc [x radius = 0.3em, y radius = 0.3em, start angle = -180, end angle = 0]
        (7.1,2.1) -- (7.1,0.6);
        \draw[color=red, line width=0.25mm] (7,2.1) ellipse (0.3em and 0.1em);
    \draw[color=red, line width=0.25mm] (5.0,0.1) -- (6.4,0.1)
        arc [x radius = 0.3em, y radius = 0.3em, start angle = 90, end angle = -90]
        (6.4,-0.1) -- (5.0,-0.1);
        \draw[color=red, line width=0.25mm] (5,0) ellipse (0.1em and 0.3em);
    \draw[color=red, line width=0.25mm] (9.0,0.1) -- (7.6,0.1)
        arc [x radius = 0.3em, y radius = 0.3em, start angle = -270, end angle = -90]
        (7.6,-0.1) -- (9.0,-0.1);
        \draw[color=red, line width=0.25mm] (9,0) ellipse (0.1em and 0.3em);
    \draw[color=red, line width=0.25mm] (6.9,-2.1) -- (6.9,-0.6)
        arc [x radius = 0.3em, y radius = 0.3em, start angle = 180, end angle = 0]
        (7.1,-2.1) -- (7.1,-0.6);
        \draw[color=red, line width=0.25mm] (7,-2.1) ellipse (0.3em and 0.1em);
    \node[label=below: \textrm{star with radius} $r_{0}>2M$] at (0,-2) {};
    \node[label=above: \textrm{collapses}] at (3.2,0.1) {};
    \node[label=right: \textrm{There is no interior}] at (8.0,1) {};
    \draw[->,bend right=30] (8.0,1) to (7.0,0);
    \node[label=right: \textrm{compactified extra dimension}] at (8,-1) {};
    \draw[->,bend right=30] (9.5,-0.8) to (9.0,0);
    \node[label=left: \textrm{smooth joining of spacetime}] at (7,-1) {};
    \draw[->,bend left=30] (5.5,-0.9) to (6.9,-0.5);
\end{tikzpicture}
\end{center}
\caption{A schematic illustration of the fuzzball paradigm. A star collapses and tunnels to a fuzzball state with no spacetime geometry interior of $r=2M$, where classical event horizon would have formed. Instead the spacetime joins smoothly on $r=2M$ through extra spatial dimensions. These smooth joinings introduce additional curvatures which acts as a proxy for curvature generating black hole mass. Since there is no interior the region near the $r=2M$ surface is no longer vacuum and hence there is no information paradox. See text for a detailed discussion.}\label{fig_Fuzzball}
\end{figure}
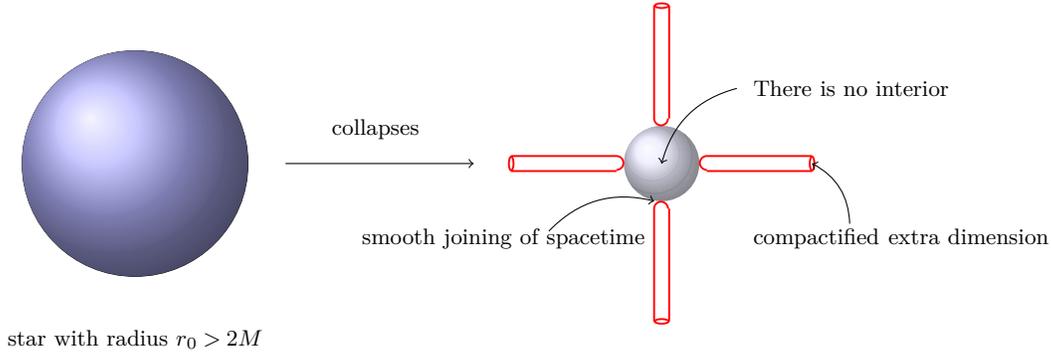

The fuzzball proposal has its origin in the calculation of the black hole entropy in the context of string theory, where counting of microscopic states of a particular system of strings and branes \emph{without gravity} resulted in $S=\textrm{area}/4$, exactly obtaining the Bekenstein entropy for general relativity \cite{Strominger:1996sh}. Surprisingly, it was also found that the size of such a string system when gravitational interactions are included are of the same order as the horizon size of a black hole, casting doubt whether horizons really exist or not \cite{Mathur:1997wb}. However due to mathematical complexity one can not perform the analysis in general, but for some simplistic systems where analytical calculations can be performed one again confronts similar situations, the black hole states in string theory seemingly do not have a horizon \cite{Lunin:2002qf,Kraus:2015zda}. This prompted the idea as radical as the fuzzball to emerge. The picture of black hole spacetime that is being proposed 
by fuzzball paradigm is as follows: There is no event horizon and also no notion 
of regular spacetime beyond some radius $r\sim 2M$. The spacetime is completely smooth and ``caps off'' at $r\sim 2M$ through some compactified extra spacetime dimensions (see \ref{fig_Fuzzball}). Since there is no spacetime beyond $r\sim 2M$, the region near it is no longer a vacuum  (as a consequence the no-hair theorem does not work). Thus the string states or ``fuzzball'' states representing traditional black holes have non-trivial hairs and the information is not lost. Further the spacetime curvature due to the ``capping off'' behavior in extra dimensions act as a proxy for the ADM mass of the spacetime \cite{Mathur:2008kg}. 
\subsubsection{The no no-hair!}\label{Fuzzball_00}

Among the three questions posed earlier, the resolution to the no-hair theorems can be explained right away. It traces back to the assumptions inherent in the no-hair theorems, which even though are natural ones in $(3+1)$ dimensions, do not remain so in presence of compactified extra spatial dimensions. In particular, one can demonstrate that starting from some gravitational theory in ten 
dimensions, one can arrive at a four dimensional spacetime which would violate no-hair theorems \cite{Gibbons:2013tqa}. Since, fuzzballs require extra spacetime dimensions for its existence, it is of no surprise that there are non-trivial structures at the horizon. Given this basic picture let us try to answer the rest of the questions raised above, which will help to understand the ideas presented here in a much better and comprehensive way. These extra structures at the horizon also exonerates the horizon from the burden of being regular and semi-classically valid and approachable. The fuzzball picture turns the semi-classical analysis on its face.
\subsubsection{Violation of semi-classical approximation}\label{Fuzzball_01}

The picture in the fuzzball paradigm is peculiarly striking i.e., the horizon of a black hole is very far from the singularity, it may even be astronomical distance away (for a supermassive black hole of course), yet small scale behavior of the spacetime somehow affects the horizon structure, a place nothing to do with quantum gravity! If quantum effects can pile up at a place as unsuspecting as the horizon, how safe are the other regular points in spacetime are?

Put in another way, if a shell of matter collapses, ultimately forming and crossing the event horizon following classical dynamics, how come quantum effects change all these? This puzzling feature can be addressed by noting that though there does not seem to be a classical configuration emerging out from the collapse, one can, in principle, tunnel to such a classically forbidden fuzzball configuration. This might look absurd, since quantum tunneling probability for macroscopic bodies into weird quantum configurations are known to be very small. {\it This is how the standard non-horizon regular points are protected against turning into something awkwardly non-classical}. However, it is the entropy of the black hole which comes to change the game for the points near horizon. The tunneling probability of any quantum system is given by $\exp(-\mathcal{A})$, where $\mathcal{A}$ is the action of the system. For gravity if one uses the Einstein-Hilbert action, 
then near the horizon, $\mathcal{A}_{\rm grav}\simeq \alpha G M^{2}$, 
where $\alpha \sim \mathcal{O}(1)$ \cite{Mathur:2008kg}. Thus the tunneling probability becomes,
\begin{align}
P_{\rm tunnel}\simeq \exp(-\mathcal{A}_{\rm grav})=\exp(-\alpha GM^{2})\ll 1
\end{align}
However there are a large number of fuzzball states which are available to tunnel into. In fact, in the string picture, these are the microstates which give the black hole the entropy, it demands classically. Therefore, the total number of such states correspond to, 
\begin{align}
\mathcal{N}=\exp(S_{\rm Bekenstein})=\exp(A/4G)\simeq\exp(GM^{2})
\end{align}
Hence total probability would depend on, total number of available states multiplied by probability of tunneling to a single such state, leading to,
\begin{align}
P_{\rm total}=\mathcal{N}P_{\rm tunnel}\sim \mathcal{O}(1)
\end{align}
Thus the collapsing shell have a finite probability to tunnel into one of the fuzzball states. We would like to emphasize that this is an order of magnitude estimate and since exponentials are involved one must exercise extra caution. But the smallness of tunneling probability and largeness Bekenstein entropy leads to a macroscopically finite probability of tunneling to a fuzzball configuration \cite{Mathur:2008kg}. 

However there is a subtle issue associated with the fuzzball picture, which has to do with the exact location of the fuzzball surface. Since exact calculations are really difficult to perform, one can ask, whether there is any possibility of forming the fuzzball surface a macroscopic distance away from the ``would be'' event horizon? If there exists such a possibility then one will have naked fuzzballs. This has the destructive potential to affect all the exterior observers, including the asymptotic ones, which is similar in spirit to the recent findings about the existence of a naked firewall \cite{Chen:2015gux}.
\subsubsection{Free fall: Fuzzball complementarity}\label{Fuzzball_02}

The experience of an in-falling observer has always been a key probe of the horizon structure irrespective of the scenario being firewall or fuzzball. In case of fuzzball this leads to the notion of fuzzball complementarity. Classically, when a particle falls into a black hole, for an outside observer it perturbs the spacetime geometry and subsequently the black hole stabilizes to a new configuration by emitting some characteristic frequencies, known as {\it the quasi-normal modes} of a black hole. Moreover, as the particle approaches the horizon, its speed increases and as a consequence it enters the event horizon with an energy $E\gg T$, where $T$ is the Hawking temperature of the black hole.

In case of fuzzball paradigm one {\it conjectures} that for such in-falling particles the characteristic frequencies of fuzzball oscillations, as measured by the exterior observers, mimic that of quasi-normal modes in general relativity. Whereas, for the outgoing Hawking quanta, $E\simeq T$ and as a consequence various fuzzball states contribute differently resulting in an information retrieval from black holes. Thus the fuzzball complementarity tunes the behavior of particles having energy $E\sim T$ such that the information paradox is resolved, while forces an universal behavior for $E\gg T$ mimicking in-falling process of a traditional black hole from the point of view of an static observer residing far away from the horizon \cite{Mathur:2011wg,Mathur:2012zp,Chowdhury:2007jx}. Therefore, any probe, with $E\gg T$ will be oblivious to the fundamentally different structure of the horizon as far as the exterior observers are concerned. Note that this is completely different from the 
complementarity introduced earlier and hence bypasses the firewall argument. However the radially in-falling observer will definitely be destroyed at the fuzzball surface, but unfortunately she really has no way to convey this horror story to her exterior colleague.

To sum up, if we are willing to give up the interior of the hole, the smoothness of the horizon and string theory is the theory of gravity at the fundamental level, the fuzzball seems like a viable option.
\subsection{A stable and a proportionate remnant?}

If we are willing to live with the fact that somehow, effects of quantum gravity can become important on scales, way larger than the Planck scale, then several possibilities can open up. In fact, some of the approaches of dealing with black hole information also seemingly demand that the effect of quantum theory may better be large scaled, on the account of symmetries in the approach, which allows them to trump the Hawking calculation on a scale much larger than the Planck scale (where it was initially supposed to get away) and hence allowing the possibility of a larger remnant, whose size hopefully will be proportional to the initial data fed in. We list a few of them below.
\begin{itemize}

\item One bound on the size of the remnant can be found from appealing to the maximum compressibility of information in a given region. Giddings \cite{Giddings:1992hh} argued that if there exists an upper bound on entropy density, it should better be that of a Planck scale sized remnant, i.e.,
\bea
s_{max} = \frac{S_{Pl}}{V_{pl}} \sim \frac{M_{Pl}^2}{l_{pl}^3}. \label{MaxEntDens}
\eea
Now, once some matter enters a macroscopic black hole horizon, it inevitably gets drifted towards the central spacelike singularity. Classically all such matter would ultimately reach the central singularity, making it also an infinite entropy density place. However, since we have already committed that the maximum entropy density should always be given by \ref{MaxEntDens}, it is reasonable to expect that a quantum theory of gravity, respecting the entropy bound will replace the central singularity by a region of maximum entropy density. The size of the region of such a maximum entropy density corresponding to a black hole of mass $M$ would be
\bea
r_M \sim l_{pl} \left( \frac{M}{M_{Pl}}\right)^{2/3}.\label{MaxEntDensSize}
\eea
This depicts the region of a {\it core} where the entropy of mass $M$ black hole is stored with the maximum capacity. However the fact that the spacetime behind event horizon is not static is in tension with the concept of a core. To remove the spacelike singularity one needs to introduce additional matter fields modifying the ``going to be'' singular point within the spacelike region. This possibly requires the core to have some non-trivial matter fields, e.g., non-linear electrodynamics may be one viable candidate \cite{Chen:2014jwq}.

From  \ref{MaxEntDensSize}, it is clear that the size of the core scales with the initial data (aka the mass). Once the black hole starts evaporating, the event horizon starts shrinking. But since the core is packed with the maximum density allowed, it does not have a capacity to further accommodate any bits carrying the information of evaporation \footnote{recall the core is saturated in sense of entropy! Any additional bit will decrease the mass but increase its entropy, since the information of the mass lost by the radiation has not gone outside, as we deduced in previous sections.}. Thus, while the horizon shrinks, the core persists. Therefore, there will come a stage when the horizon meets the core and then the Hawking process loses its credibility. What happens next is up for speculation as it caters to regime of Planck scale physics, which we hardly have any clue of (note that the entropy density is at the Planck scale, while the physical size can still be macroscopic). There can be a plethora of 
various exotic possible outcomes when the horizon comes crashing down to the Planck density core, which has replaced the classical singularity \cite{Aharonov:1987tp}. It can also be thought that the gravity shows some phase transition at this leg of evaporation, and end up in some sort of crystal as a leftover remnant \cite{Nikolic:2015vga}. It is well known that once systems undergo a phase transition into a crystalline phase, quantum correlations become pretty long range. Therefore, it is possible that effects of quantum gravity, in the similar spirit, become important even at large length scales leading to a mass proportionate remnant. Either we need to know the physics inside these remnants or we will have to further wait for the time scale which a remnant of this kind lives for, before jumping back to the revival of (or declaring a convincing triumph over) the problem of information loss.

\item  In canonical splitting of gravity, we end up with various constraints on the variables of the theory, we use to describe the same. These constraints appear as operator equations at the quantum level, if we keep using these variables in quantum gravity. Using the method of discretization of spacetime lattice \cite{Vaz:2012zq}, one can argue that the solution space, consistent with diffeomorphism constraints, with positive frequency, is obtained through states of the kind
\bea
\psi_i = e^{\omega_i b_i} \times \exp{\left(\frac{-i\omega_i}{\hbar}\left[a_i \tau_i \pm \int^{R_i} d R\frac{\sqrt{1-a_i^2 {\cal F}_i}}{{\cal F}_i}\right] \right)}, \label{WFforWDQG}
\eea
representing a collapsing shell marked as $i$, with $\tau_i$ giving its proper time, $R_i(\tau_i)$ being its area radius (in the case of spherically symmetric collapse) at proper time  $\tau_i$ and ${\cal F}_i $ is related to the mass function $F_i$ of the shell. $a$ and $b$ are functions of $F_i$. Now, vanishing of ${\cal F}_i$ determines the location of the {\it apparent horizon}, classically. So the classical location of the apparent horizon (${\cal F}_i= 0$) serves as a point where the wave functional in \ref{WFforWDQG} develops a discontinuous behavior \footnote{It has also been argued that during collapse of a matter shell the equations governing the collapse process becomes non-local, resulting in a wavefunction with non-singular probability density at the classical black hole singularity, indicating removal of the classical singularity \cite{Saini:2014qpa,Greenwood:2008ht}.}. The argument of the exponential has a discontinuous (by an amount of residue at the pole) across the ``apparent horizon''. 
Moreover for a particular branch (i.e., with choosing one sign of the $\pm$ in the exponent), we see that the directions of the shells across the horizons are opposite.
 
Using this approach of canonical quantization, it can be argued  \cite{Vaz:2012zq} that any collapse of matter in the exterior of the apparent horizon of the collapsing data is accompanied by an out going radiation in the interior of the apparent horizon (and vice versa)\footnote{this is also argued to be a reason of genesis of Hawking radiation through quantum gravity, interior collapse will always be accompanied by an outgoing radiation in the exterior \cite{Vaz:2012zq}}\cite{Vaz:2012zq}. It is further argued \cite{Vaz:2014rya,Vaz:2014era} that, the absence of firewall must require that only one branch of solutions exists, i.e., a collapsing shell in the exterior accompanied by a thermal radiation in the interior. There should not be any branch that has a collapsing shell in the interior accompanied by a Hawking radiation in the exterior. Therefore, the profile of mass distribution of the shell should be such that its collapse is effectively counterbalanced and gets supported by the interior thermal 
radiation. Such a mass profile, the size of a resulting ``stable structure'' and its relation to gravitational mass of the cloud is estimated for different cases \cite{Vaz:2014era, Sarkar:2016kqt}. Therefore, in this scenario, the collapse halts at the apparent horizon and the black hole never forms in the first place. All what is left is a dust cloud-ball hinging at the apparent horizon, as a stable remnant, being supported by a radiation pressure from the inside necessitated by quantum gravity. This cloud-ball also hides in its inside, a negative mass singularity which gives rise to the radiation flux in the interior.

\end{itemize}

One may have another interesting possibility as well, in which case non-trivial interior geometry may result in a macroscopic remnant. In particular it is possible to glue a non-trivial geometry to the Schwarzschild interior in such a way that asymptotic observers will not be able to ``feel'' the corresponding geometry change. If this glued spacetime have large volume or is connected to other universes, the remnant can still have a large volume appearing as a small one to outside observers. For more details on this perspective, see \cite{Chen:2014jwq}.

If any of these constructs are actually realized in nature, they indeed make very fascinating objects of study: macroscopic in size, while fundamentally quantum in nature. It is very unlikely that effects of quantum character near the horizon, will be washed out in careful tests and observations of strong gravity (e.g., \cite{Sahu:2015dea,Chakraborty:2016lxo}). Apart from aesthetically modeling a possible solution to a paradox, these approaches lead to falsifiable predictions (albeit for a bit futuristic set of observations) as well. 
\section{The quantum framework: A re-look required(?)}

Till this point, all we have discussed firmly sits within the standard quantum theory framework. We discussed how the interpretations of pertaining to black hole geometry, event horizon or the state of matter on a black hole spacetime should be played around to obtain a consistent picture. In this game, what really at stake was the equivalence principle. Either the horizon has to be dubbed as a special place or somehow the correlations thought to be totally captured in the interior of the black hole, were needed to be transported to a different geometrical configuration (wormhole, fuzzball interior etc.). All this was attempted to help the quantum theory retain the structure and charm it ever had. However, a fair analysis of the problem will also require us to wonder how much we can tweak around the universally accepted structure of the quantum theory. Let us try to see in this section what has the quantum theory got to trade for resolving the paradox.
\subsection{Can non-locality save the day?}\label{Nonlocality}

We, by now, have discussed and explored multiple interesting possibilities which, if pursued in rigor, hold the potential to churn out a possible resolution to the information paradox. Even though the various models have different structures to evade the original argument by Hawking, there is one crucial similarity --- all of them are local. Then a natural question arises, is there any way to introduce {\it non-locality}, while preserving causality? It turns out that indeed one can achieve the same by two very different non-local scenarios. The first one, due to Maldacena and Susskind \cite{Maldacena:2013xja} invokes wormhole like structures, while Papadodimus and Raju advocate \cite{Papadodimas:2013wnh} for state dependent operators. Both of these pictures rely heavily upon {\it the AdS/CFT correspondence} \cite{Ramallo:2013bua,Maldacena:1997re,Witten:1998qj,Aharony:1999ti,Hartnoll:2009sz,DHoker:2002nbb,Sachdev:2010ch,Nishioka:2009un}, discussion of which being out of the scope of this review, we will 
content ourselves with a broad overview of the ideas involved. 
\subsubsection{ER=EPR}\label{Nonlocality_01}

It is remarkable that non-locality appears in both quantum mechanics and general relativity, albeit in a completely different form, as Einstein-Podolsky-Rosen (henceforth EPR) correlations and as Einstein-Rosen (henceforth ER) bridge, respectively. In the case of EPR, two quantum systems which are spatially separated can have quantum correlations between them \cite{Einstein:1935rr}, while in the classical general relativity two apparently disjoint regions of spacetime can be connected by relatively short ER bridges (a more common terminology being wormhole) \cite{Einstein:1935tc,Fuller:1962zza,Visser:1995cc}. It turns out that even though both these effects are non-local, neither of them violates causality \cite{Visser:1995cc}. In \cite{Maldacena:2013xja} Maldacena and Susskind conjectured that these two disjoint non-local effects are actually identical, in particular the ER bridge between two spacetime regions is due to EPR like correlations between microstates of these regions. Hence they conjectured 
ER=EPR. One can also turn the argument around --- any two EPR correlated particles are also connected by some sort of ER bridge (see also \cite{Marolf:2012xe,Israel:1976ur,Maldacena:2001kr,Bryan:2016wzx}). 

We will now try to explain the phenomenon of Hawking radiation in this light. Since any two systems which have quantum correlations (or, in other words are entangled), by the above conjecture must be connected by a ER bridge among them. Exactly same situation appears in the case of Hawking radiation. The early Hawking radiation is entangled with the remains of the black hole spacetime and thus by the above conjecture the early Hawking radiation must be connected to the remains of the hole by a ER bridge. Thus there is no loss of information, since the black hole interior is connected to the Hawking radiation by ER bridge and the information is being passed on to the radiation, restoring unitarity. However there are subtle issues, like stability of the ER bridge for eternal black holes, which may pose a problem.

At this stage it is important to argue how this conjecture overcomes the firewall puzzle. According to the firewall proposal, the black hole is entangled with early Hawking radiation and as a consequence, smoothness at the horizon will be destroyed. Consider now an observer `Alice' who collects the early Hawking radiation and by some means (may be a quantum computer) coverts it to another black hole. Since the early Hawking radiation is entangled with the remnant hole, due to ER=EPR conjecture, they should also be connected by a ER bridge. Then Alice can throw in some message to the new hole which will transmit through the ER bridge and will affect the other entangled remnant hole. If some other observer `Bob' jumps into the remnant of the black hole, he will encounter this message as he enters the horizon. Depending on the mood of Alice (whether she wants to send flowers or fire), Bob will either encounter a smooth horizon or a firewall. Thus action of Alice on the early Hawking radiation can be non-trivial,
 due to their quantum correlations with the remnant black hole interior an ER bridge develops. 

Put in another way, let Alice observes an early Hawking quanta which is entangled with a late Hawking quanta. Since these are maximally entangled, by strong sub-additivity the entanglement with modes inside the horizon is tampered with. Hence if Bob falls in the horizon, he might observe high energy particles --- a firewall. Then \cite{Almheiri:2012rt} argues that the above statement is independent of whether Alice has measured the early quanta or not and hence the firewall must exist. This is what Maldacena and Susskind rules out \cite{Maldacena:2013xja}, there can be firewalls or not depending on what Alice is doing with the early Hawking radiation. But one can of course ask the respective probability for Bob to detect a firewall or travel through a smooth horizon. This seemed far out from our present technical knowledge. Thus the above conjecture presents a very nice non-local explanation of both the information loss and the firewall puzzle by invoking both general relativity and quantum mechanics 
together. We will now concentrate on the Papadodimas-Raju proposal.    
\subsubsection{Bargaining micro-causality: State-dependent operators}\label{Nonlocality_02}

Another completely new and novel proposal for resolving the information paradox was proposed by Papadodimas and Raju \cite{Papadodimas:2012aq,Papadodimas:2013wnh,Papadodimas:2013jku,Papadodimas:2015jra,Papadodimas:2015xma,
Banerjee:2016mhh,Ghosh:2016fvm} using the notion of state-dependent operators, which naturally leads to non-locality. The construction has been performed in the context of AdS/CFT correspondence, by which local gravitational operators in the bulk AdS are mapped to operators on the boundary CFT (the other way around as well). In particular it turns out that in order to construct local bulk operators describing the black hole interior, one must acknowledge that the mapping of CFT operators to the bulk will depend on the state of the quantum field in CFT.  

The motivation for the above scenario originates from complementarity proposal itself. Complementarity demands fields inside the event horizon to be mapped to the fields outside. Due to this mapping the early Hawking radiation, the late Hawking radiation and interior of black hole horizon are no longer independent and as a consequence the strong sub-additivity fails. This implies that even small corrections to the Hawking radiation can restore unitarity. One such possible explicit construction of complementarity is being provided by empty AdS spacetime \cite{Banerjee:2016mhh}, where one can explicitly demonstrate that operators in a given region of spacetime can be written in terms of operators at space-like separated events. To quickly glance through such a construction, let us write down some operator $\phi$ in a spacetime region $\mathcal{V}$ in terms of a complete set of basis $\lbrace |n\rangle \rbrace$,
\begin{align}\label{Eq_PR_01}
\phi(\mathcal{V})=\sum _{m,n}c_{mn}|n\rangle \langle m|
\end{align}
By Reeh-Schlieder theorem \cite{schlieder1965}, the complete set of states can be written in terms of the vacuum $|0\rangle$ as,
\begin{align}
|n\rangle =\mathcal{P}_{n}\left[\phi(\bar{\mathcal{V}})\right]|0\rangle
\end{align}
where $\bar{\mathcal{V}}$ is space-like separated from $\mathcal{V}$ and $\mathcal{P}_{n}$ are simple polynomials. One can also write the projection operator $|0\rangle \langle 0|$ in terms of the operators in the disjoint region $\bar{\mathcal{V}}$, as $P_{0}=|0\rangle \langle 0|\simeq \mathcal{C}[\phi(\bar{\mathcal{V}})]$, where $\mathcal{C}$ is a complicated polynomial. This can be achieved by writing
\begin{align}
\mathcal{C} =\lim_{\alpha \rightarrow 0} e^{-\alpha H} \approx \sum_{n=0}^{n_c}\frac{(\alpha H)^n}{n !},
\end{align}
where, again $H$ can be expressed in terms of field operators $\phi(\bar{\mathcal{V}})$ in $\bar{\mathcal{V}}$, through some complicated polynomial. In empty AdS, the parameters $\alpha $ and $n_c$ get determined in terms of Planck length $\ell _{\rm Pl}$ and the AdS length scale $\ell_{\rm AdS}$ as
\begin{align}\label{Eq_Raju_01}
\alpha = \log\left(\frac{\ell_{\rm AdS}}{\ell_{\rm Pl}}\right),\qquad \textrm{and } \qquad n_c =\alpha \exp{\alpha}.
\end{align}
Thus \ref{Eq_PR_01} can be written as,
\begin{align}
\phi(\mathcal{V})&=\sum _{m,n}c_{mn}\left\lbrace \mathcal{P}_{n}\left[\phi(\bar{\mathcal{V}})\right]|0\rangle \right\rbrace
\left\lbrace \langle 0|\mathcal{P}_{n}^{\dagger}\left[\phi(\bar{\mathcal{V}})\right] \right\rbrace
\nonumber
\\
&=\sum _{m,n}c_{mn}\mathcal{P}_{n}\left[\phi(\bar{\mathcal{V}})\right]\mathcal{C}\left[\phi(\bar{\mathcal{V}})\right]
\mathcal{P}_{n}^{\dagger}\left[\phi(\bar{\mathcal{V}})\right]
\end{align}
Thus one is able to write down the operators in region $\mathcal{V}$ in terms of operators in the region $\bar{\mathcal{V}}$, explicitly realizing complementarity. Now, assuming this scheme sails through for a AdS-black hole solution, the operators corresponding to the interior of the hole will be mapped to the operators in the exterior of the hole. At this stage one can use the AdS/CFT conjecture to gain the information about the interior through boundary CFT operators, which essentially tell us about the bulk (exterior) operators.

Keeping the above scheme in mind, when one tries to find the local operators in the CFT that will be mapped to bulk operators within the event horizon, {\it one is led to state dependent operators in the CFT} \cite{Papadodimas:2015jra}. The same also seems to hold for eternal black holes as well. The natural question corresponds to: Will this lead to possible violation of local quantum mechanics, specially for a radially infalling observer? Surprisingly, it turns out that the commutator of operators evaluated outside and inside the horizon vanishes for low point correlators but does not vanish identically as an operator \cite{Papadodimas:2015jra}. If we consider n-point correlation function of the field operator it will vanish outside the light cone provided $n$ is not \emph{too} large. This suggests that locality might brake down if one probes the spacetime at a large number of points\footnote{To provide a quantitative estimate regarding the number of spacetime points, one notes that the 
only non-trivial number associated with this problem is $n_{c}$, defined in terms of the AdS length scale in \ref{Eq_Raju_01}. Thus by large number of points we mean $n>n_{c}$. Similarly two spacetime points will be close enough if the proper distance $\ell$ between them satisfies $\ell <\ell_{P}$.}. This is sort of identical to the fact the locality also breaks down if one probes spacetime points which are close 
enough \cite{Banerjee:2016mhh,Papadodimas:2013jku}. Hence such state-dependent modifications of CFT enables one to obtain operators corresponding to black hole interior and a possible resolution of the information paradox at the cost of micro-causality. 
\subsubsection{A final state for black holes}

As emphasized several times in the earlier sections, the main crux of the information loss paradox hides in the fact that, the particles falling inside the black hole event horizon are lost from the point of view of outside observers, compelling them to trace over  the field configurations in the interior region and hence arrive at a thermal density matrix. In standard quantum theory, typically what we do is to evolve an initial state and obtain the final state.  To restore time symmetry in the context of quantum mechanics, there have been several proposals, e.g., inclusion of final state density matrix in the decoherence functional (see \cite{GellMann:1991ck,Bennett:1992tv}), where one studies the evolution given the final state.

Following these ideas it has been postulated in \cite{Horowitz:2003he} that there is a {\bf unique} configuration (corresponding to the interior of the black hole) on a part of final data, describing the quantum state of matter that has formed the black hole and the interior. In particular, one takes the part of the final state, living in the interior Hilbert space (which corresponds to $H_{\rm matter}\times H_{\rm in}$) to be a linear combination of the basis states $|m\rangle _{\rm matter}$ and $|i\rangle_{\rm in}$ corresponding to the matter Hilbert space $H_{\rm matter}$ and  interior Hilbert space $H_{\rm in}$ respectively, as \cite{Horowitz:2003he}
\begin{align}
|\textrm{BH Interior}\rangle =\sum _{i,m}S_{mi}\Big\{|m\rangle _{\rm matter}\otimes |i\rangle _{\rm in}\Big\}.
\end{align}
Boldly, this interior state is proposed to be the unique final configuration any black hole  interior settles into.
In tune with the unitary quantum physics the standard assumption of $S_{mi}$ being unitary (but completely random besides being non-local in the coordinates near the singularity), is adopted. Following \cite{Horowitz:2003he} one can demonstrate that even with this final quantum state defined in the black hole interior, the corresponding density matrix will lead to a thermal state in the micro-canonical ensembles in accordance with the original result from Hawking (see also \cite{Gottesman:2003up}). At this stage one might wonder about the arrow of time, since if the black hole settles into a unique quantum state, it will be seemingly settling into a more organized state (i.e. zero entropy), which is opposite to the flow of time. However as far as measurements are concerned, any observer has access to only a very small portion of the Hilbert space, let alone the dimension of the Hilbert space responsible for black hole entropy. This will force the observer to trace over the unobserved part of the Hilbert 
space, leading to non-zero entropy \cite{Horowitz:2003he}. Interestingly, the above proposal also has consequences in relation to the firewall puzzle. 
In particular, it is possible to show that if the black hole has a final state, then the sub-additivity of entropy no longer holds and hence there is no need to have a firewall at the horizon \cite{Bousso:2012as}.
\subsection{Bargaining unitarity: Modifications of quantum physics }

All this while, we have discussed problems and remedies within the framework which was inherently unitary. We have covered much ground within this and have realized that the notion of unitary evolution comes in conflict with black hole evaporation one way or another. There are sharp tussles with theorems in quantum information when the notion of causal horizons separating causal supports of events on the spacetime are carried over to the Hilbert space.

May be it is time to come to terms with possibility that quantum theory may not be unitary after all\footnote{However it was 
advocated in \cite{Unruh:1995gn} that the transition from pure state to mix state does not necessarily lead to bizarre consequences for laboratory physics, thereby asserting no information paradox for black holes.}! It's not only about the confrontation with black hole dynamics which must force us into adopting such a radical idea with goes against the spirit of the {\it absolutely successful} theory of quantum mechanics. There always have been a discomfort with the postulates involved in quantum theory. The expectation values obtained from quantum theory appeals to statistical interpretation. That is, a quantum system gives an average output when some observables is being measured on many copies of a system, matching with the expectation value. Due to its statistical interpretational feature the theory is not well suited to explain observation made on a single copy. (No matter how smart we are in probability theory, one can not tell an output for sure, when a gambler rolls a dice for a single time). The 
output in a single copy inherits quantum fluctuations which we have no control on. So this inability of a theory, 
which has probabilistic description at its core, reflects majorly in two prominent contexts in physics.
\begin{itemize}
\item {\bf Output of a measurement}\vspace{0.5 em}

The standard quantum theory we are familiar with tells us that the time evolution of a closed system is unitary in nature,
$$ |\psi(t)\rangle = e^{i Ht} |\psi(0)\rangle.$$
where $H$ is the Hamiltonian of the system. However, when a measurement is done on such system, it {\it collapses} into one of the eigenstates corresponding to the observable being measured, i.e.,
$$|\psi\rangle = \sum_n c_n |n\rangle \underrightarrow{\textrm{Measurement~}} ~|i\rangle,$$
with $|i\rangle$ being one of the eigenstates $\lbrace |n\rangle \rbrace$. Therefore, such a process is seemingly non-unitary and information destroying. (Once a measurement is done and the system ``collapses'' to one of the eigenstates, information is destroyed, about what is was before the measurement is done.) The system evolves unitarily between measurements but measurement itself is a seemingly non unitary process. Although it is also a debatable point of view \cite{BOHM:1966zz,Weinberg:1989us,Weinberg:2011jg,Zurek:2003zz,Zeh:1970zz} and the debate and discussion is quite hot on this. So if we subscribe to the idea that measurement itself is a non unitary process, we should also reflect on what is so special about a measurement process. In a measurement process a quantum system interacts with an apparatus (which itself is a collection of many small quantum systems) and collapses. So, it looks like the combined system of (quantum system + apparatus) is effectively a collection of large number of quantum 
systems which 
evolves non unitarily! This prompted the proponents of non unitary quantum theory to say that quantum mechanics fundamentally should be non-unitary in nature, with non-unitarity becomes apparent only when the number of degrees of freedom becomes very large \cite{Adler:1993hm,Bohm:1951xw,Bohm:1951xx,Diosi:1986nu,Everett:1957hd,Bassi:2012bg,Bassi:2003gd}.  A system with small number of degrees of freedom has a dynamics which is very well approximated by a unitary theory. In that sense protecting information does not remain a principle of nature any more, but only an effective feature when the number of degrees of freedom in the system is very small. A macroscopic black hole, which has very large number of degrees of freedom (as suggested by its entropy) is a fit case where such non-unitarity, if present, should be visible.\\
\item {\bf A single copy of a closed system}\vspace{0.5em} 

Another possible explanation of measurement process can come through the idea of decoherence. If we consider only a subsystem of a large system, the effective dynamics of the subsystem may be non unitary such that the evolution of the full system is still unitary. Also making many copies of subsystem, we can make sense of a statistical interpretation of quantum theory. However, this becomes a problem when we look at a closed system, which has only a single realization. In this case, we have neither system-subsystem division  nor many copies of this system to apply statistical description. Therefore, the quantum theory, as it is, fails to give any deterministic prediction. Problem becomes more prominent if such a system evolves on its own to a classical description. Then we are really in a fix. This is a setting where quantum theory is not designed to be applied upon. Our universe makes a concrete example of such a system. Thus, non-unitary modifications to quantum theory derives motivation 
also from quantum cosmology \cite{Perez:2005gh,Sudarsky:2009za,Canate:2012ua,Martin:2012pea,Das:2013qwa,Okon:2013lsa,Lochan:2014dca}.
\end{itemize}
\subsubsection{How much of non-unitarity in quantum mechanics is tolerable ?}

Years of research towards a feasible model of non-unitary quantum theory has led us to a single one, which is non-relativistic as well as stochastic in nature (in fact this is the only possible non-unitary generalization, which fits the experimental scrutiny and is consistent with causality). This model is known as \CSL model of quantum theory (however for a possible relativistic generalization see \cite{Tumulka:2005ki,Bedingham:2010hz,Pearle:2014tda}). The non-unitarity in such a model is controlled by the strength parameter $\lambda_0$. We have a strong temptation for believing that $\lambda_0=0$, as fit for a unitary theory. The regime where quantum mechanics has been tested suggests that in microscopic system the strength parameter better be below the resolutions of the probe, while an estimated lower bound is obtained from theoretical considerations of the regimes where the quantum theory is anticipated to break down\footnote{Evidently, macroscopic systems do not follow quantum mechanics. Whether it is 
due to decoherence or the breakdown of unitarity is the topic of debate.}.  Thus, there are phenomenological constraints on the parameter $\lambda_0$ \cite{Bassi:2012bg}. In this 
model an unknown stochastic field always keeps perturbing systems non-unitarily and causes them to collapse into a particular eigenstate of the 
driving operator which appears in the evolution equation, given by
\begin{equation}\label{CSL1}
|\psi,t\rangle_w = \hat{\cal T}\exp\Big[-\int_{0}^{t}dt'\left\lbrace i\hat H+\frac{1}{4\lambda_{0}}\left(w(t')-2\lambda_{0}\hat A\right)^{2}\right\rbrace\Big]|\psi,0\rangle
\end{equation}
where $\hat {\cal T}$ is the time-ordering operator, $w(t)$ is a random, white noise type classical function of time (representing the stochastic field) with its probability distribution accounted by 
\begin{equation}\label{CSL2}
\textrm{P.D.}\lbrace w(t)\rbrace \equiv{}_w \langle\psi,t|\psi,t\rangle_w \prod_{t_{i}=0}^{t}\frac{dw(t_{i})}{\sqrt{ 2\pi\lambda_{0}/dt}},
\end{equation}
where the time has been divided in $N$ steps and the probability density of the stochastic field is specified with the specification at at step $i$, the stochastic field takes value between $w(t_i)$ and $w(t_i) + dw(t_i)$, specifying the functional behavior over the time domain in the process.
The parameter $\lambda_{0}$ (the \CSL parameter) essentially decides the rate of collapse of a system. In multi-particle systems this parameter gets rescaled and becomes large  such that collapse is faster. This evolution gives a description of nonlinear, non unitary evolution of a quantum system when probed with an observable $\hat A $ such that the dynamics does not allow any faster than light communication. 
Such an evolution drives the system to an eigenstate of $\hat A$, stochastically, with probability density as allowed by \ref{CSL2}. In the  non-relativistic settings, e.g., for a single particle,  this  scheme of evolution ensures that  there is a   spontaneous and steadfast  reduction of the state (popularly called the wavefunction collapse),  driven  by $\hat A \equiv \hat A[\hat {\vec X}]$, where  $\hat{\vec X}$ is the position operator\footnote{In fact $\hat A$ is a smeared $\hat{\vec X}$ operator, with the smearing telling us about the resolution of different positions} (since non-relativistic macroscopic particles appear to be localized in position basis). Recently Modak et al. \cite{Modak:2014vya} attempted a speculative generalization of \ref{CSL1} to 
relativistic (and field theoretic) settings and apply it in the black hole context. This is what we will elaborate next.
\subsubsection{\CSL evolution of the quantum state}

In standard \CSL theory, a quantum state is evolved according to \ref{CSL1}. As we discussed, for multi-particle system (where macroscopic physics begins to hold grasp) the \CSL parameter becomes large and the unitarity is effectively broken. Therefore, the parameter should depend on system parameters which in turn decide the effective strength of unitarity. In the black hole context, Modak et. al. \cite{Modak:2014vya} chose the parameter to become a  function of time parameter $\tau$ through its conjectured dependence on the local curvature of spacetime. In order to see these ideas in a clear setting, the \CSL evolution is applied to the modes on the CGHS (short form for Callan-Giddings-Harvey-Strominger \cite{Callan:1992rs,Giddings:1992ff}) black hole spacetime, with 
\begin{equation}
\lambda(\mathcal{R}) = \lambda_0\left(1+ (\mathcal{R}/\mu)^{\gamma}\right),
\label{lr}
\end{equation}
where $\mathcal{R}$ stands for the Ricci scalar evaluated for the CGHS black hole, with $\gamma\ge 1$ \cite{Modak:2014vya}. Parameter $\mu$ typically depicts a parameter of the theory. In the CGHS black hole case, this parameter is the cosmological constant in the theory. 

\begin{figure}
\sidecaption[t]
\begin{tikzpicture}[scale=1]
           \draw[fill=red!20]  (0,-3) -- (-3,0) -- (-1.5,1.5) -- (1.5,-1.5) -- cycle;
           \draw[fill=green!20]  (0,0) -- (1.5,-1.5) -- (3,0) -- (1.5,1.5) -- cycle;
           \draw[cyan!50, line width=1mm] (0,-3) -- (3,0) -- (1.5,1.5);
           \draw[cyan!50, line width=1mm] (0,-3) -- (-3,0) -- (-1.5,1.5);
           \draw[black, line width=0.1mm] (0,-3) -- (3,0) -- (1.5,1.5);
           \draw[black, line width=0.1mm] (0,-3) -- (-3,0) -- (-1.5,1.5);
           \draw[red!80, snake it, line width=1mm] (-1.5,1.5) -- (1.5,1.5);
           \draw[black, snake it, line width=0.1mm] (-1.5,1.5) -- (1.5,1.5);
           \draw[yellow, line width=1mm] (1.5,1.5) -- (0,0) -- (-1.5,-1.5);
           \draw[black,line width=0.5mm] (1.5,1.5) -- (0,0);
           \draw[black, dashed, line width=0.1mm] (0,0) -- (-1.5,-1.5);
           \draw[blue, line width=1mm] (1.5,-1.5) -- (0,0) -- (-1.5,1.5); 
           \draw[magenta!80, line width=0.5mm, bend right=30] (-3,0) to (3,0);
           \draw[black, line width=0.1mm, bend right=30] (-3,0) to (3,0);
           \draw[magenta!80, line width=0.5mm, bend left=30] (-3,0) to (3,0);
           \draw[black, line width=0.1mm, bend left=30] (-3,0) to (3,0);
           \draw[magenta!80, line width=0.5mm, bend left=50] (-3,0) to (3,0);
           \draw[black, line width=0.1mm, bend left=50] (-3,0) to (3,0);
           \node[label=right:$\mathcal{J}_{R}^{+}$] at (2.0,1.0) {};
           \node[label=right:$\mathcal{J}_{R}^{-}$] at (2.0,-1.0) {};
           \node[label=right: $x_{p}^{+}$] at (1.5,-1.6) {};
           \node[label=below: $x^{-}$] at (-1.0,-2.0) {};
           \node[label=left:$\mathcal{J}_{L}^{+}$] at (-2.0,1.0) {};
           \node[label=left:$\mathcal{J}_{L}^{-}$] (8) at (-2.0,-1.0) {};
           \node[label=below: $x^{+}$] at (1.0,-2.0) {};
           \node[label=above: $i_{L}^{+}$] at (-1.5,1.5) {};
           \node[label=above: $i_{R}^{+}$] at (1.5,1.5) {};
           \node[label=below: $i^{-}$] at (0,-3) {};
           \node[label=above: $\textrm{singularity}$] at (0,1.5) {};
           \draw[->,black, line width=0.5mm] (1.0,-3.0) to (1.5,-2.5);
           \draw[->,black, line width=0.5mm] (-1.0,-3.0) to (-1.5,-2.5);
           \node[label=above: $\Sigma _{i}$] at (0,-1.0) {};
           \node[label=below: $\Sigma_{0}$] at (-1.5,0.8) {};
           \node[label=below: $\Sigma_{f}$] at (-1.5,1.2) {};
           \node[label=below: $\textrm{linear}$] at (-0.5,-1.0) {};
           \node[label=below: $\textrm{dilaton}$] at (-0.4,-1.3) {};
           \node[label=below: $\textrm{vacuum}$] at (-0.3,-1.65) {};
           \node[label=below: $\textrm{BH exterior}$] at (1.5,0.5) {};
           \node[label=below: $\textrm{BH interior}$] at (0,1.4) {};
           \end{tikzpicture}
\caption{The Penrose diagram depicts a CGHS black hole formed by a null shell (shown as blue thick left moving line) collapse. There are three spacetime regions --- (a) the linear dilaton vacuum, (b) the black hole exterior and (c) the black hole interior. Also three Cauchy surfaces in this spacetime is also illustrated. See text for more discussions.}
\label{fig_Non_Unitary}
\end{figure}
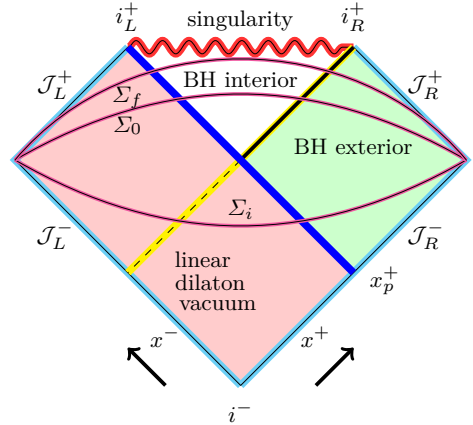

The initial state of the system is one in which there is a leftwards traveling pulse, collapsing to form a black hole and vacuum elsewhere.  The state can be viewed as a direct product of left-moving and right-moving fields. For the right moving modes, which would be traveling from ${\cal I}^{-}_{L}$ to ${\cal I}^{+}_{R}$ (see \ref{fig_Non_Unitary}) the state is ``in'' vacuum (vacuum state at past infinity in time), which can also be expressed in the ``out'' basis (late time future) as a superposed state (precisely due to the fact the out basis differs from the in basis, evident by non trivial Bogoliubov transformation coefficients \cite{Lochan:2016cxt}). The state using the out basis of right-moving modes is expressed as
\begin{equation}
|\Psi\rangle_{\rm in} =|\Psi\rangle_R \otimes \kets{\rm Pulse}_{L} = N \displaystyle\sum_{F_{n_j}} C_{F_{n_j}} \kets{F_{n_j}}^{\rm ext} \otimes \kets{F_{n_j}}^{\rm int}\otimes \kets{\rm Pulse}_{L}, \label{inst}
\end{equation}
where  $F_{n_j}$ marks the eigenvalue of the operator $\hat{N}_{n_j}^{\rm int}$, when acted upon the state  $|F\rangle^{\rm int}$, in the Fock basis. Indices ``int'' and ``ext'' mark the region interior and exterior to the horizon and $\kets{\rm Pulse}_{L}$ depicts the state of the matter pulse which moves leftwards and collapses to form the black hole. The evolution equation given by \ref{CSL1} drives the the state vector for right moving modes to become
\begin{equation}
|\Psi, \tau\rangle_R = N\sum_F C_{F_{n_j}} e^{-\int_o^\tau d\tau' [\frac{1}{4\lambda} \sum_{n,j}(w_{n_j} - 2\lambda F_{n_j})^2]} |F_{n_j}\rangle^{\rm int}_R\otimes|F_{n_j}\rangle^{\rm ext}_R,
\label{stev}
\end{equation}
where $w_{n_j}$ is the stochastic field, the mode $n_j$ is subject to.

The Cauchy surfaces are parameterized with proper time $\tau$, and the Ricci curvature $\mathcal{R}$ 
is a function of $\tau$ \cite{Modak:2014vya}. As $\tau$ approaches $\tau_s$, the Cauchy slices (such as $\Sigma_f$) graze very close to the spacetime singularity, and $\mathcal{R}$ starts diverging on a portion of the Cauchy surface. This divergence in $\mathcal{R}$ and hence the subsequent divergence in the parameter $\lambda_0$, makes the integral in the exponent of \ref{stev} divergent, and  causes the initial state to collapse into a state with definite quantum numbers $n,j$ corresponding to the out-basis. Thus, a superposed state in out basis makes a transition to a definite state
\begin{equation}\label{Review_Rev_01}
\lim_{\tau \rightarrow \tau_s} |\Psi, \tau\rangle_R = N C_{F_{n_{0} j_{0}}} |F_{n_{0}j_{0}}\rangle^{\rm int}_R\otimes|F_{n_{0}j_{0}}\rangle^{\rm ext}_R,
\end{equation}
on the hypersurface $\Sigma_f$ (see \ref{fig_Non_Unitary}) corresponding to $\tau \rightarrow \tau_s$.  Further, the entanglement between interior and the exterior modes causes the modes outside the horizon also to collapse. Thus, due to the effectiveness of the non unitary process on $\Sigma_f,$ due to diverging curvature on a portion of it, the mode functions make a non unitary jump destroying appreciable information very rapidly, almost instantly, something they would have done over a large time otherwise. The resultant out mode's  excitation (i.e. the number eigenstate the wavefunction settles into) depends on the realization of the stochastic value $w_{n_j} (\tau_s)$ in a collapse process, the final state after collapse, although remains pure (a number eigenstate), but is undetermined (due to a stochastic selection of the number eigenstate). Therefore, the ensemble level, we have a density matrix description which is not obtainable from a single wave function, since the system lands up into different 
number eigenstates in different realizations. It has further been argued \cite{Modak:2014vya} that the choice of driving operator of the collapse does not really matter in such a process, the collapse driven by other operators gives rise to effectively the same physical consequences at the ensemble level. Therefore, as an oxy-moronic outcome, the state of the matter field makes a jump from pure to pure state, (which was not necessarily required by a non unitary theory, but \CSL does it for free) though not unitarily but stochastically!  And hence we have a mixed thermal density matrix at the ensemble level.

It can further be shown that the character of the pure state on $\Sigma_f$ is such that the reduced ensemble description (through density matrix) for the modes moving rightwards in the exterior region of the black hole is thermal.
Through time evolution \ref{CSL1}, the density matrix (for the right-moving modes) at a subsequent time, can be  written as
\begin{equation}\label{rhot}
\rho_R (\tau) = N^2 \sum_{F,G} e^{-\frac{\pi}{\Lambda} (E_F +E_G )} e^{- \sum_{n_j} (F_{n_j}-G_{n_j})^2\int_{\tau_0}^{\tau}d\tau'\frac{\lambda(\tau')}{2}} 
\kets{F}^{\rm int}_R\otimes\kets{F}^{\rm ext}_R \bras{G}^{\rm int}_R \otimes\bras{G}^{\rm ext}_R,
\end{equation}
where $\kets{F}$ and $\kets{G}$ give states of Fock bases.
Near the singularity, the \CSL parameter $\lambda_0$ grows very large in the exponential factor, causing a collapse to a definite number state, thereby diagonalizing the density matrix (omitting $n,j$ from subscript for convenience) in number eigen-basis:
\begin{equation}
\lim_{\tau\to\tau_s} \rho_R (\tau) = N^2 \sum_{F} e^{-\frac{2\pi}{\Lambda} E_F}  \kets{F}^{\rm int}_R\otimes\kets{F}^{\rm ext}_R \bras{F}^{\rm int}_R \otimes\bras{F}^{\rm ext}_R, 
\end{equation}
where $E_F = \sum_{n_j} \omega_{n_j}F_{n_j}$  is the total energy of the final  excited state and $\Lambda$ is cosmological constant in the theory (which defines the parameter $\mu$ discussed above in  \ref{lr}), which also plays the role of temperature of the radiation in 2-dimensions.
We include the left moving matter pulse as well to give the state of the field on the full Cauchy surface, so that
\begin{eqnarray}
\lim_{\tau\to\tau_s} \rho (\tau) =N^2 \sum_{F} e^{-\frac{2\pi}{\Lambda} E_F}  \kets{F}^{\rm int}_R\otimes\kets{F}^{\rm ext}_R \bras{F}^{\rm int}_R \otimes\bras{F}^{\rm ext}_R \otimes \kets{\rm Pulse}_L\bras{\rm Pulse}_L.
\label{rhot3}
\end{eqnarray}
Hopefully, when a quantum theory of gravity becomes available, it will cure the singularity, which the black hole carefully hides. 
If the singularity of the black hole is resolved through some quantum theory of gravity, one hopes that the spacetime will be extended and we can think of many more Cauchy surfaces passing through the extended region. In that case the spacetime can be extended beyond the singularity and the portion of initial data which fell inside the horizon can well be available in some modified form (see \cite{Ashtekar:2008jd} for a demonstration in two dimensions). Though any  fully  workable theory of that kind is not available till date, one can make a  few  natural assumptions regarding such a futuristic possible theory ---  (a) The boundary of spacetime in the extended region are trivial and (b) No large violations of energy conservation are allowed.  In that case,  quantum gravity combines the negative energy state $|F_0\rangle^{\rm int}_R$ (describing negative energy modes falling inside complementing the positive energy flux of the Hawking radiation) with the matter forming the hole, characterized by $|{\rm Pulse}
\rangle _{L}$ to arrive at
\begin{equation}
|F_0\rangle^{\rm int}_R \otimes |{\rm Pulse}\rangle_L \rightarrow |{\rm p.s}\rangle,
\end{equation}
where $F_0$ is thought of as a specific  particle excited state $F_{n_{0}j_{0}}$ in the Fock basis and $|{\rm p.s}\rangle$ is a post-singularity quantum state which has negligible energy (marking that of the Planck scale `remnant', i.e., the place where quantum gravity kicks in). Thus, on the final hypersurface $\Sigma_f$, the quantum state (in a single realization) is a 
number eigenstate $|F_0\rangle$ times the post singularity state $|{\rm p.s}\rangle$ and the density matrix (in the ensemble of stochastically selected number eigenstates) has an effective thermal description,
\begin{eqnarray}
&&|\Psi \rangle = N e^{-\frac{\pi}{\Lambda} E_{F_0}} |F_0\rangle^{\rm ext}_R \otimes |{\rm p.s}\rangle, \hspace{1 cm} \text{in a particular realization;}\\
&&\rho =N^2 \sum_{F} e^{-\frac{2\pi}{\Lambda} E_F}  \kets{F}^{\rm ext}_R \bras{F}^{\rm ext}_R \otimes \kets{\rm p.s} \bras{\rm p.s}, \\
&&~~ = \rho^{\rm ext}_{\rm thermal}\otimes \kets{\rm p.s} \bras{\rm p.s}.
\label{rhotC}
\end{eqnarray}
Therefore, at the ensemble level, the description  on $\Sigma _{f}$, effectively becomes that of {\it a mixed  state} (with the density matrix being  clearly thermal for the exterior region modes in the right-moving section) with a very low energy contribution from post singular region characterizing the remaining portion of $\Sigma_f$ ( taken to  be  a  portion of  flat  spacetime, due to vanishingly small energy associated with the post singularity state). Thus, in this non-unitary evolution, leading to a wave-function collapse, {\it information is essentially lost}, mainly in the interior of the black hole. The abandoning of conservation of information, as a principle of nature, suggests that 
nothing paradoxical happens in a black hole evaporation \cite{Modak:2016uwr,Okon:2016qlh,Bedingham:2016aus,Modak:2014vya,Okon:2014dpa,Modak:2014qja}.
\section{Black holes have (soft) hairs}\label{Section_Soft}

One of the key concepts associated with the information loss paradox is the existence of no hair theorems, according to which, the black holes do not carry any additional charges except mass, electric charge and angular momentum. If, one can show somehow that this is not really true, and the black holes may contain many more additional charges present, then that would be a very economical plausible solution to the information loss paradox. May be those extra charges inherit all the additional informations about what has fallen into the black hole. This is precisely what was proposed in \cite{Hawking:2016msc,Hawking:2016sgy} by Hawking, Perry and Strominger. They argued that certain symmetries of the horizons allow black holes to contain soft hairs which carry the information of the in-fallen matter to the asymptotia.

The existence of such additional hairs (actually infinitely many) for the event horizons can open up new directions of exploration. In fact, this phenomenon is really not peculiar to black hole horizons only, but is a generic property of any null surface \cite{Chakraborty:2016dwb,Chakraborty:2015hna}. To understand any new concept, it is best to illustrate it in the simplest context. We will follow exactly the same route here and discuss the situation and existence of these hairs for asymptotically flat spacetimes at null infinity. 

The fact that asymptotically flat spacetimes inherit non-trivial hairs were known from the much earlier works by Bondi, Metzner and Sachs (for short BMS) \cite{Bondi:1962px,Sachs:1962wk}. In \cite{Bondi:1962px,Sachs:1962wk} they have observed that in addition to the Poincar\'{e} symmetry, there exist an infinite number of diffeomorphisms (in modern language supertranslations) that change physical data at past or future null infinity non-trivially\footnote{Any such symmetry of spacetime which enhances the global symmetry group under consideration are known as large diffeomorphisms. Thus supertranslation is an example of a large diffeomorphism due to enhancement of global Poincar\'{e} symmetry.}. Since any symmetry must have an associated conserved charge, it follows as an immediate consequence, that the asymptotic future or past null infinity are endowed with an infinite set of supertranslation charges (or supertranslation hairs). The conservation of these charges was observed more explicitly,
 as it was recently demonstrated that{\it  a  particular (antipodal) combination of past and future supertranslation charges is an exact symmetry of gravitational scattering} \cite{Strominger:2013jfa,Weinberg:1965nx,Cachazo:2014fwa}. This suggests 
that total incoming energy of gravitational radiation at some angle is related to the outgoing energy at some other angle with antipodal connection. In the context of quantum field theory such conservation relations were derived by Weinberg \cite{Weinberg:1965nx} and is known as the soft graviton theorem. This leads to another realization of spontaneous symmetry breaking, albeit in a gravitational context. The infinite number of supertranslation symmetries introduce an infinite number of inequivalent vacua differing by exchange of one or more soft gravitons, which have zero energy but non-zero angular momentum. When one vacuum is chosen over the others, it breaks the asymptotic symmetry spontaneously, leading to Goldstone bosons, which in this case, are just the soft gravitons \cite{Hawking:2016msc,Hawking:2016sgy}. 

This has two associations with the information paradox --- (a) The derivation of the thermal Hawking radiation depends crucially on the existence of a unique vacuum. However in light of the discussion we had above, it follows that after the black hole formation (or evaporation), the quantum field may settle in one of the infinite degenerate vacua, such that quantum purity is maintained; (b) Black holes have an infinite number of soft hairs. During the evaporation of black holes the supertranslation charge will be radiated away to infinity. However due to conservation of supertranslation charge, total charge in the radiation and the remains of the hole would be conserved. The same must work even after the black hole has fully evaporated and may result in correlations between early and late Hawking radiation. It must be emphasized that in order for the soft hairs to turn the table around, i.e, to prove unitarity in black hole evaporation, one has to demonstrate that all informations are 
essentially conserved charges. Otherwise, soft hairs will not able to account for all the informations thrown into the black hole and unitarity would still be lost.

To make the reader familiar with supertranslation symmetry we will work with asymptotically flat spacetimes and shall introduce supertranslations there, before jumping into discussions on black holes.
\subsection{Supertranslation}\label{Supertranslations}

BMS supertranslations are an infinite set of diffeomorphisms transforming one asymptotically flat solution to another, albeit a unitarily inequivalent one. At the classical level these supertranslations can be divided into two parts --- (a) diffeomorphisms keeping the structure of future null infinity $\mathcal{J}^{+}$ invariant, (b) diffeomorphisms that preserve the past null infinity $\mathcal{J}^{-}$. It turns out that in the quantum domain the supertranslation symmetry manifests in the result that outgoing states at $\mathcal{J}^{+}$ live in the Hilbert space on which the BMS generators act upon \cite{Ashtekar:1978zz,Ashtekar:1981sf,Weinberg:1965nx,Kulish:1970ut,Ware:2013zja,Barnich:2009se,Barnich:2011ct,Barnich:2011mi}. However originally, the symmetries at $\mathcal{J}^{+}$ and those at $\mathcal{J}^{-}$ were treated differently, with no canonical identifications between them. This shortcoming was removed in \cite{Strominger:2013jfa}, where the symmetries both at future and past 
null infinities were connected through an antipodal matching. The same exercise also leads to the result that total incoming energy flux integrated over $\mathcal{J}^{-}$ with some null generator is equal to the integral at $\mathcal{J}^{+}$ with a continuation of the previous null generator \cite{Strominger:2013jfa}. This happens precisely due to the existence of soft gravitons carrying zero total energy but can contribute \emph{locally}. In the quantum version this conservation takes the form of a Ward identity which is essentially the soft graviton theorem of Weinberg \cite{Strominger:2013jfa,Weinberg:1965nx,Cachazo:2014fwa}. These results were derived in the context of four dimensional spacetime, possible generalizations to higher spacetime dimensions were carried out in \cite{Kapec:2015vwa} (see however \cite{Hollands:2016oma} as well).

To understand the supertranslation invariance, let us start with the metric in asymptotically flat spacetime, which can be written in terms of retarded null coordinates at future null infinity (with $r\rightarrow \infty$) to leading order as,
\begin{align}\label{Eq_Asymp_Flat}
ds^{2}&=-du^{2}-2dudr+2r^{2}\gamma _{z\bar{z}}dzd\bar{z}
\nonumber
\\
&+\frac{2m_{B}}{r}du^{2}+rC_{zz}dz^{2}+rC_{\bar{z}\bar{z}}d\bar{z}^{2}+D^{z}C_{zz}dudz+D^{\bar{z}}C_{\bar{z}\bar{z}}dud\bar{z}.
\end{align}
Here, we have introduced a set of complex coordinates $(z,\bar{z})$ related to the original $(\theta,\phi)$ coordinates such that, $z=\cot (\theta/2)\exp(i\phi)$ and $\bar{z}=\cot (\theta/2)\exp(-i\phi)$. The two-metric $\gamma _{z\bar{z}}$ in the coordinate system $(z,\bar{z})$ takes the following form: $2/(1+z\bar{z})^{2}$. Further one obtains, $D^{z}C_{zz}=\gamma ^{z\bar{z}}\partial _{\bar{z}}C_{zz}$ and $D^{\bar{z}}C_{\bar{z}\bar{z}}=\gamma ^{\bar{z}z}\partial _{z}C_{\bar{z}\bar{z}}$ respectively. Note that the derivative operator $D_{z}$ is actually the covariant derivative acting on the transverse space. It turns out that the above form of the metric remains invariant under a diffeomorphism $x^{a}\rightarrow x^{a}-\zeta ^{a}$ having the following structure (see \ref{App_Supertranslation} for a detailed, first principle derivation),
\begin{equation}\label{Eq_Diff_Super}
\zeta^{a}(f)\partial _{a}=f(z,\bar{z})\partial _{u}+D^{z}D_{z}f(z,\bar{z})\partial _{r}-\frac{1}{r}\left(D^{z}f\partial _{z}+D^{\bar{z}}f\partial _{\bar{z}}\right)
\end{equation}
This is equivalent to a coordinate transformation which reads,
\begin{align}
u\rightarrow u-f(z,\bar{z});\qquad r\rightarrow r-D^{z}D_{z}f(z,\bar{z})
\nonumber
\\
z\rightarrow z+\frac{1}{r}D^{z}f;\qquad \bar{z}\rightarrow \bar{z}+\frac{1}{r}D^{\bar{z}}f
\end{align}
and is a symmetry of the asymptotically flat spacetime. The diffeomorphism vector $\zeta ^{a}$ is called the supertranslation. Interestingly, supertranslation is not the only symmetry associated with the asymptotically flat spacetime as in \ref{Eq_Asymp_Flat}, there exist two more such diffeomorphisms that keep the asymptotic structure of the spacetime invariant, which we will discuss next.
\subsection{Lorentz transformation and superrotation}\label{Superrotations}

Another set of vectors keeping the asymptotic structure of the manifold invariant are the standard Lorentz transformations, since they satisfy a $\textrm{SL}(2,C)$ Lie algebra. This should not come as a surprise, since the spacetime is asymptotically flat and Lorentz transformation is a symmetry of any flat spacetime. It turns out that the following vector does the job (see \ref{App_LT} for a justification)
\begin{equation}\label{Eq_Lorentz}
\xi ^{a}\partial _{a}=\frac{u}{2}D_{z}\zeta ^{z}\partial _{u}-\frac{r}{2}\left(1+\frac{u}{r}\right)D_{z}\zeta ^{z}\partial _{r}
+\left(1+\frac{u}{2r}\right)\zeta ^{z}\partial _{z}-\frac{u}{2r}\gamma ^{z\bar{z}}D_{z}^{2}\zeta ^{z}\partial _{\bar{z}},
\end{equation}
where, $\zeta ^{z}$ is a complex function of $z$ alone, treated as the $z$ component of a two dimensional vector $(\zeta ^{z},0)$ and hence, $D_{z}\zeta ^{z}=\partial _{z}\zeta ^{z}-(2\bar{z}/(1+z\bar{z}))\zeta ^{z}$, is a scalar. 

Recently, it has also been found that there exists another class of vectors, called superrotations \cite{Barnich:2009se,deBoer:2003vf,Banks:2003vp,Hogan:1993xj,Nutku:1977wp,Penrose:1972xrn,Strominger:2016wns}, which also keeps the structure of the asymptotic metric invariant. However there is one crucial difference. In the case of supertranslation or Lorentz transformation, the key role was played by the $(u,r)$ sector, while for superrotations the key role is being played by the angular coordinates. It turns out that the superrotations are also infinite in number and they act as the global conformal group on the celestial sphere at infinity. However finite superrotations are plagued with singularities at isolated points on the celestial sphere and in particular they map globally asymptotically flat spacetime to locally asymptotically flat spacetime \cite{Barnich:2009se,deBoer:2003vf,Banks:2003vp,Hogan:1993xj,Nutku:1977wp,Penrose:1972xrn}. Another physical interpretation for superrotation appears from decay 
and subsequent evolution of cosmic strings through black hole pair creation. To be precise, the initial and final metrices differs in such a scenario by a finite superrotation \cite{Strominger:2016wns}.

The superrotation involves transformation of the angular coordinates on the two sphere, such that, $z\rightarrow w(z)$, with $w$ being a locally holomorphic function. Then it turns out that if one starts with the global flat spacetime one ends up in getting a non-zero $C_{zz}$ and becomes locally asymptotically flat, given by \ref{Eq_Asymp_Flat}. The corresponding $C_{zz}$ generated due to the above transformation corresponds to,
\begin{align}\label{Eq_Superrot}
C_{zz}\sim \left\lbrace w,z\right\rbrace ;\qquad \left\lbrace w,z\right\rbrace =\frac{w'''}{w'}-\frac{3}{2}\left(\frac{w''}{w'}\right)^{2}.
\end{align}
Thus the quantity $C_{zz}$ changes by the Schwarzian derivative which has found extensive use in two dimensional conformal field theory. As already mentioned earlier, such a change of $C_{zz}$ between early and late spacetime can be achieved by the physical process of snapping of cosmic string \cite{Strominger:2016wns}. However we would like to mention that the coordinate transformation generating superrotations satisfying \ref{Eq_Superrot} are not simply $z\rightarrow w(z)$ as one can immediately verify, but a rather complicated one. We provide an explicit demonstration of the coordinate transformation as well as its connection with \ref{Eq_Superrot} in \ref{App_superrotation}.
\subsection{Soft photon hair on black holes: An illustration}

In this section we will try to explore how one can infer soft photon hairs on a black hole as an illustration of the fact that black holes must carry infinitely many soft hairs. To start with, one notes that the metric at future null infinity $\mathcal{J}^{+}$ can be written in the retarded coordinates $(u,r,z,\bar{z})$ as in \ref{Eq_Asymp_Flat} and the metric at past null infinity $\mathcal{J}^{-}$ must be written in advanced coordinates $(v,r,z,\bar{z})$. Here $u=t-r$ and $v=t+r$ are the standard flat spacetime null coordinates. Further we relate the angular coordinates on $\mathcal{J}^{+}$ to those on $\mathcal{J}^{-}$ antipodally, i.e., $z\rightarrow -(1/\bar{z})$. One can easily verify that the two-sphere metric remains invariant under such a transformation.

\begin{figure}[h]
\sidecaption[t]
\begin{tikzpicture}[scale=1]
           \draw[dashed,line width=0.5mm] (0,1.5) --(0,-3); 
           \draw[orange!60,snake it,line width=1mm] (-2.5,1.5) -- (2.5,1.5);
           \draw[black,snake it,line width=0.1mm] (-2.5,1.5) -- (2.5,1.5);
           \draw[yellow, line width=1mm] (0,-3) -- (3.5,0.5) -- (2.5,1.5);
           \draw[black,line width=0.1mm] (0,-3) -- (3.5,0.5) -- (2.5,1.5);
           \draw[yellow, line width=1mm] (0,-3) -- (-3.5,0.5) -- (-2.5,1.5);
           \draw[black, line width=0.1mm] (0,-3) -- (-3.5,0.5) -- (-2.5,1.5);
           \draw[blue!70,line width=1mm] (2.25,-0.75) -- (1.25,0.25) -- (0,1.5); 
           \draw[blue!70, line width=1mm] (-2.25,-0.75) -- (0,1.5);
           \draw[red, dashed, line width=0.5mm]  (2.5,1.5) -- (0,-1);
           \draw[red, dashed, line width=0.5mm]  (-2.5,1.5) -- (0,-1);             
           \node[label=right:$\mathcal{J}^{+}$] at (2.8,1.2) {};
           \node[label=right:$\mathcal{J}^{-}$] at (1.8,-1.4) {};
           \node[label=above:$\mathcal{J}^{+}_{+}$] at (-2.5,1.5) {};
           \node[label=above:$\mathcal{J}^{+}_{-}$] at (-3.5,0.5) {};
           \node[label=left: $v$] at (1.75,-1.2) {};
           \draw[->,black,line width=0.2mm] (1.6,-1.0) -- (2.0,-0.60);
           \draw[->,black, line width=0.2mm] (1.4,-1.4) -- (1.0,-1.8);
           \node[label=below: $\mathcal{J}^{-}_{+}$] at (-3.5,0.5) {};
           \node[label=below: $\mathcal{J}^{-}_{-}$] at (0,-3.0) {};
           \node[label=below: $u$] at (2.8,1.2) {};
           \draw[->,line width=0.2mm] (2.8,1.0) to (2.5,1.3);
           \draw[->,line width=0.2mm] (2.9,0.8) to (3.3,0.4);
           \node[label=left: $\mathcal{H}$] at (-0.5,-0.5) {};
           \end{tikzpicture}
\caption{The Penrose diagram depicts the collapse of a null shell in isotropic coordinates forming a black hole. The future null infinity is being depicted as $\mathcal{J}^{+}$, while the past null infinity as $\mathcal{J}^{-}$. The future and past of $\mathcal{J}^{+}$ are denoted as $\mathcal{J}^{+}_{+}$ and as $\mathcal{J}^{+}_{-}$ respectively. The same for past null infinity is being denoted by $\mathcal{J}^{-}_{\pm}$ respectively. The coordinate $u$ describes the future null infinity and the coordinate $v$ explores the past null infinity. The black hole horizon is being denoted by $\mathcal{H}$.}\label{fig_Soft_Hair}
\end{figure}
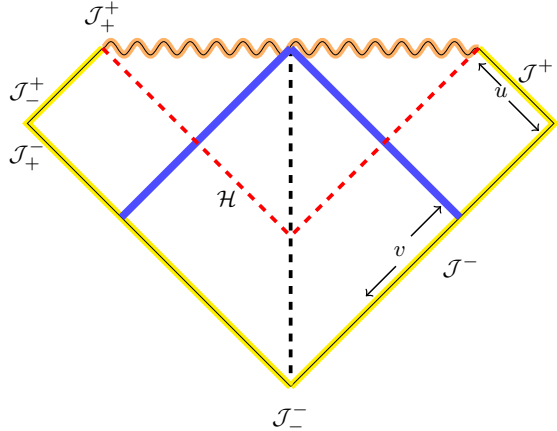

It is useful to first analyze the case with the electromagnetic filed, before jumping to gravitational field. The Maxwell's equation in the asymptotically flat spacetime regions reads $\nabla _{a}F^{ab}=e^{2}J^{b}$, where $e$ is the electric charge, $J^{b}$ is the electromagnetic four-current and $F^{ab}$ being the electromagnetic field tensor. One also introduces the future and past of $\mathcal{J}^{+}$ by $\mathcal{J}^{+}_{\pm}$ and that for $\mathcal{J}^{-}$ being $\mathcal{J}^{-}_{\pm}$ respectively. Since the past of $\mathcal{J}^{+}$ and the future of $\mathcal{J}^{-}$ are essentially the same spacetime point (see \ref{fig_Soft_Hair}) the corresponding Maxwell fields at those two points must be identical. Since at asymptotic infinity the field strength varies as $\sim r^{-2}$, one obtains the only non-vanishing component of the field tensor varying as $\sim r^{-2}$ to satisfy $F_{ru}(\mathcal{J}^{+}_{-})=F_{rv}(\mathcal{J}^{-}_{+})$. This relation is associated with conservation of an infinite number 
of charges $Q_{\epsilon}$, associated with an arbitrary function $\epsilon (z,\bar{z})$ on 
the two sphere, such that $Q_{\epsilon}^{+}=Q_{\epsilon}^{-}$, with
\begin{eqnarray}
Q_{\epsilon}^{+}&=\frac{1}{e^{2}}\int _{\mathcal{J}^{+}_{-}}\epsilon *\mathbf{F},\\
Q_{\epsilon}^{-}&=\frac{1}{e^{2}}\int _{\mathcal{J}^{-}_{+}}\epsilon *\mathbf{F}.
\end{eqnarray}
Here $Q_{\epsilon}^{+}$ is defined on the two surface $\mathcal{J}^{+}_{-}$ and the charge $Q_{\epsilon}^{-}$ defined on $\mathcal{J}^{-}_{+}$. Thus by Gauss theorem one can convert them to integrals over the total future and past null infinity respectively, such that,
\begin{align}
Q_{\epsilon}^{+}&=\frac{1}{e^{2}}\int _{\mathcal{J}^{+}_{-}}\epsilon *\mathbf{F}=\frac{1}{e^{2}}\int _{\mathcal{J}^{+}}\mathbf{d}\epsilon \wedge *\mathbf{F}+\frac{1}{e^{2}}\int _{\mathcal{J}^{+}}\epsilon \mathbf{d}*\mathbf{F}
=\frac{1}{e^{2}}\int _{\mathcal{J}^{+}}\mathbf{d}\epsilon \wedge *\mathbf{F}+\int _{\mathcal{J}^{+}}\epsilon *\mathbf{J}
\end{align}
where in the last step we have used the Maxwell's equations to relate derivatives of $F^{ab}$ with the matter current $J^{a}$. A similar result holds for $Q_{\epsilon}^{-}$ as well. However there is one caveat. The integrals must be performed over Cauchy surfaces which has $\mathcal{J}^{+}_{-}$ or $\mathcal{J}^{-}_{+}$ as boundary. The future null infinity forms a Cauchy surface \emph{only} when the spacetime inherits no black holes. In presence of black holes $\mathcal{J}^{+}$ ceases to be Cauchy surfaces while the surface $\mathcal{J}^{+}\oplus \mathcal{H}$ is still a Cauchy surface. Hence one obtains,
\begin{align}
Q_{\epsilon}^{+}&=Q^{\mathcal{H}}_{\epsilon}+\frac{1}{e^{2}}\int _{\mathcal{J}^{+}}\mathbf{d}\epsilon \wedge *\mathbf{F}+\int _{\mathcal{J}^{+}}\epsilon *\mathbf{J};\qquad Q_{\epsilon}^{\mathcal{H}}=\frac{1}{e^{2}}\int _{\mathcal{H}}\mathbf{d}\epsilon \wedge *\mathbf{F}+\int _{\mathcal{H}}\epsilon *\mathbf{J}
\end{align}
Thus one naturally ends up with soft photons on the horizon as dictated by the global structure of the spacetime. 

Let us explore the classical consequences of $Q_{\epsilon}^{\mathcal{H}}$. By no-hair theorem one must have the charge to be vanishing on the horizon except for the monopole term $\ell =0$. Thus, classically the charge on the horizon vanishes. However in quantum domain these classical charges must act as operators on states in the Hilbert space. Even though the classical charge vanishes, the quantum operators are definitely non-zero. In particular, returning back to the Minkowski spacetime without any source term, $Q_{\epsilon}^{+}$ when acted on a vacuum state yields,
\begin{align}
Q_{\epsilon}^{+}|0\rangle =\left(\frac{1}{e^{2}}\int _{\mathcal{J}^{+}}\mathbf{d}\epsilon \wedge *\mathbf{F}\right)|0\rangle
=|\mathbf{d}\epsilon \rangle.
\end{align}
Thus the final state is a new vacuum \emph{but} with an additional soft photon with polarization $\mathbf{d}\epsilon$. Since the vacuum expectation value of $Q_{\epsilon}^{+}$ is still zero, the above result is consistent with vanishing classical value of the charge. An identical scenario applies to the horizon as well. Assuming neutral matter forming the hole and the quantum state of the field being $|M\rangle$ it follows that,
\begin{align}
Q_{\epsilon}^{\mathcal{H}}|M\rangle=\left(\frac{1}{e^{2}}\int _{\mathcal{H}}\mathbf{d}\epsilon \wedge *\mathbf{F}\right)|M\rangle =|M+\mathbf{d}\epsilon \rangle,
\end{align}
which is a new quantum state with an additional soft photon with polarization $\mathbf{d}\epsilon$. Thus two stationary black holes having the same mass and angular momentum are distinguished by their soft electric hairs. This can, in principle, be a measurable quantity and for black holes the best procedure to do that is to look at its evaporation. The state $|M\rangle$ evaporates to another quantum state $|X\rangle$, while the state $|M+\mathbf{d}\epsilon \rangle$ evolves to some other state $|X'\rangle$ due to evaporation. Even though the states $|X\rangle$ and $|X'\rangle$ are energetically degenerate and hence classically indistinguishable, nevertheless these two states are quantum mechanically very different --- they differ by existence of soft electric hairs. Hence there exist no no-hair theorem in a quantum world.

One can also devise physically realizable processes by which such soft electric hairs can be implanted on a black hole. The most straightforward one is to throw a charge with $\ell >0$ towards future infinity generating $Q_{\epsilon}^{+}$, or throw the charge to the black hole interior leading to nonzero $Q_{\epsilon}^{\mathcal{H}}$. This can also be achieved by using a null shock wave (for such an explicit construction see \cite{Hawking:2016msc}). The above has an interesting implication vis-a-vis entropy area relation. Due to position momentum uncertainty relation, if such a null shock wave is too much localized, it will have energy larger compared to the Planck scale. Thus the minimum area covered by such soft photon wave function seems to be $\sim L_{P}^{2}$, such that total number of soft electric hair becomes, $\sim A/L_{P}^{2}$ resembling Bekenstein's area law.  

We can briefly account for such gravitational hairs as well, in a similar spirit. At asymptotic infinity which is locally flat, one has the BMS supertranslations --- an infinite number of vector fields preserving the asymptotic structure of the spacetime. There is also an additional conservation of energy related to soft graviton theorem \cite{Weinberg:1965nx,He:2014laa} (for an elaborate discussion see \ref{Supertranslations}). One can follow an identical route to derive the corresponding charge at the black hole horizon. This can be achieved by requiring invariance of the near horizon metric under a general diffeomorphism. It turns out that such diffeomorphism vector fields besides yielding supertranslation charges at the horizon also help in many other ways. For example, using the Gaussian null coordinates \cite{Parattu:2015gga} to represent the metric around any null surface, one can compute the corresponding diffeomorphism that keeps the structure of the metric invariant. It turns out that for a 
subclass of this diffeomorphism vector field, one obtains a BMS-like algebra in both three and four dimensions. This puts the horizon diffeomorphisms in direct connection with diffeomorphisms keeping asymptotic structure invariant \cite{Donnay:2015abr,Compere:2016hzt,Eling:2016xlx}. Use of these horizon diffeomorphisms also leads to the area entropy relation for a class of arbitrary null surfaces by exploiting the Virasoro algebra of the generators of these diffeomorphisms \cite{Chakraborty:2016dwb}. Thus, this particular method of attaching new hairs to the horizon, introduces several new ideas which have varied implications on a much broader scale.

\section{A simple way to extract information --- Non-vacuum distortions}

\subsection{Hard (quantum) hairs on black holes?}

One of the most prominent intuition about the black hole is that whatever happens outside its horizon is really insulated from what falls inside, in the form of no hair theorem. In fact this expectation fails at quantum scrutiny, not only due to existence of soft hairs we just discussed above, but otherwise as well. Even the semiclassical analysis of the matter field outside the horizon, shows its dependence on the part which resides inside the black hole \cite{Lochan:2015oba, Lochan:2016nbs}, thus carrying the imprint outside and also questioning the factorizability of the Hilbert space as usually done : outside $\otimes$ inside.

The dependence on the interior can be studied through the dependence of the outside observable on the state fallen in.
The initial state of the field is specified at the past null infinity (${\cal J}^{-}$). The geometry at ${\cal J}^{-}$ is Minkowski-like and therefore the modes necessary for describing the quantum field at past asymptotic will be the flat spacetime modes. For the flat spacetime free-field theory, the in-falling field decomposition is given as
\bea
\hat{\phi}({x}) = \int \frac{d^3 {\bf k}}{\sqrt{2 \omega_{\bf k}}}(\hat{a}_{{\bf k}}e^{ik \cdot x}+\hat{a}_{{\bf k}}^{\dagger}e^{-ik \cdot x}) = \int d^3 {\bf k} \hat{\phi}_{\bf k}(t)e^{i{\bf k}\cdot {\bf x}},
\eea
where
\bea
 \hat{\phi}_{\bf k}(t) = \frac{(\hat{a}_{{\bf k}}e^{-i\omega_{\bf k} t}+\hat{a}_{{\bf-k}}^{\dagger}e^{i\omega_{\bf k}t})}{\sqrt{2 \omega_{\bf k}}},
\eea
satisfies $\hat{\phi}_{\bf k}=\hat{\phi}_{-{\bf k}}^{*}$ for a real scalar field (which we use for a conveniently simple discussion). One further defines $\hat{\bar{\phi}}_{\bf k}=\hat{a}_{{\bf k}}e^{-i\omega_{\bf k} t}$, such that
\bea
\sqrt{2 \omega_{\bf k}} \hat{\phi}_{\bf k} = \hat{\bar{\phi}}_{\bf k} + \hat{\bar{\phi}}^{\dagger}_{\bf -k}.
\eea
Therefore, specifying $\hat{\bar{\phi}}_{\bf k}$ is equivalent to specifying $\hat{\phi}_{\bf k}$, which describes the field configuration in terms of the Fourier momentum modes at $\mathcal{J}^{-}$. An observable (Hermitian operator) of the momentum correlation at $\mathcal{J}^{-}$ can further be  introduced as
\bea
\hat{N}_{{\bf k_1}{\bf k_2}} &\equiv& \hat{\bar{\phi}}_{\bf k_1} \hat{\bar{\phi}}^{\dagger}_{\bf k_2} + \hat{\bar{\phi}}_{\bf k_2} \hat{\bar{\phi}}^{\dagger}_{\bf k_1} \nonumber\\
&=& \hat{a}_{\bf k_1}\hat{a}^{\dagger}_{\bf k_2}e^{-i(\omega_{\bf k_1}-\omega_{\bf k_2})t}+\hat{a}_{\bf k_2}\hat{a}^{\dagger}_{\bf
k_1}e^{i(\omega_{\bf k_1}-\omega_{\bf k_2})t}.  \label{Cor-Op}
\eea
For a massless field in a spherically symmetric configuration, this operator can also measure the frequency correlation if we suppress the angular dependence. For demonstration purpose, we consider a field undergoing spherically symmetric (s-wave) collapse, slowly \cite{Visser:2014ypa,Gray:2015pma} to form a large mass black hole. The relevant  positive frequency modes describing the initial state will be
\bea
u_{\omega}(t,r,\theta,\phi) \sim \frac{1}{r\sqrt{\omega}}e^{-i\omega(t+r)}S(\theta,\phi)
\eea
where $S(\theta,\phi)$ gives a combination of spherical harmonics $Y_{lm}(\theta,\phi)$. For this collapsing case, the initial state will have to be in-moving  at ${\cal J}^{-}$ which is totally spherically symmetric, i.e., $l=0$. Once the event horizon is formed, the full state can again be described using a combined description at the event horizon ${\cal H}$ and the future null infinity (${\cal J}^{+}$) \cite{Padmanabhan:2009vy,Parker:2009uva} jointly as we discussed in the beginning. For an asymptotic observer, the out-state, will be out moving, which can be related to modes at ${\cal J}^{+}$ and which are again flat spacetime modes owing to the asymptotic flatness of the spacetime.  The field content of the out-state can be obtained by constructing suitable operators which are connected to operators on the ${\cal J}^{+}$ using the Bogoliubov 
coefficients between the 
modes at ${\cal J}^{-}$ and ${\cal J}^{+}$ \cite{Padmanabhan:2009vy, Parker:2009uva}. The asymptotic form of these Bogoliubov coefficients are obtained as \ref{BT} (see \cite{Hawking:1974sw}).
Though these Bogoliubov coefficients are actually accurate only for large values of $\omega$ we are interested in  the late-time radiation at future null infinity (${\cal J}^{+}$). One can show that the dominating spectra will come from those modes which have just narrowly escaped the black hole, i.e. which came out just before the formation of the event horizon. Such modes graze the horizon with extremely high frequencies. So the calculations done with \ref{BT} will be accurate to the leading order.
The infalling matter is thought (for simplicity, for generalizations see \cite{Lochan:2016nbs}) of as states which are eigenstates of the number operator defined with the help of the in-modes, but are not the energy or momentum eigenstates. Let the state of the field undergoing collapse in a black hole spacetime, be a (superposition of) single particle excitation state, write as
\bea
|\Psi\rangle_{in} = \int_0^{\infty} \frac{d \omega}{\sqrt{4 \pi \omega}}f(\omega)\hat{a}^{\dagger}(\omega)|0\rangle. \label{state1}
\eea
The information of the state completely resides in $f(\omega) $, which is turned into a new function of a variable related to the frequency $\omega$ of mode functions at ${\cal J}^{-}$, as  
\bea
\log{\frac{\omega}{C}} =z \Rightarrow f(C e^z)=g(z).  \label{vartrans}
\eea
In order to specify the state, we need to specify $f(\omega)$ or equivalently $g(z)$. One needs to check how much of the outgoing radiation really depends on $f(\omega)$ in violation of no-hair theorems. The black hole emission spectra will be characterized by the Fourier transform of $g(z)$
\bea
F(y)= \int_{-\infty}^{\infty} dz g(z) e^{i y z}. \label{F_FT}
\eea
It is useful to  construct yet another function with proper rescaling
\bea
\tilde{F}\left(\frac{\Omega}{\kappa}\right) = \exp{\left[\frac{\pi\Omega}{2\kappa}\right]} F\left(\frac{\Omega}{\kappa}\right), \label{Ftilde}
\eea
starting from \ref{F_FT}. Then one  can obtain the distribution function $g(t)$ from \ref{Ftilde} as
\bea
g(z)=\frac{1}{2\pi} \int_{-\infty}^{\infty} d \left(\frac{\Omega}{\kappa}\right) \tilde{F}\left(\frac{\Omega}{\kappa}\right) e^{-i \frac{\Omega}{\kappa}z}e^{-\frac{\pi \Omega}{2\kappa}}. \label{Def_gz}
\eea
We will now discuss the information about $\tilde{F}\left(\Omega/\kappa\right)$  available  in the out-going modes, by analyzing the frequency space correlations available to the late time observers. 
\subsection{Information of black hole formation: Correlation function}\label{Sec_02A}

We wish to obtain the information regarding the quantum states falling into the black hole, or more elaborately, regarding the quantum states which formed the black hole itself (by interpolating this exercise). The annihilation operator $\hat{b}_{\Omega}$ associated with the outgoing modes at  $\mathcal{J}^{+}$ is related to the creation and annihilation operator $\hat{a}_{\omega}$ and $\hat{a}^{\dagger}_{\omega}$ of the ingoing mode as,
\begin{align}
\hat{b}_{\Omega}=\int d\omega \left(\alpha _{\Omega \omega}^{*}\hat{a}_{\omega}-\beta _{\Omega \omega}^{*}\hat{a}_{\omega}^{\dagger}\right).
\label{bgt}
\end{align}
Now following \ref{Cor-Op}, the frequency space correlation for the outgoing modes yields
\begin{align}
\hat{N}_{\Omega _{1}\Omega _{2}}=\hat{b}^{\dagger}_{\Omega _{1}}\hat{b}_{\Omega _{2}}e^{-i(\Omega _{1}-\Omega _{2})t}+\hat{b}^{\dagger}_{\Omega _{2}}\hat{b}_{\Omega _{1}}e^{i(\Omega _{1}-\Omega _{2})t}
\label{frequ-cor}.
\end{align}
For the  initial state (in-state) of the field, being vacuum $|0\rangle$ or one with a definite momentum $|{\bf k}\rangle$ (and hence for all the Fock basis states), the expectation value of this correlation operator vanishes identically. We now consider this quantum correlation of the field in the out-going modes. If there is a test field which falls into the hole, starts in the in-vacuum state $|0\rangle_{\text{in}}$, then the expectation value of the frequency correlation becomes,
\begin{align}
{}_{\text{in}}\langle 0 |\hat{N}_{\Omega_{1}\Omega_{2}}|0\rangle_{\text{in}}&=\delta (\Omega_1 - \Omega_2) \times e^{\frac{-\pi(\Omega_1 + \Omega_2)}{2\kappa}}\frac{\sqrt{\Omega_1 \Omega_2}}{4 \pi^2 \kappa^2}
\left\lbrace \Gamma \left[-i \frac{\Omega_1}{\kappa} \right] \Gamma \left[i \frac{\Omega_2}{\kappa}\right] e^{-i(\Omega_1 - \Omega_2)t} + c.c.\right\rbrace 
\end{align}
which vanishes identically for the off-diagonal elements and hence {\it the asymptotic future observer will also measure no frequency correlation} in the out-going modes.  The diagonal elements of this observable gives the number spectrum. Such an observer measures the out-going spectrum to be a thermal one which can be verified by taking $\Omega_1 = \Omega_2$. This operator turns out to be a crucial discriminator of whether something has fallen inside a black hole with zero energy or not.
When the test field starts in a non-vacuum state, the out-going modes will develop a {\it non-zero frequency correlation} and the expectation of \ref{frequ-cor} will become non-zero. The correction to the expectation of the frequency correlator $\hat{N}_{\Omega _{1}\Omega _{2}}$ in a non-vacuum in-state leads to a state dependency,
\begin{align}
&{}_{\text{in}}\langle \psi |\hat{N}_{\Omega_{1}\Omega_{2}}|\psi\rangle_{\text{in}}=\Bigg[\left(\int \frac{d\omega}{\sqrt{4\pi\omega}}f(\omega)\alpha ^{*}_{\Omega _{2}\omega}\right)
\left(\int \frac{d\bar{\omega}}{\sqrt{4\pi\bar{\omega}}}f^{*}(\bar{\omega})\alpha _{\Omega _{1}\bar{\omega}}\right)
+\left(\int \frac{d\omega}{\sqrt{4\pi\omega}}f^{*}(\omega)\beta ^{*}_{\Omega _{2}\omega}\right)
\nonumber
\\
&\times \left(\int \frac{d\bar{\omega}}{\sqrt{4\pi\bar{\omega}}}f(\bar{\omega})\beta _{\Omega _{1}\bar{\omega}}\right)\Bigg]e^{-i(\Omega _{1}-\Omega _{2})t}
+\textrm{c.c}
\end{align}
Using the expressions for $\alpha _{\Omega \omega}$ and $\beta _{\Omega \omega}$ from \ref{BT}, one obtains,
\begin{align}
{}_{\text{in}}\langle \psi |\hat{N}_{\Omega_{1}\Omega_{2}}|\psi\rangle_{\text{in}}&=\frac{1}{4\pi}\left[A(\Omega_1)A(\Omega_2)^* +c.c.\right]
+\frac{1}{4\pi}\left[B(\Omega _{1})B(\Omega _{2})^{*}+c.c.\right]. \label{FrqCorr}
\end{align}
with
\bea
A(\Omega)= e^{-\frac{\pi \Omega}{2\kappa}}\frac{\sqrt{\Omega}}{2 \pi \kappa}\Gamma \left[-i \frac{\Omega}{\kappa} \right] F\left(\frac{\Omega}{\kappa}\right)e^{-i\Omega t}, 
\eea
and 
\begin{align}
B(\Omega)=e^{\frac{\pi \Omega}{2\kappa}}\frac{\sqrt{\Omega}}{2\pi \kappa}\Gamma\left[-i\frac{\Omega}{\kappa}\right]F^{*}\left(-\frac{\Omega}{\kappa}\right)e^{-i\Omega t}
\end{align}
expressed in terms of the time co-ordinate as defined by observers living close to $\mathcal{J}^{+}$. The frequency correlation for two distinctly separated frequencies remains zero for all the field configurations which were in the vacuum (incidentally, also for configurations in individual Fock basis elements) in the in-state. However, as for the out-state, the frequency correlation remains zero {\it only if} the in-state were a vacuum. {\it The out-going modes develop non-zero frequency correlation, even if the in-state was a non-vacuum Fock basis state with zero correlation.} Alternatively, those fields which carried some amount of stress-energy into the black hole definitely develop some non-zero correlation (or a non vanishing operator of such a correlator) at future asymptotia, while only those fields which were in vacuum state develop no late time frequency correlation dependence. Therefore, by just measuring this operator, a late time observer will be able to tell if some non-zero stress-energy has 
entered the black hole.

Additionally, the observer can also decipher the state that entered into the black hole, by reconstructing $F(\Omega/\kappa)$ from this non-zero expectation value of the correlation with clever analysis of the correlation (s)he measures, which we review next.
\subsection{Radiation from Black Hole: Information about the initial state}\label{Sec_03}

From the off-diagonal elements of \ref{FrqCorr}, one can construct a complex, yet a simple, quantity
\bea
{\cal D}_{\Omega_1 \Omega_2}\equiv N_{\Omega_1 \Omega_2}+\frac{i}{\Delta \Omega}\frac{\partial}{\partial t}N_{\Omega_1 \Omega_2},\label{InfoExtract1} 
\eea
where $\Delta \Omega = \Omega_1 - \Omega_2$ and $  N_{\Omega_1 \Omega_2} = {}_{\text{in}}\langle \psi |\hat{N}_{\Omega_{1}\Omega_{2}}|\psi\rangle_{\text{in}}$. From here onwards, the symmetry profile of the initial data plays a pivotal role \footnote{this is analogous to the classical channel in the EPR terminology.}.
For example, for a real initial state, we have $F(\Omega/\kappa) = F^*(-\Omega/\kappa)$ and therefore, \ref{InfoExtract1} can be used to extract the function $F(\Omega/\kappa)$ from  a particular observable,
\begin{align}
S_{\Omega_1 \Omega_2}&\equiv \frac{4 \pi^3 \kappa^2 }{\sqrt{\Omega_1 \Omega_2}}\frac{{\cal D}_{\Omega_1 \Omega_2}e^{i(\Omega _{1}-\Omega _{2})t}}{\Gamma \left[-i \frac{\Omega_1}{\kappa} \right] \Gamma \left[i \frac{\Omega_2}{\kappa}\right]\cosh \left(\frac{\pi(\Omega _{1}+\Omega _{2})}{2\kappa}\right) } 
\nonumber
\\
&= F\left(\frac{\Omega_1}{\kappa}\right)F^{*}\left(\frac{\Omega_2}{\kappa}\right). \label{InfoExtract2}
\end{align}
Clearly, left hand side of \ref{InfoExtract2} can be determined by observing the emission spectrum as the left hand side is under complete control of the late time observer. From the above relation, we learn that this quantity has to be separable as a product in terms of the frequencies $\Omega_1$ and $\Omega_2$. Using this property, one can obtain the function $F(\Omega/\kappa)$, (upto an irrelevant constant phase), from the symmetric sum
\bea
\log{S_{\Omega_1 \Omega_2}} = \log{F\left(\frac{\Omega_1}{\kappa}\right)} +\log{F^{*}\left(\frac{\Omega_2}{\kappa}\right)},
\eea
Therefore, for the real initial state, the state can be identically and completely reconstructed from correlations in the out-going modes.

In \cite {Lochan:2014xja}, a formalism is developed to deal with the analysis of field content of a non-vacuum pure states corresponding to a particular observer with respect to another set of observers, using the correlation functions. The information about the state through the function $f(\omega)$, together with the Bogoliubov coefficients, completely characterize the deviations form the standard vacuum response. The analysis of the spectrum operator \cite{Page:1993wv,Birrell:1982ix} also captures this distortion.  The extraction of information about initial data using the spectral distortion (from the diagonal elements) $\hat{N}_{\Omega} = \langle \Psi | \hat{b}^{\dagger}_{\Omega}\hat{b}_{\Omega}|\Psi\rangle -\langle 0 | \hat{b}^{\dagger}_{\Omega}\hat{b}_{\Omega}|0 \rangle$, can as well be done, as reported in \cite{Lochan:2015oba}. This also shows that the distortion, therefore, the information rich semiclassical hairs are present at non-zero energy but become rarer with increasing energy. Therefore, 
such hairs will be predominantly visible in low energy probes. 

We also learn that if we are aware of the symmetries of the system which is going to form a black hole, from some general principles,  we will know exactly how the non-vacuum response would look like. We can measure particle content for different test fields in the black hole spacetime. The test fields which contribute infinitesimal energy to the formation will reflect their non-vacuum characters in the late time radiation. That is to say, the  spectra will show deviations from the expected vacuum response, corresponding to the symmetries of initial data. Measurements of such non-vacuum distortion will reveal partial or complete character of the state of the field depending on the knowledge of the symmetry of initial profile, something  which would have been missed classically. In fact, this analysis can be extended for states which correspond to configurations analogous to classical picture,(e.g. the coherent state), where the correlations do not spill over the classical allowed region in spacetime, i.e., 
are confined in 
a compact region as expected classically. Yet the Fourier space correlation do care about the full spatial correlation and becomes aware of this. An exercise with a simple state will illustrate the above point quiet naturally and is presented in \ref{NVDwithCompact}.  A similar strategy can be developed for quantum bits as well \cite{Chatwin-Davies:2015hna} by observing the emitted radiation and a set of subsequent measurements done  on a hole.
\section{Information regain?}

In order to preserve the equivalence principle until late times in unitarily evaporating
black holes, the thermodynamic entropy of a black hole must be primarily entropy of entanglement across
the event horizon. For such black holes, the information entering a black hole becomes
encoded in correlations within a tripartite quantum state and is
only decoded into the outgoing radiation very late in the evaporation. If, somehow we come to terms with, without 
lighting up a firewall, we could ask when will the information start leaking in this tripartite system. Standard arguments suggest that it will be done very late in time, {\it but once black hole starts emitting information, it will do so very quickly.} The next question would be in which form the correlation will appear ? In this section we consider ``when and where'' of the information retrieval.
\subsection{Late time Flash}

Tunneling seems a viable mechanism which is consistent with unitarity as well as Equivalence principle.
Nothing from inside region of a black hole, can come out classically ever. However, it can quantum mechanically tunnel, in principle, into classically forbidden region. So, if we have such a trustworthy proposal which respects both unitarity and the Equivalence principle, then it becomes the question of characterizing the release process of the black hole. 

If the process of mapping information initial state into the outgoing radiation is unitary, then we can apply {\it Decoupling theorems} on this set up. Decoupling theorems \cite{Dupuis} tell us precisely how many bits of the encoded state (i.e. radiation) must be released in order to reconstruct the original 
uncoded state, with required precision. 

We can use generalized decoupling theorem \cite{Braunstein:2009my} to any evaporative dynamics which is unitary, such as tunneling. If we believe that almost all of the black hole thermodynamic entropy $S_{BH}$ is the entanglement entropy in the exterior, we can account for number of bits providing thermodynamic entropy to the hole but which are not entangled with outside, through ``excess'', defined as
\begin{equation}
\chi^{(q)}\equiv S_{\text{BH}} - S_{\text{matter}}
 -H^{(q)}(\rho_{\text{ext}})\ge 0,  \label{chiq}
\end{equation}
with $H^{(q)}$ being the R\'{e}nyi entropy of order $q$, defined as
\bea 
H^{(q)} =\frac{\log_2{Tr \rho^q}}{1-q},
\eea
when the state of the system is described by the density matrix $\rho$. Clearly, for the case when entanglement entropy is the primary source of the thermodynamic entropy, the excess
is negligibly small 
$\chi^{(1/2)} \ll S_{\text{BH}}$, since $ S_{\text{BH}} \gg S_{\text{matter}}$.
Using decoupling theorem on this tripartite system, one obtains that
\begin{itemize}

\item Between the initial and final $S_{\text{matter}} +  \chi^{(2)}/2$ bits radiated, the information about the
in-fallen matter is effectively deleted from each individual subsystem (interior or early or late time radiation considered individually). 

\item Prior to $(\chi^{(2)} -2c_1)/2$ bits are radiated (for any positive $c_1$) the information of infalling matter resides in the interior only with fidelity $1-2^{-c_1}$.

\item It is in only the final $( \chi^{(2)} -2c_2)/2$ bits (for any positive $c_2$), the information is imprinted in the late time radiation considered separately  with fidelity $1-2^{-c_1}$.

\end{itemize}
Thus, we have to wait for final $\chi^{(2)}/2$ bits to be emitted for the radiation to be seemingly full of information, which is the final ${\cal O }(S_{BH}^{-1/4})$ \cite{tHooftTS} units of time.

Also once the black hole starts imprinting information in the radiation it does that very-very quickly. It was shown \cite{Hayden:2007cs} that if we throw something small inside a black hole, the outgoing radiation once it starts carrying information gets the information of the object out very quickly.
\subsection{Old enough black hole behaves as a ``mirror''}

 If black holes were to evaporate unitarily, such a process would ensure that anything thrown into the black hole which is old enough (is in the entangled phase, i.e. after the Page time or after first  $(\chi^{(2)} -2c_1)/2 $ bits are emitted, would essentially ``bounce back immediately''. That is to say, if black hole is in the information leaking phase, it releases any correlation in very short amount of time after eschewing it.
 
 Preskill \cite{Hayden:2007cs} evaluated the time required by the black hole to bring out the correlations of $k$ quantum bits (qubits), if the black hole has lived long such that the interior is maximally entangled with the exterior. The idea here is to focus on a pure state described in the exterior which is further subdivided into subsystem $N$ and $M$ (with the same Hilbert space dimensions). The state on the exterior is given as
 \begin{equation}
 | \Psi\rangle = \frac{1}{\sqrt{|M|}}\sum_{a=1}^{|M|}|a\rangle^M \otimes |a\rangle^N.
 \end{equation}
The reduced description of $M$ or $N$ is maximally entangled with the other. If $M$ is thrown into the black hole, then $N$ will be maximally entangled with the interior. If the black hole evaporates unitarily the correlations across the horizon, which purify $N$ come into the radiation. Consequently, the correlations between the remaining interior and $N$ get erases. These correlations now lie in the radiation. If we calculate the difference of the interior and $N$ from being in a maximally entangled to completely separable (no correlations) configuration,  is obtained by
\begin{equation}
\int d V^B || \sigma^{NB'}(V^B) - \sigma^N(V^B)\otimes \sigma^{B'} || \leq \frac{|N|^2}{|R|^2}, \label{DensityMatrixMeasure}
\end{equation}
where $V^B$ is a unitary transformation on all the qubits in the interior of black hole (to scramble them), $dV^B$ is a Haar measure on unitary matrices and $R$ is the dimensionality of Hilbert space of emitted qubits (after Page time). The distance  operation  is $||A|| = Tr \sqrt{A^{\dagger}A}$; $B'$ is the remaining interior after it has lost $s-$qubits; $\sigma^N = Tr_{B'}(\sigma^{NB'})$ and $\sigma^{B'}$ is maximally mixed interior state.

If the number of qubits are few larger than those thrown in (say $c$ extra) then the two configurations (maximally mixed versus purely separable) are distinguishable only by $2^{-2c}$ as could be seen from \ref{DensityMatrixMeasure}. Thus, as the dimension of emitted qubits' Hilbert space grows larger, the description evolves into a configuration where interior and $N$ have hardly any correlations between them. Almost all the correlations have appeared outside!
\section{Black hole scattering matrix}\label{Scattering}

A very interesting proposal was put forward recently by t' Hooft \cite{Hooft:2015jea}, using which one can relate the quantum state of particles fallen into the black hole to the quantum state of the outgoing particles participating in the Hawking radiation (see also \cite{tHooft:1984kcu,tHooft:1996rdg,tHooft:1986cfy} for related ideas). This can be achieved using a very interesting result in classical gravitational theory, namely {\it the Aichelburg-Sexl boost} \cite{Aichelburg:1970dh,Dray:1984ha}. This represents the spacetime of a fast moving particle, which surprisingly has a discontinuous behavior along the direction of motion of the particle. To be precise, let the transverse coordinates be denoted by $x_{\perp}$ and we introduce two null coordinates $u,v$. The coordinates before the particle has passed is given by $x^{\mu}_{\rm in}=(u_{\rm in}, v_{\rm in},x_{\perp})$ while that after are $x^{\mu}_{\rm out}=(u_{\rm out},v_{\rm out},x_{\perp})$. These two sets of coordinates are related through the 
following transformation,
\begin{align}
x^{\mu}_{\rm out}=x^{\mu}_{\rm in}-4G p^{\mu}\log (|x_{\perp}|/C)
\end{align}
where, $p^{\mu}$ stands for the four momentum of the massless particle and $x_{\perp}$ is the transverse distance of the spacetime event from the particle. Note that the flat spacetime metric in these coordinates takes the form $ds^{2}=-2dudv+dx_{\perp}^{2}$ \cite{Dray:1984ha,Hooft:2015jea}.
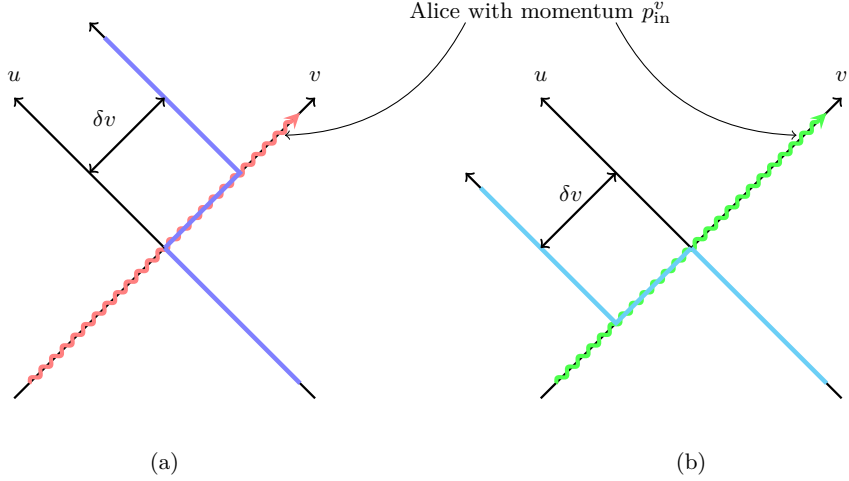
\begin{figure}
\begin{center}
\begin{tikzpicture}[scale=1]   
            \draw[->,black,line width=0.3mm] (-2,-2) -- (0,0) -- (2,2);
            \draw[-stealth,decoration={snake,amplitude =0.4mm,segment length = 2mm, post length=0.9mm},decorate,color=red!50, line width=0.55mm] (-1.8,-1.8) -- (0,0) -- (1.8,1.8);       
            \draw[->,black,line width=0.3mm] (2,-2) -- (0,0) -- (-2,2);
            \draw[blue!50,line width=0.6mm] (1.8,-1.8) -- (0,0);
            \draw[->,black,line width=0.3mm] (1,1) -- (-1,3);
            \draw[blue!50,line width=0.6mm] (0,0) -- (1,1) -- (-0.8,2.8);
            \draw[->,black,line width=0.3mm] (5,-2) -- (9,2);
            \draw[-stealth,decoration={snake,amplitude =0.4mm,segment length = 2mm, post length=0.9mm},decorate,color=green!70, line width=0.55mm] (5.2,-1.8) --  (8.8,1.8);       
            \draw[->,black,line width=0.3mm] (9,-2) -- (5,2);
            \draw[cyan!50,line width=0.6mm] (8.8,-1.8) -- (7,0);
            \draw[->,black,line width=0.3mm] (6,-1) -- (4,1);
            \draw[cyan!50,line width=0.6mm] (7,0) -- (6,-1) -- (4.2,0.8);
            \node[label=above:$v$] at (2,2) {};
            \node[label=above:$u$] at (-2,2) {};
            \node[label=above:$v$] at (9,2) {};
            \node[label=above:$u$] at (5,2) {};
            \draw[<->,black,line width=0.3mm] (-1,1) to (-0.0,2.0);
            \draw[<->,black,line width=0.3mm] (5.0,0.0) to (6,1.0);
            \draw[->,black,bend left=30] (4,3.0) to (1.6,1.5);
            \draw[->,black,bend right=30] (6,3.0) to (8.4,1.5);  
            \node[label=above: $\delta v$] at (-0.8,1.4) {};
            \node[label=above: $\delta v$] at (5.4,0.4) {};
            \node[label=below: Alice with momentum $p^{v}_{\rm in}$] at (5.0,3.5) {};
            \node[label=below: (a)] at (0,-2.5) {};
            \node[label=below: (b)] at (7,-2.5) {};
\end{tikzpicture}
\end{center}
\caption{The above presents a schematic representation of the shift suffered by any particle along the trajectory of a null ray. The null ray for illustration is taken along the $v$ direction with momentum $p^{v}_{\rm in}$. Two cases are depicted --- (a) in which $x_{\perp}/C<1$, such that $\delta v>0$, (b) in which $x_{\perp}/C>1$, leading to $\delta v<0$. All the dependence on angular coordinates, $x_{\perp}$ has been suppressed. See text for more discussions.}\label{fig_Scattering}
\end{figure}
It has serious implications for geodesics crossing the particle. For example let Alice has a momentum $p^{\mu}=(0,p_{\rm in}^{v},0,0)$ traveling along the $v$ direction. Then if Bob was at rest at location $(0,0,x_{\perp})$, after Alice has passed through $(0,0,0,0)$, the location of Bob will be modified to $(0,-4Gp_{\rm in}^{v}\log(|x_{\perp}|/C),x_{\perp})$. Thus Bob will be dragged along the direction of motion of Alice (see \ref{fig_Scattering}). 

Let us now consider a situation in which a black hole is being formed in some initial quantum state $|\psi _{\rm in}^{0}\rangle$. Then one throws a massless particle with momentum $\delta p^{v}$ in the black hole and thus changed the state. If the particle enters the horizon at a transverse position $x_{\perp}$, then all other particles at different transverse location $x_{\perp}'$ will get shifted by an amount,
\begin{align}
\delta v=-4G\delta p^{v}\log(|x_{\perp}-x_{\perp}'|/C)
\end{align}
This must be a property of the unitary matrix $U$ (also called black hole scattering matrix), such that if
\begin{align}
|\psi _{\rm out}^{0}\rangle =U|\psi _{\rm in}^{0}\rangle
\end{align}
the out state corresponding to $|\psi _{\rm in}^{0}+\delta p^{v}\rangle$ is essentially the operation of translation on $|\psi _{\rm out}^{0}\rangle$ by an amount $\delta v$. Thus we obtain,
\begin{align}
U|\psi _{\rm in}^{0}+\delta p^{v}\rangle &=\exp(ig^{\mu \nu}p_{\mu}x_{\nu})|\psi _{\rm out}^{0}\rangle
=\exp(-ip_{\rm out}^{u}(x_{\perp}')\delta v)|\psi _{\rm out}^{0}\rangle
\nonumber
\\
&=\exp\left[4iGp_{\rm out}^{u}(x_{\perp}') \delta p^{v}\log (|x_{\perp}'-x_{\perp}|/C) \right]|\psi _{\rm out}^{0}\rangle
\end{align}
The above result follows since in double null coordinates the flat spacetime line element such that $g^{uv}=-1$. One can now repeat this procedure as long as one wishes by adding more and more particles. Thus one can reach any other state $|\psi _{\rm in}\rangle$, such that,
\begin{align}
|\psi _{\rm in}\rangle =\lim _{N\rightarrow \infty}|\psi _{\rm in}^{0}+\sum _{i=1}^{N}\delta p^{v}_{i}\rangle
\end{align}
and the total momentum going in being $p_{\rm in}^{v}(x_{\perp})$. Thus one have a distribution of momentum of the outgoing particles $p_{\rm out}^{u}(x_{\perp}')$ over the transverse region. Hence to achieve a final state one must average over all the final transverse locations, leading to,
\begin{align}
U|\psi _{\rm in}\rangle =\exp\left[4iG\int d^{2}x_{\perp}'~\hat{p}_{\rm out}^{u}(x_{\perp}') p^{v}_{\rm in}(x_{\perp})\log (|x_{\perp}'-x_{\perp}|/C) \right]|\psi _{\rm out}^{0}\rangle \label{Scatter2}.
\end{align}
We can further postulate that the entire Hilbert space of black hole evaporation can have these ingoing particles and the outgoing particles as the basis, then the in state is expressible in terms of $|p_{\rm in}^{v}\rangle$ and the out state can be expanded in terms of $|p_{\rm out}^{u}\rangle$. Since the translation operators lead to conjugacy between position and momentum, in this case, $p_{\rm out}^{u}$ and $v_{\rm out}$ are conjugate, as well as $p_{\rm in}^{v}$ and $u_{\rm in}$ are conjugate, thanks to the appearance of $g^{\mu \nu}p_{\mu}x_{\nu}$ in the phase factor. Thus we must have,
\begin{align}
\left[v_{\rm out}(x_{\perp}),p_{\rm out}^{u}(x_{\perp}')\right]=\left[u_{\rm in}(x_{\perp}),p_{\rm in}^{v}(x_{\perp}')\right] =i\delta ^{2}\left(x_{\perp}-x_{\perp}'\right).
\end{align}
The phase factor in \ref{Scatter2} can be written as, $ip^{v}_{\rm in}u_{\rm in}$ and thus,
\begin{align}\label{Eq_Hooft_02}
u_{\rm in}(x_{\perp})=4G\int d^{2}x_{\perp}'~p_{\rm out}^{u}(x_{\perp}')\log (|x_{\perp}'-x_{\perp}|/C)
\end{align}
One can further invert the situation and the integral will now be on $p_{\rm in}$, such that,
\begin{align}
v_{\rm out}(x_{\perp})=-4G\int d^{2}x_{\perp}'~p_{\rm in}^{v}(x_{\perp}')\log (|x_{\perp}'-x_{\perp}|/C)
\end{align}
Let us use \ref{Eq_Hooft_02} in the commutation relations, such that,
\begin{align}
i\delta ^{2}\left(x_{\perp}-x_{\perp}'\right)&=\frac{i}{2\pi}\partial ^{2}_{\perp '}\log(|x_{\perp}-x_{\perp}'|/C)
=\frac{i}{2\pi}\int d^{2}x_{\perp}''\delta ^{2}\left(x_{\perp}''-x_{\perp}'\right)\partial ^{2}_{\perp ''}\log(|x_{\perp}-x_{\perp}''|/C)
\nonumber
\\
&=4G\int d^{2}x_{\perp}''~\left[p_{\rm out}^{u}(x_{\perp}''),p_{\rm in}^{v}(x_{\perp}')\right]\log (|x_{\perp}''-x_{\perp}|/C)
\end{align}
Hence it immediately follows that,
\begin{align}
\left[p_{\rm out}^{u}(x_{\perp}'),p_{\rm in}^{v}(x_{\perp})\right]=\frac{i}{8\pi G}\partial _{x_{\perp}'}^{2}\delta ^{2}\left(x_{\perp}-x_{\perp}'\right)
\end{align}
Finally use of the same leads to,
\begin{align}
\left[u_{\rm in}(x_{\perp}),v_{\rm out}(x_{\perp}')\right]&=-16G^{2}\int d^{2}x_{\perp}''d^{2}x_{\perp}'''~\log (|x_{\perp}'-x_{\perp}'''|/C)\log (|x_{\perp}-x_{\perp}''|/C)\left[p_{\rm out}^{u}(x_{\perp }''),p_{\rm in}^{v}(x_{\perp}''') \right]
\nonumber
\\
&=-\frac{2iG}{\pi}\int d^{2}x_{\perp}''d^{2}x_{\perp}'''~\log (|x_{\perp}'-x_{\perp}'''|/C)\log (|x_{\perp}-x_{\perp}''|/C)
\partial _{x_{\perp}''}^{2}\delta ^{2}\left(x_{\perp}''-x_{\perp}'''\right)
\nonumber
\\
&=-4iG\int d^{2}x_{\perp}''d^{2}x_{\perp}'''~\log (|x_{\perp}'-x_{\perp}'''|/C)
\delta ^{2}\left(x_{\perp}''-x_{\perp}'''\right)\delta ^{2}\left(x_{\perp}''-x_{\perp}\right)
\nonumber
\\
&=-4iG \log (|x_{\perp}-x_{\perp}'|/C)
\end{align}
Thus one can work with the Hilbert space spanned by the operators, $(p^{v}_{\rm in},u_{\rm in})$ for the in states and $(p^{u}_{\rm out}, v_{\rm out})$ for the out states. Thus the wave function of the incoming state will be $\psi(u_{\rm in})$. In the near horizon regime one can introduce local Rindler observers and hence we will have Rindler time coordinate $\tau$ and Rindler tortoise coordinate $\rho$, such that, $u_{\rm in}=e^{\rho}$. Thus the wave function in position space becomes $\phi (\rho)=e^{\rho/2}\psi(e^{\rho})$, courtesy the Jacobian of the transformation. A similar consideration applies to corresponding Fourier transformed version in the momentum space as well. It turns out that the wave function in Tortoise coordinates can be diagonalized and as a consequence the in states gets coupled to the out states \cite{Hooft:2015jea}. 

This leads to another possibility of retrieving information from black holes, where the incoming particles, which have energy as the only information, shares the same with outgoing Hawking radiation by changing their coordinates. This is how the process becomes unitary and the information of what has fallen in the black hole gets stored in the Hawking radiation.
\section{What have we learnt?}

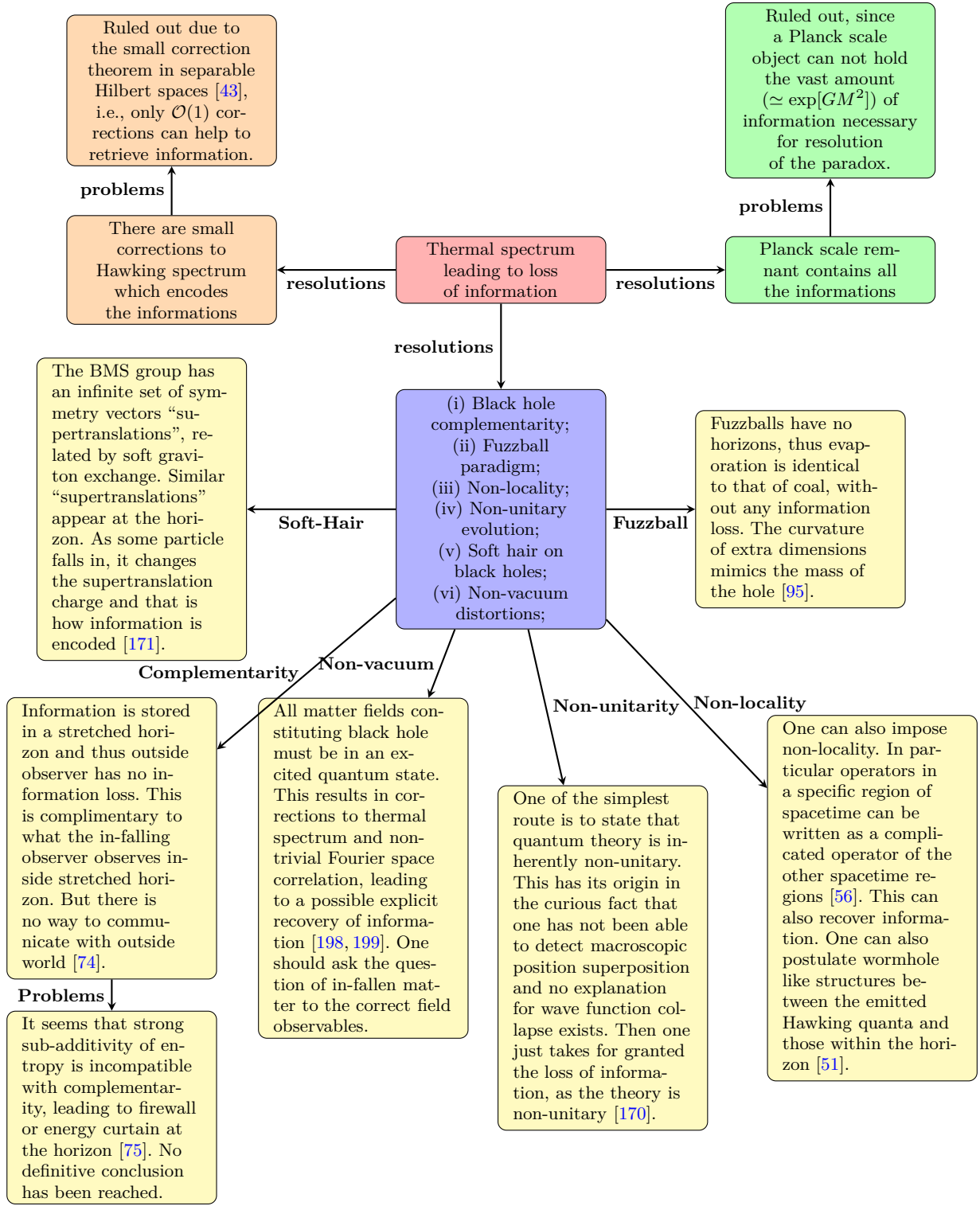
\begin{figure}[t]
\begin{center}
\begin{tikzpicture}[node distance=3cm]
             \node (start) [startstop] {Thermal spectrum leading to loss of information};
             \node (R12) [io, below of=start, yshift=-1cm] { 
             (i) Black hole complementarity;\\
             (ii) Fuzzball paradigm; \\
             (iii) Non-locality;\\
             (iv) Non-unitary evolution;\\
             (v) Soft hair on black holes;\\
             (vi) Non-vacuum distortions;};
             \node (R11) [process, left of=start, xshift=-2.5cm] {There are small corrections to Hawking spectrum which encodes the informations};
             \node (R13) [decision, right of=start, xshift=2.5cm] {Planck scale remnant contains all the informations};
             \draw [arrow] (start) -- node[anchor=east] {\bf resolutions} (R12);
             \draw [arrow] (start) -- node[anchor=north] {\bf resolutions} (R11);
             \draw [arrow] (start) -- node[anchor=north] {\bf resolutions} (R13);
             \node (R21) [process, above of=R11] {Ruled out due to the small correction theorem in separable Hilbert spaces \cite{Mathur:2009hf}, i.e., only $\mathcal{O}(1)$ corrections can help to retrieve information.};
             \node (R23) [decision, above of=R13] {Ruled out, since a Planck scale object can not hold the vast amount ($\simeq \exp[GM^{2}]$) of information necessary for resolution of the paradox.};
             \draw [arrow] (R11) -- node[anchor=east] {\bf problems} (R21);
             \draw [arrow] (R13) -- node[anchor=east] {\bf problems} (R23);
             \node (R34) [is, left of=R12, xshift=-3cm] {The BMS group has an infinite set of symmetry vectors ``supertranslations'', related by soft graviton exchange. Similar ``supertranslations'' appear at the horizon. As some particle falls in, it changes the supertranslation charge and that is how information is encoded \cite{Hawking:2016msc}.};
             \node (R35) [is, right of=R12, xshift=2cm] {Fuzzballs have no horizons, thus evaporation is identical to that of coal, without any information loss. The curvature of extra dimensions mimics the mass of the hole \cite{Mathur:2016ffb}.};
             \node (R33) [is, below of=R12, yshift=-3cm, xshift=-2.3cm] {All matter fields constituting black hole must be in an excited quantum state. This results in corrections to thermal spectrum and non-trivial Fourier space correlation, leading to a possible explicit recovery of information \cite{Lochan:2015oba,Lochan:2016nbs}. One should ask the question of in-fallen matter to the correct field observables.};
             \node (R32) [is, below of=R34, xshift=-0.5cm, yshift=-2.5cm] {Information is stored in a stretched horizon and thus outside observer has no information loss. This is complimentary to what the in-falling observer observes inside stretched horizon. But there is no way to communicate with outside world \cite{Susskind:1993if}.};
             \node (R31) [is, below of=R35, xshift=1.2cm, yshift=-3.5cm] {One can also impose non-locality. In particular operators in a specific region of spacetime can be written as a complicated operator of the other spacetime regions \cite{Papadodimas:2013wnh}. This can also recover information. One can also postulate wormhole like structures between the emitted Hawking quanta and those within the horizon \cite{Maldacena:2013xja}.};
             \node (R36) [is, left of=R31, xshift=-1.5cm, yshift=-1cm] {One of the simplest route is to state that quantum theory is inherently non-unitary. This has its origin in the curious fact that one has not been able to detect macroscopic position superposition and no explanation for wave function collapse exists. Then one just takes for granted the loss of information, as the theory is non-unitary \cite{Modak:2014qja}.};
             \draw [arrow] (R12) -- node[anchor=east] {\bf Non-vacuum} (R33);
             \draw [arrow] (R12) -- node[anchor=north] {\bf Fuzzball} (R35);
             \draw [arrow] (R12) -- node[anchor=north] {\bf Soft-Hair} (R34);
             \draw [arrow] (R12) -- node[anchor=east] {\bf Complementarity} (R32);
             \draw [arrow] (R12) -- node[anchor=west] {\bf Non-locality} (R31);
             \draw [arrow] (R12) -- node[anchor=west] {\bf Non-unitarity} (R36);
             \node (R41) [is, below of=R32, yshift=-1.5cm] {It seems that strong sub-additivity of entropy is incompatible with complementarity, leading to firewall or energy curtain at the horizon \cite{Almheiri:2012rt}. No definitive conclusion has been reached.};
             \draw [arrow] (R32) -- node[anchor=east] {\bf Problems} (R41);
\end{tikzpicture}
\end{center}
\caption{A schematic illustration of the current status of various resolutions of the black hole information loss paradox. This ranges from drastic modifications like removal of spacetime inside horizon in fuzzball paradigm, introduction of non-locality, non-unitarity, soft hairs to simple non-vacuum distortions. Further one such proposal, complementarity, has been challenged recently by the firewall argument. There are people advocating both against and for this proposal and no single consensus has been arrived at. Most likely, exploration of the information paradox will continue to be our guiding principle in search for merging quantum theory with relativity. }\label{fig_Summary}
\end{figure}
After reviewing many different and radical proposals towards resolution of the black hole information paradox it is now time to summarize the current status and the possible future directions of explorations (see \ref{fig_Summary}). The models discussed in this review, not only showed the problems general relativity presents in the face of quantum mechanics, but also bring about intriguing and fundamentally new ideas on the canvas, not only from the point of view of black hole geometry but also the structure of quantum theory, thereby acting as pathfinders towards new physics which could eventually lead to the search of quantum gravity. We will now summarize the various proposals we covered in this review, with possible directions of openings and challenges faced, for each of them.
\begin{itemize}

\item Initial belief was that a black hole is no different than a black body and after a certain time (Page time), i.e., when the Hilbert spaces of the early Hawking radiation and the remaining black hole have similar dimensionality the information starts leaking out of the hole \cite{Page:1993wv}. This view was challenged by the no-hiding theorem of Brustein \textit{et. al.}, \cite{Braunstein:2006sj} where they showed that there exists virtually no information in the correlations of a bipartite system. It was subsequently realized that black hole can actually be thought as a tripartite system, and this tripartite system can have correlations storing information. But then one can ask the question, which of these correlations have the full information of what has fallen in the black hole. It turns out the attempt to answer this challenge leads to the  complementarity proposal \cite{Hayden:2007cs}.\\

\item In the complementarity proposal, one assumes that the same copy of information is present both for the static and infalling observer, but since none of them can access the both the copies, there is no violation of the no-cloning theorem. This virtually reduces the no-cloning theorem, for a single causal patch. For the static observer the information is assumed to be stored in a stretched horizon, Planck length away from the hole, while for the infalling observer the information is residing inside the horizon. However the above scenario invites the firewall puzzle. Since the region near the stretched horizon is entangled with both black hole interior and the early Hawking radiation, thus violating the strong sub-additivity of entropy. One possible resolution being breakdown of equivalence principle, resulting in a non-vacuum structure at the horizon --- the firewall \cite{Almheiri:2012rt}.\\

\item There have been numerous works, both supporting and opposing the firewall argument. Many have agreed on the violation of strong sub-additivity of entropy, but suggesting some third alternative rather than leaving equivalence principle or quantum theory at stake. There have also been claims that the firewall argument is wrong --- possibly it does not describe a classical world. Till now no unique consensus has been reached.\\

\item A very straightforward plausible resolution of the information puzzle is to suggest that the black holes themselves do not exist. This seemingly is the case in the context of certain string theory approach, leading to the fuzzball paradigm \cite{Mathur:2008kg}. According to this picture, there is no spacetime region inside the event horizon, instead the spacetime caps off through compactified extra dimensions and the resulting curvature acts as a proxy for the gravitating mass. Then evaporation of black holes becomes exactly like burning of coal (nothing separated by causal curtain) and as a result, there is no paradox. As of future, one might want to study the geometry of a fuzzball state, and see if there is any deviation from the Schwarzschild solution. Such a modification have to confront the solar system tests and many more, which the \gr\ allowed geometries have done successfully. It will be interesting to see, what the fuzzball paradigm has to offer when the horizons have no tussle with quantum 
theory. For example, 
it would be interesting to see what happens for other horizons in \gr\ (e.g., Rindler horizon, de Sitter horizon) because they also hide information and have entropy so as to give rise to large number of microscopic configurations. There are also other ideas \cite{Giddings:1992hh, Nikolic:2015vga, Vaz:2014era} which call for a drastic changes to the classical understanding of black holes to fight the information loss they seemingly cause.\\

\item Non-local character of quantum correlations can also play a very crucial role towards resolving both the information paradox and the firewall puzzle. Quantum correlations, themselves, are not constrained by causality and spill over from the interior part as well. One may postulate that any two systems, which have quantum correlations (like EPR correlations) are connected by an wormhole like structure in spacetime, famously known as Einstein-Rosen bridge. Thus the early and late Hawking quanta are connected by ER bridge and hence can exchange information. Also the existence of firewall depends on the static observer, who is playing with the early Hawking radiation. If she decides to decipher the correlations in the early Hawking radiation, the infalling observer will experience firewall, otherwise it will be a smooth transition through the horizon. It would be interesting if one can arrive at an estimate of the probability for existence of a firewall.\\
 
\item In order to describe the black hole interior using local bulk operators in the AdS/CFT correspondence, it seems essential for the mapping of CFT operators corresponding to the bulk, to depend on the state of the quantum field in CFT. This also leads to an explicit realization of complementarity, where operators in two causally disconnected regions are related to each other. These state dependent operators lead to violation of micro-causality, i.e., commutator between operators inside and outside the event horizon vanishes for low point correlation but \emph{not} for high point ones. The above realization of complementarity was performed in the context of empty AdS, it would be interesting to investigate the situation for more general setups with non-maximal symmetric cases. Moreover, the loss of micro-causality at high point correlations should be thoroughly investigated.\\

\item One can also, though very reluctantly, think of giving up unitarity, which is the main point of tussle between black hole and the quantum theory. One has to tackle this situation with extreme care, since then one has to confront the very successful unitary quantum theory as well. However it turns out that there exist two features which are not well explained in the context of standard quantum theory --- (a) the measurement process, resulting in collapse of wave function itself appears non-unitary and (b) the probabilistic interpretation fails when we have a single copy of a closed system, e.g., the universe as a whole. Till date, the attempts to obtain a consistent non-unitary theory explaining the quantum world have yielded a single non-unitary and non-relativistic generalization, known as the \CSL theory with a stochastic, classical source field driving the quantum system helping in collapse of the wave function. A variation of this model when applied to evaporating black hole results in collapse of 
the wave function to a number eigenstate, such that the resultant density matrix is thermal. Thus in this picture one has a thermal radiation and loss of information, but there is nothing paradoxical as fit for a standard quantum system, where measurement leads to collapse of the wave function.\\

\end{itemize}
These approaches list few of the (of course many) prominent pathways towards tackling the apparent loss of information which the black hole gallops in. The hope would be for one of them to succeed in the future, in which case it will be worthwhile to ask, in particular, if the information indeed comes out of the black hole, when it does and in what form? Is information, rather than the local curvature the quantity, which guides quantum gravity in at a point ? 
\begin{itemize}

\item It has been more or less an agreed upon stance that in unitarily evaporating black hole, most of the information comes very late in time and comes out as a flash. In fact, calculations also show that once the black hole enters the ``revealing phase'' it literally becomes incapable of holding anything which is then thrown in. Therefore, after a certain time it actually adapts the blackbody character of radiation.\\

\item As advocated recently, in the semi-classical and the quantum domain the no-hair theorems fail and black holes do have additional hairs, storing information about what has fallen into the hole. There can be hairs due to non-vacuum character of the matter fallen in, leading to non-vacuum characters of quantum correlations and also there can be soft hairs, due to infinitely many diffeomorphisms the horizon can afford. The non-vacuum hairs encode the departure of the in-fallen state to be different from the vacuum state. Whenever a matter field carry non-zero stress energy into the black hole, that clearly gets reflected in its late time frequency correlation. With added information of the initial symmetry of the in-data, it is even possible to fully reconstruct the state.\\ 

\item The soft hairs originate due an infinite number of degenerate vacuum states with the same energy but differing in the existence of additional soft photons or gravitons. These are the Goldstone bosons originating from additional symmetries at the black hole horizon, similar to the supertranslations and superrotations for the asymptotically flat spacetimes. The conservation of charges associated with these symmetries following the soft graviton theorem of Weinberg leads to these infinitely inequivalent vacuum states. Understanding of these symmetries and the resulting soft quanta for an arbitrary null surface (of which the black hole horizon is a special case) could be a very insightful exercise for progress.\\

\item One can also invoke more subtle effects to recover the information thrown in the black hole. One such possibility being the Aichelburg-Sexl boost, in which a test particle is dragged along the trajectory of a passing by photon and the displacement suffered depends on the energy of the photon. This shows that as a particle falls into the black hole, it drags the outgoing Hawking quanta along it and hence pass over the information to the quanta, resulting in recovery of information fallen into the hole.\\ 

\end{itemize}
The approaches we discuss in the review have helped sharpening the paradox into a more information theoretic language in modern times. The tussle between unitarity and Equivalence, in modern times has been more sharply framed as a tussle between the idea of holography of black holes and concepts of sub-additivity of entropy in quantum information. The future explorations in the area will most definitely lead to a more concrete as well as convincing merger of principles of quantum theory with those of spacetime geometry, even possibly paving a totally new way forward of understanding physics, not only of black holes, but also of the things which we seemingly think to know pretty well.


\section*{Acknowledgements} 

The authors gratefully acknowledge fruitful discussions with Thanu Padmanabhan, Daniel Sudarsky, Suvrat Raju and Raphael Sorkin. Useful inputs from Samir Mathur, Daniel Sudarsky and Krishnamohan Parattu on the manuscript are also thankfully acknowledged. The authors express gratitude towards ICTS, Bangalore for hosting them during the program - Fundamental Problems of Quantum Physics (ICTS/Prog-fpqp/2016/11), the discussions during the meeting have helped a lot to shape this review. Research of S.C. is supported by the SERB-NPDF grant (PDF/2016/001589) from SERB, Government of India. Research of K.L. is supported by the INSPIRE Faculty Grant from the Department of Science and Technology, Government of India.
\appendix
\labelformat{section}{Appendix #1} 
\labelformat{subsection}{Appendix #1}
\section{Appendix: Derivation of supertranslation diffeomorphism}\label{App_Supertranslation}

In this appendix we will derive the diffeomorphism vector field $\zeta ^{a}$ that keeps the asymptotically flat metric structure (as in \ref{Eq_Asymp_Flat}) invariant. For that we require $\pounds _{\zeta}g_{ab}=0$. The explicit evaluation of the Lie variation of the asymptotically flat metric is given by,
\begin{equation}
\pounds _{\zeta}g_{ab}=\zeta ^{c}\partial _{c}g_{ab}+g_{ac}\partial _{b}\zeta ^{c}+g_{bc}\partial _{a}\zeta ^{c}
\end{equation}
For the metric given in \ref{Eq_Asymp_Flat} we have,
\begin{align}
\pounds _{\zeta}g_{uu}&=\zeta ^{c}\partial _{c}g_{uu}+2g_{uc}\partial _{u}\zeta ^{c}
\nonumber
\\
&=-2\partial _{u}\zeta ^{u}-2\partial _{u}\zeta ^{r}+\mathcal{O}(r^{-1})
\end{align}
Equating this to zero, at the leading order we obtain, $\zeta ^{u}=\zeta ^{u}(r,z,\bar{z})$ and $\zeta ^{r}=\zeta ^{r}(r,z,\bar{z})$ respectively. Then we obtain Lie variation of other metric components to be,
\begin{align}
\pounds _{\zeta}g_{ur}&=g_{uc}\partial _{r}\zeta ^{c}+g_{rc}\partial _{u}\zeta ^{c}
\nonumber
\\
&=-\partial _{r}\zeta ^{u}-\partial _{r}\zeta ^{r}
\end{align}
Equating this to zero as well, we obtain, $\zeta ^{u}=\zeta ^{u}(z,\bar{z})$ and $\zeta ^{r}=\zeta ^{r}(z,\bar{z})$ respectively. Since, $g_{rz}=0=g_{r\bar{z}}$, their lie variation should also vanish, which implies,
\begin{align}
0=\pounds _{\zeta}g_{rz}&=g_{rc}\partial _{z}\zeta ^{c}+g_{zc}\partial _{r}\zeta ^{c}
\nonumber
\\
&=-\partial _{z}\zeta ^{u}+r^{2}\gamma _{z\bar{z}}\partial _{r}\zeta ^{\bar{z}}
\end{align}
Thus choosing, $\zeta ^{u}=f(z,\bar{z})$, the above equation would be satisfied provided, $\zeta ^{\bar{z}}=-(1/r)\gamma ^{\bar{z}z}\partial _{z}f$. Along identical lines we obtain, $\zeta ^{z}=-(1/r)\gamma ^{z\bar{z}}\partial _{\bar{z}}f$. Finally, Lie variation of $g_{z\bar{z}}$ leads to,
\begin{align}
\pounds _{\zeta}g_{z\bar{z}}&=\zeta ^{z}\partial _{z}g_{z\bar{z}}+\zeta ^{\bar{z}}\partial _{\bar{z}}g_{z\bar{z}}+\zeta ^{r}\partial _{r}g_{z\bar{z}}+g_{z\bar{z}}\partial _{\bar{z}}\zeta ^{\bar{z}}+g_{\bar{z}z}\partial _{z}\zeta ^{z}
\nonumber
\\
&=-r\gamma ^{z\bar{z}}\partial _{\bar{z}}f\partial _{z}\gamma _{z\bar{z}}-r\gamma ^{\bar{z}z}\partial _{z}f\partial _{\bar{z}}\gamma _{z\bar{z}}+2r\zeta ^{r}\gamma _{z\bar{z}}
\nonumber
\\
&-r\gamma _{z\bar{z}}\partial _{\bar{z}}\left(\frac{1}{\gamma _{z\bar{z}}}\partial _{z}f\right)
-r\gamma _{z\bar{z}}\partial _{z}\left(\frac{1}{\gamma _{z\bar{z}}}\partial _{\bar{z}}f\right)
\nonumber
\\
&=2r\zeta ^{r}\gamma _{z\bar{z}}-2r\partial _{z}\partial _{\bar{z}}f
\end{align}
Equating this to zero we finally obtain,
\begin{equation}
\zeta ^{r}=\gamma ^{\bar{z}z}\partial _{\bar{z}}\partial _{z}f=D^{z}D_{z}f 
\end{equation}
Then we obtain to leading order, using expressions for the diffeomorphism vector as obtained above, the following Lie variation,
\begin{align}
\pounds _{\zeta}g_{zz}&=\zeta ^{c}\partial _{c}g_{zz}+2g_{za}\partial _{z}\zeta ^{a}
\nonumber
\\
&=-2r\gamma _{z\bar{z}}\partial _{z}\left(\gamma ^{z\bar{z}}\partial _{z}f\right)
\nonumber
\\
&=-2r\partial _{z}^{2}f+2r\partial _{z}f\gamma ^{z\bar{z}}\partial _{z}\gamma _{z\bar{z}}
\end{align}
which can be accommodated in the metric, by defining, a new $C_{zz}$ as,
\begin{equation}
\tilde{C}_{zz}=C_{zz}-2\partial _{z}^{2}f+2\partial _{z}f\gamma ^{z\bar{z}}\partial _{z}\gamma _{z\bar{z}}
\end{equation}
Hence we obtain,
\begin{align}\label{App_Eq_01}
\frac{1}{2}D^{z}\tilde{C}_{zz}&=\frac{1}{2}D^{z}C_{zz}+\gamma ^{z\bar{z}}\partial _{\bar{z}}\left(-\partial _{z}^{2}f+\partial _{z}f\gamma ^{z\bar{z}}\partial _{z}\gamma _{z\bar{z}}\right)
\nonumber
\\
&=\frac{1}{2}D^{z}C_{zz}-\gamma ^{z\bar{z}}\partial _{\bar{z}}\partial _{z}^{2}f+(\gamma ^{z\bar{z}})^{2}\partial _{z}\gamma _{z\bar{z}}\left(\partial _{\bar{z}}\partial _{z}f\right)
\nonumber
\\
&+(\gamma ^{z\bar{z}})^{2}\partial _{z}f\left(\partial _{\bar{z}}\partial _{z}\gamma _{z\bar{z}}\right)
-\partial _{z}f\partial _{z}\gamma _{z\bar{z}}\partial _{\bar{z}}\gamma _{z\bar{z}}(\gamma ^{z\bar{z}})^{3}
\nonumber
\\
&=\frac{1}{2}D^{z}C_{zz}-\gamma ^{z\bar{z}}\partial _{\bar{z}}\partial _{z}^{2}f+(\gamma ^{z\bar{z}})^{2}\partial _{z}\gamma _{z\bar{z}}\left(\partial _{\bar{z}}\partial _{z}f\right)-\partial _{z}f
\end{align}
In order to arrive at the last expression we have used the following results,
\begin{equation}
\partial _{\bar{z}}\partial _{z}\gamma _{z\bar{z}}=-\frac{4}{(1+z\bar{z})^{3}}+\frac{12 z\bar{z}}{(1+z\bar{z})^{4}};
\qquad
\partial _{z}\gamma _{z\bar{z}}\partial _{\bar{z}}\gamma _{z\bar{z}}=\frac{16z\bar{z}}{(1+z\bar{z})^{6}}
\end{equation}
as well as,
\begin{equation}
(\gamma ^{z\bar{z}})^{2}\left(\partial _{\bar{z}}\partial _{z}\gamma _{z\bar{z}}\right)
-\partial _{z}\gamma _{z\bar{z}}\partial _{\bar{z}}\gamma _{z\bar{z}}(\gamma ^{z\bar{z}})^{3}
=-(1+z\bar{z})+3z\bar{z}-2z\bar{z}=-1
\end{equation}
Finally the Lie variation of the only remaining metric element becomes,
\begin{align}
\pounds _{\zeta}g_{uz}&=g_{uu}\partial _{z}\zeta ^{u}+g_{ur}\partial _{z}\zeta ^{r}+g_{z\bar{z}}\partial _{u}\zeta ^{\bar{z}}
\nonumber
\\
&=-\partial _{z}f-\gamma ^{z\bar{z}}\partial _{\bar{z}}\partial _{z}^{2}f+(\gamma ^{z\bar{z}})^{2}\partial _{z}\gamma _{z\bar{z}}\partial _{\bar{z}}\partial _{z}f
\nonumber
\\
&=\frac{1}{2}D^{z}\tilde{C}_{zz}-\frac{1}{2}D^{z}C_{zz}
\end{align}
where we have used \ref{App_Eq_01}. These results are used to write down the supertranslation vector as in \ref{Eq_Diff_Super}.
\section{Appendix: Lorentz transformation at infinity}\label{App_LT}

Using the expression for the Lie derivative of the spacetime metric one can immediately verify whether \ref{Eq_Lorentz} keeps the metric structure of asymptotically flat spacetime invariant. Thus we obtain the following Lie variation terms, 
\begin{align}
\pounds _{\xi}g_{uu}&=-2\partial _{u}\xi ^{u}-2\partial _{u}\xi ^{r}=-D_{z}\zeta ^{z}+\frac{2}{r}\frac{r}{2}D_{z}\zeta ^{z}=0
\\
\pounds _{\xi}g_{ur}&=-\partial _{r}\xi ^{u}-\partial _{r}\xi ^{r}-\partial _{u}\xi ^{u}=-\frac{1}{2}D_{z}\zeta ^{z}+\frac{1}{2}D_{z}\zeta ^{z}=0
\\
\pounds _{\xi}g_{uz}&=-\partial _{z}\xi ^{u}-\partial _{z}\xi ^{r}+r^{2}\gamma _{z\bar{z}}\partial _{u}\xi ^{\bar{z}}
\nonumber
\\
&=-\frac{u}{2}\partial _{z}D_{z}\zeta ^{z}+\left(1+\frac{u}{r}\right)\frac{r}{2}\partial _{z}D_{z}\zeta ^{z}-\frac{r}{2}D_{z}^{2}\zeta ^{z}=0
\\
\pounds _{\xi}g_{u\bar{z}}&=-\partial _{\bar{z}}\xi ^{u}-\partial _{\bar{z}}\xi ^{r}+r^{2}\gamma _{z\bar{z}}\partial _{u}\xi ^{z}
=-\frac{u}{2}\partial _{\bar{z}}D_{z}\zeta ^{z}+\left(1+\frac{u}{r}\right)\frac{r}{2}\partial _{\bar{z}}D_{z}\zeta ^{z}
+\frac{r}{2}\gamma _{z\bar{z}}\zeta ^{z}
\nonumber
\\
&=\frac{r}{2}\partial _{\bar{z}}\left[\partial _{z}\zeta ^{z}-\frac{2\bar{z}}{(1+z\bar{z})}\zeta ^{z}\right]
+\frac{r}{2}\frac{2}{(1+z\bar{z})^{2}}\zeta ^{z}=0
\nonumber
\\
\pounds _{\xi}g_{rz}&=-\partial _{z}\xi ^{u}+r^{2}\gamma _{z\bar{z}}\partial _{r}\xi ^{\bar{z}}
=-\frac{u}{2}\partial _{z}D_{z}\zeta ^{z}+\frac{u}{2}D_{z}^{2}\zeta ^{z}=0
\\
\pounds _{\xi}g_{r\bar{z}}&=-\partial _{\bar{z}}\xi ^{u}+r^{2}\gamma _{z\bar{z}}\partial _{r}\xi ^{\bar{z}}
=-\frac{u}{2}\partial _{\bar{z}}D_{z}\zeta ^{z}-\frac{u}{2}\frac{2}{(1+z\bar{z})^{2}}\zeta ^{z}
=0
\end{align}
as well as,
\begin{align}
\pounds _{\xi}g_{z\bar{z}}&=\xi ^{c}\partial _{c}g_{z\bar{z}}+g_{zc}\partial _{\bar{z}}\xi ^{c}+g_{\bar{z}c}\partial _{z}\xi ^{c}
\nonumber
\\
&=-\frac{r}{2}\left(1+\frac{u}{r}\right)D_{z}\zeta ^{z}\partial _{r}\left(\frac{2r^{2}}{(1+z\bar{z})^{2}}\right)
+\left(1+\frac{u}{2r}\right)\zeta ^{z}\partial _{z}\left(\frac{2r^{2}}{(1+z\bar{z})^{2}}\right)
\nonumber
\\
&-\frac{u}{4r}(1+z\bar{z})^{2}D_{z}^{2}\zeta ^{z}\partial _{\bar{z}}\left(\frac{2r^{2}}{(1+z\bar{z})^{2}}\right)
-\frac{ur}{2}\gamma _{z\bar{z}}\partial _{\bar{z}}\left[\frac{(1+z\bar{z})^{2}}{2}D_{z}^{2}\zeta ^{z}\right]
\nonumber
\\
&+r^{2}\gamma _{z\bar{z}}\partial _{z}\left[(1+\frac{u}{2r})\zeta ^{z}\right]
\nonumber
\\
&=-\frac{2r^{2}+2ur}{(1+z\bar{z})^{2}}\left(\partial _{z}\zeta ^{z}-\frac{2\bar{z}}{1+z\bar{z}}\zeta ^{z}\right)
-\frac{2\bar{z}(2r^{2}+ur)}{(1+z\bar{z})^{3}}\zeta ^{z}
\nonumber
\\
&+\frac{urz}{(1+z\bar{z})}\left[\partial _{z}^{2}\zeta ^{z}+\frac{2\bar{z}^{2}}{(1+z\bar{z})^{2}}\zeta ^{z}-\frac{2\bar{z}}{1+z\bar{z}}\partial _{z}\zeta ^{z}\right]+\frac{2r^{2}}{(1+z\bar{z})^{2}}(1+\frac{u}{2r})\partial _{z}\zeta ^{z}
\nonumber
\\
&-\frac{ur}{(1+z\bar{z})^{2}}\partial _{\bar{z}}\left[\frac{(1+z\bar{z})^{2}}{2}\left(\partial _{z}^{2}\zeta ^{z}+\frac{2\bar{z}^{2}}{(1+z\bar{z})^{2}}\zeta ^{z}-\frac{2\bar{z}}{1+z\bar{z}}\partial _{z}\zeta ^{z}\right) \right]
\nonumber
\\
&=\frac{urz}{(1+z\bar{z})}\partial _{z}^{2}\zeta ^{z}-\left[\frac{2r^{2}}{(1+z\bar{z})^{2}}+\frac{2ur}{1+z\bar{z}}\right]\partial _{z}\zeta ^{z}+\frac{2ur \bar{z}}{(1+z\bar{z})^{2}}\zeta ^{z}
\nonumber
\\
&-\frac{urz}{(1+z\bar{z})}\partial _{z}^{2}\zeta ^{z}+\left[\frac{2r^{2}}{(1+z\bar{z})^{2}}+\frac{2ur}{1+z\bar{z}}\right]\partial _{z}\zeta ^{z}-\frac{2ur \bar{z}}{(1+z\bar{z})^{2}}\zeta ^{z}
=0
\end{align}
Thus the vector field keeps the asymptotic form of the metric invariant. Further one computes Lie derivative of $C_{zz}$ as, 
\begin{align}
\pounds _{\xi}C_{zz}&=\frac{1}{r}\pounds _{\xi}g_{zz}=\frac{1}{r}\left[\xi ^{c}\partial _{c}g_{zz}+2g_{zc}\partial _{z}\xi ^{c}\right]
\nonumber
\\
&=\frac{1}{r}\left[r\frac{u}{2}D_{z}\zeta ^{z}\partial _{u}C_{zz}+r\zeta ^{z}\partial _{z}C_{zz}-\frac{r}{2}D_{z}\zeta ^{z}C_{zz}
+2rC_{zz}\partial _{z}\zeta ^{z}\right]
\nonumber
\\
&=\zeta ^{z}\partial _{z}C_{zz}+2C_{zz}\partial _{z}\zeta ^{z}+\frac{1}{2}D_{z}\zeta ^{z}\left(u\partial _{u}-1\right)C_{zz}
\end{align}
Hence, $D_{\bar{z}}C_{zz}=\partial _{\bar{z}}C_{zz}$, but for $D_{z}C_{zz}$ one have to use $\Gamma ^{z}_{zz}$ as one should for covariant derivative of a tensor. 
\section{Appendix: The structure of superrotation}\label{App_superrotation}

The asymptotic structure of a spacetime has a further symmetry, called superrotation which is more non-trivial. To see this, let us start with the standard flat Minkowski spacetime,
\begin{equation}
ds^{2}=-\left(dx^{0}\right)^{2}+\left(dx^{1}\right)^{2}+\left(dx^{2}\right)^{2}+\left(dx^{3}\right)^{2}
\end{equation}
Introduce a set of new coordinates $(u,v,\zeta,\bar{\zeta})$, where $(u,v)$ are null and real, while $(\zeta,\bar{\zeta})$ are the complex coordinates, such that,
\begin{equation}
u=\frac{1}{\sqrt{2}}(x^{0}+x^{1});\qquad v=\frac{1}{\sqrt{2}}(x^{0}-x^{1});\qquad \zeta =\frac{1}{\sqrt{2}}(x^{2}+ix^{3});\qquad \bar{\zeta} =\frac{1}{\sqrt{2}}(x^{2}-ix^{3})
\end{equation}
Then the flat spacetime line element turns out to yield,
\begin{equation}
ds^{2}=-2dudv+2d\zeta d\bar{\zeta}
\end{equation}
Let us now apply another coordinate transformation, such that,
\begin{equation}
u\rightarrow u;\qquad v\rightarrow v+\zeta \bar{\zeta}u;\qquad \zeta \rightarrow u\zeta;\qquad \bar{\zeta}\rightarrow u\bar{\zeta}
\end{equation}
Then the above line element becomes,
\begin{align}
ds^{2}&=-2du\left(dv+du \zeta \bar{\zeta}+u\bar{\zeta}d\zeta+u\zeta\bar{\zeta}\right)+2\left(ud\zeta +\zeta du \right)\left(ud\bar{\zeta}+\bar{\zeta}du\right)
\nonumber
\\
&=-2dudv+2u^{2}d\zeta d\bar{\zeta}
\end{align}
Let us now introduce a set of new coordinates such that,
\begin{align}
\zeta &=w(z)-\frac{V}{2U}\alpha \bar{\beta}\left(1+\frac{V}{4U}|\beta|^{2} \right)^{-1}
\\
\bar{\zeta}&=\bar{w}(\bar{z})-\frac{V}{2U}\bar{\alpha}\beta\left(1+\frac{V}{4U}|\beta|^{2} \right)^{-1}
\\
u&=\frac{U}{|\alpha|}\left(1+\frac{V}{4U}|\beta|^{2} \right)
\\
v&=|\alpha|V\left(1+\frac{V}{4U}|\beta|^{2} \right)^{-1}
\end{align}
Then it follows that (defining $A=1-(V/4U)|\beta|^{2}$),
\begin{align}
du&=\frac{AdU}{|\alpha|}+\frac{|\beta|^{2}}{4U|\alpha|}(UdV-VdU)+dz\left[-\frac{UA\beta}{2|\alpha|}+\frac{V\bar{\beta}}{4|\alpha|}\frac{d\beta}{dz}\right]
\nonumber
\\
&+d\bar{z}\left[-\frac{UA\bar{\beta}}{2|\alpha|}+\frac{V\beta}{4|\alpha|}\frac{d\bar{\beta}}{d\bar{z}}\right]
\\
dv&=\frac{|\alpha|dV}{A}-\frac{|\alpha||\beta|^{2}V}{4A^{2}U^{2}}(UdV-VdU)+dz\left[\frac{|\alpha|V\beta}{2A}-\frac{|\alpha|V^{2}\bar{\beta}}{4UA^{2}}\frac{d\beta}{dz} \right]
\nonumber
\\
&+d\bar{z}\left[\frac{|\alpha|V\bar{\beta}}{2A}-\frac{|\alpha|V^{2}\beta}{4UA^{2}}\frac{d\bar{\beta}}{d\bar{z}} \right]
\\
d\zeta&=\frac{\alpha \bar{\beta}(UdV-VdU)}{8U^{3}A}(V|\beta|^{2}-4U)+dz\left[\alpha-\frac{V}{2U}\frac{\alpha |\beta|^{2}}{A}
+\frac{V^{2}}{8U^{2}}\frac{\alpha \bar{\beta}^{2}}{A^{2}}\frac{d\beta}{dz}\right]
\nonumber
\\
&-\frac{V}{2U}\frac{\alpha}{A^{2}}\frac{d\bar{\beta}}{d\bar{z}}d\bar{z}
\end{align}
where, $\alpha=dw/dz$ and $\alpha \beta=d^{2}w/dz^{2}$. Thus we finally obtain, the coefficient of $dz^{2}$ to yield,
\begin{align}
\textrm{Coeff.}dz^{2}&=-2\left[-\frac{UA\beta}{2|\alpha|}+\frac{V\bar{\beta}}{4|\alpha|}\frac{d\beta}{dz}\right]
\left[\frac{|\alpha|V\beta}{2A}-\frac{|\alpha|V^{2}\bar{\beta}}{4UA^{2}}\frac{d\beta}{dz} \right]
\nonumber
\\
&-2\frac{U^{2}}{|\alpha|^{2}}A^{2}\left[\frac{|\alpha|V\beta}{2A}-\frac{|\alpha|V^{2}\bar{\beta}}{4UA^{2}}\frac{d\beta}{dz} \right]
\frac{V}{2U}\frac{\bar{\alpha}}{A^{2}}\frac{d\beta}{dz}
\nonumber
\\
&=\frac{UV\beta ^{2}}{2}-\frac{V^{2}}{2A}|\beta|^{2}\frac{d\beta}{dz}+\frac{V^{3}\bar{\beta}^{2}}{8UA^{2}}(\frac{d\beta}{dz})^{2}
-UV\left[\frac{d\beta}{dz}-\frac{V}{2U}\frac{|\beta|^{2}}{A}\frac{d\beta}{dz}+\frac{V^{2}}{8U^{2}}\frac{\bar{\beta}^{2}}{A^{2}}(\frac{d\beta}{dz})^{2}\right]
\nonumber
\\
&=-UV\left(\frac{d\beta}{dz}-\frac{\beta ^{2}}{2}\right)
\end{align}
Note that,
\begin{align}
\frac{d\beta}{dz}-\frac{\beta ^{2}}{2}=\frac{d}{dz}\left(\frac{w''}{w'} \right)-\frac{1}{2}\left(\frac{w''}{w'}\right)^{2}
=\frac{w'''}{w'}-\frac{3}{2}\left(\frac{w''}{w'}\right)^{2}=\left\lbrace w,z\right\rbrace
\end{align}
This is the superrotation, in which the $C_{zz}$ changes by the Schwarzian derivative.
\section{Appendix: Coherent (like) States  with Compact Correlation Support}\label{NVDwithCompact}

We can consider the following situation in Minkowski coordinates with mode function $u_{k}(x)=e^{ikx}$. Then let us construct the following state,
\begin{equation}
|\psi\rangle =\int dx f(x)\hat{\phi}(x)|0\rangle
\end{equation}
Expanding out the field in Minkowski mode functions and noting that $a_{k}|0\rangle =0$ we obtain,
\begin{equation}
|\psi\rangle =\int dx f(x)\int dk \hat{a}_{k}^{\dagger}e^{-ikx}|0\rangle
=\int dk f(k)\hat{a}_{k}^{\dagger}|0\rangle
\end{equation}
where we have introduced, 
\begin{equation}
f(k)=\int dx~f(x)e^{-ikx}
\end{equation}
Let the function $f(x)$ has compact support, then, we can choose it of the form,
\begin{equation}
f(x)=A\Theta (x-x_{1})\Theta (x_{2}-x)
\end{equation}
such that $f(x)$ vanishes outside the region $x_{1}<x<x_{2}$. The two point correlation function being,
\begin{equation}
\langle \psi| \phi (x)\phi (y)|\psi \rangle =\langle 0|\phi (x)\phi (y)|0\rangle+f(x)f^{*}(y)+f(y)f^{*}(x)
\end{equation}
Hence correlation between any two points not in the region $x_{1}<x<x_{2}$ identically vanishes. However note that the k-space correlation still exists, as the $k$ space would not have any compact support, as,
\begin{equation}
f(k)=\int _{-\infty}^{\infty}dx f(x)e^{-ikx}=A\int _{x_{1}}^{x_{2}}dx~e^{-ikx}=\frac{iA}{k}\left(e^{-ikx_{2}}-e^{-ikx_{1}}\right)
\end{equation}
which is non-vanishing for all $k$. Let us now consider whether the correction to number expectation can vanish or not. One such term in the correction would be,
\begin{equation}
N_{\rm corr}(K)=\int dk \alpha (k,K)f(k)
\end{equation}
which is nonzero, since f(k) is non-zero for all k. One might tempt to think the situation in position space, but that leads to difficulties, since the support for the function $\alpha (K,k)$ is only in the right Rindler wedge, so its Fourier transform should involve integration over that range only, while that of $f(k)$ is throughout the Minkowski time slice. Thus one would have in position space,
\begin{equation}
N_{\rm corr}(K)=\int dk \int dz e^{ikz}\alpha (K,z)\int dx e^{ikx}f(x)
\end{equation}
which will be nonzero even if $f(x)$ have a compact support due to the non-trivial relation between $z$ and $x$ and the fact that the range $z\in (-\infty,\infty)$ will map to a finite range in $x$.

\bibliography{Information_Paradox}

\providecommand{\href}[2]{#2}\begingroup\raggedright\begin{thebibliography}{10%
0}

\bibitem{Adler:1993hm}
S.~L. Adler, ``{Generalized quantum dynamics},''
  \href{http://dx.doi.org/10.1016/0550-3213(94)90072-8}{{\em Nucl. Phys.}
  {\bfseries B415} (1994) 195--242},
\href{http://arxiv.org/abs/hep-th/9306009}{{\ttfamily arXiv:hep-th/9306009
  [hep-th]}}.

\bibitem{Bohm:1951xw}
D.~Bohm, ``{A Suggested interpretation of the quantum theory in terms of hidden
  variables. 1.},''
\href{http://dx.doi.org/10.1103/PhysRev.85.166}{{\em Phys. Rev.} {\bfseries 85}
  (1952) 166--179}.

\bibitem{Bohm:1951xx}
D.~Bohm, ``{A Suggested interpretation of the quantum theory in terms of hidden
  variables. 2.},''
\href{http://dx.doi.org/10.1103/PhysRev.85.180}{{\em Phys. Rev.} {\bfseries 85}
  (1952) 180--193}.

\bibitem{Diosi:1986nu}
L.~Diosi, ``{A Universal Master Equation for the Gravitational Violation of
  Quantum Mechanics},''
\href{http://dx.doi.org/10.1016/0375-9601(87)90681-5}{{\em Phys. Lett.}
  {\bfseries A120} (1987) 377}.

\bibitem{Everett:1957hd}
H.~Everett, ``{Relative state formulation of quantum mechanics},''
\href{http://dx.doi.org/10.1103/RevModPhys.29.454}{{\em Rev. Mod. Phys.}
  {\bfseries 29} (1957) 454--462}.

\bibitem{Bassi:2012bg}
A.~Bassi, K.~Lochan, S.~Satin, T.~P. Singh, and H.~Ulbricht, ``{Models of
  Wave-function Collapse, Underlying Theories, and Experimental Tests},''
  \href{http://dx.doi.org/10.1103/RevModPhys.85.471}{{\em Rev. Mod. Phys.}
  {\bfseries 85} (2013) 471--527},
\href{http://arxiv.org/abs/1204.4325}{{\ttfamily arXiv:1204.4325 [quant-ph]}}.

\bibitem{Bassi:2003gd}
A.~Bassi and G.~C. Ghirardi, ``{Dynamical reduction models},''
  \href{http://dx.doi.org/10.1016/S0370-1573(03)00103-0}{{\em Phys. Rept.}
  {\bfseries 379} (2003) 257},
\href{http://arxiv.org/abs/quant-ph/0302164}{{\ttfamily arXiv:quant-ph/0302164
  [quant-ph]}}.

\bibitem{Hawking:1976ra}
S.~W. Hawking, ``{Breakdown of Predictability in Gravitational Collapse},''
\href{http://dx.doi.org/10.1103/PhysRevD.14.2460}{{\em Phys. Rev.} {\bfseries
  D14} (1976) 2460--2473}.

\bibitem{Wald:1997wa}
R.~M. Wald, ``{Gravitational collapse and cosmic censorship},'' in {\em {In
  *Iyer, B.R. (ed.) et al.: Black holes, gravitational radiation and the
  universe* 69-85}}.
\newblock 1997.
\newblock \href{http://arxiv.org/abs/gr-qc/9710068}{{\ttfamily
  arXiv:gr-qc/9710068 [gr-qc]}}.
\newblock
\url{http://alice.cern.ch/format/showfull?sysnb=0259673}.
\newblock

\bibitem{Joshi:2002}
P.~S. JOSHI, ``Cosmic censorship: A current perspective,''
  \href{http://dx.doi.org/10.1142/S0217732302007570}{{\em Modern Physics
  Letters A} {\bfseries 17} no.~15n17, (2002) 1067--1079},
  \href{http://arxiv.org/abs/http://www.worldscientific.com/doi/pdf/10.1142/S0%
217732302007570}{{\ttfamily
  http://www.worldscientific.com/doi/pdf/10.1142/S0217732302007570}}.
  \url{http://www.worldscientific.com/doi/abs/10.1142/S0217732302007570}.

\bibitem{Hod:2008zza}
S.~Hod, ``{Weak Cosmic Censorship: As Strong as Ever},''
  \href{http://dx.doi.org/10.1103/PhysRevLett.100.121101}{{\em Phys. Rev.
  Lett.} {\bfseries 100} (2008) 121101},
\href{http://arxiv.org/abs/0805.3873}{{\ttfamily arXiv:0805.3873 [gr-qc]}}.

\bibitem{Hamid:2014kza}
A.~I.~M. Hamid, R.~Goswami, and S.~D. Maharaj, ``{Cosmic Censorship Conjecture
  revisited: Covariantly},''
  \href{http://dx.doi.org/10.1088/0264-9381/31/13/135010}{{\em Class. Quant.
  Grav.} {\bfseries 31} (2014) 135010},
\href{http://arxiv.org/abs/1402.4355}{{\ttfamily arXiv:1402.4355 [gr-qc]}}.

\bibitem{MTW}
C.~W. Misner, K.~S. Thorne, and J.~A. Wheeler, {\em {Gravitation}}.
\newblock W. H. Freeman and Company, 3~ed., 1973.

\bibitem{Hawking:1973uf}
S.~W. Hawking and G.~F.~R. Ellis, {\em {The Large Scale Structure of
  Space-Time}}.
\newblock Cambridge Monographs on Mathematical Physics. Cambridge University
  Press,
2011.
\newblock

\bibitem{Hawking:2010mca}
S.~W. Hawking and W.~Israel~eds, {\em {General Relativity: An Einstein
  Centenary Survey}}.
\newblock Cambridge University Press, 1979.

\bibitem{gravitation}
T.Padmanabhan, {\em {Gravitation: Foundations and Frontiers}}.
\newblock Cambridge University Press, Cambridge, UK, 2010.

\bibitem{Mukhanov-Winitzky}
V.~Mukhanov and S.~Winitzki, {\em {Introduction to Quantum Effects in
  Gravity}}.
\newblock Cambridge University Press, 1st~ed., 2007.

\bibitem{Poisson}
E.~Poisson, {\em {A Relativist's Toolkit: The Mathematics of Black-Hole
  Mechanics}}.
\newblock Cambridge University Press, 1st~ed., 2007.

\bibitem{Wald}
R.~M. Wald, {\em {General Relativity}}.
\newblock The University of Chicago Press, 1st~ed., 1984.

\bibitem{Birrell:1982ix}
N.~D. Birrell and P.~C.~W. Davies,
  \href{http://dx.doi.org/10.1017/CBO9780511622632}{{\em {Quantum Fields in
  Curved Space}}}.
\newblock Cambridge Monographs on Mathematical Physics. Cambridge Univ. Press,
  Cambridge, UK, 1984.
\newblock
\url{http://www.cambridge.org/mw/academic/subjects/physics/theoretical-physics%
-and-mathematical-physics/quantum-fields-curved-space?format=PB}.
\newblock

\bibitem{Parker:2009uva}
L.~E. Parker and D.~Toms, {\em {Quantum Field Theory in Curved Spacetime}}.
\newblock Cambridge Monographs on Mathematical Physics. Cambridge University
  Press, 2009.
\newblock
\url{http://www.cambridge.org/de/knowledge/isbn/item2327457}.
\newblock

\bibitem{Fabbri:2005mw}
A.~Fabbri and J.~Navarro-Salas, {\em {Modeling black hole evaporation}}.
\newblock
2005.
\newblock

\bibitem{Fulling:1989nb}
S.~A. Fulling, ``{Aspects of Quantum Field Theory in Curved Space-time},''
{\em London Math. Soc. Student Texts} {\bfseries 17} (1989) 1--315.

\bibitem{Bekenstein:1974ax}
J.~D. Bekenstein, ``{Generalized second law of thermodynamics in black hole
  physics},''
\href{http://dx.doi.org/10.1103/PhysRevD.9.3292}{{\em Phys. Rev.} {\bfseries
  D9} (1974) 3292--3300}.

\bibitem{Bardeen:1973gs}
J.~M. Bardeen, B.~Carter, and S.~W. Hawking, ``{The Four laws of black hole
  mechanics},''
\href{http://dx.doi.org/10.1007/BF01645742}{{\em Commun. Math. Phys.}
  {\bfseries 31} (1973) 161--170}.

\bibitem{Gibbons:1977mu}
G.~W. Gibbons and S.~W. Hawking, ``{Cosmological Event Horizons,
  Thermodynamics, and Particle Creation},''
\href{http://dx.doi.org/10.1103/PhysRevD.15.2738}{{\em Phys. Rev.} {\bfseries
  D15} (1977) 2738--2751}.

\bibitem{Hawking:1976de}
S.~W. Hawking, ``{Black Holes and Thermodynamics},''
\href{http://dx.doi.org/10.1103/PhysRevD.13.191}{{\em Phys. Rev.} {\bfseries
  D13} (1976) 191--197}.

\bibitem{Bekenstein:1973ur}
J.~D. Bekenstein, ``{Black holes and entropy},''
\href{http://dx.doi.org/10.1103/PhysRevD.7.2333}{{\em Phys. Rev.} {\bfseries
  D7} (1973) 2333--2346}.

\bibitem{Bekenstein:1972tm}
J.~D. Bekenstein, ``{Black holes and the second law},''
\href{http://dx.doi.org/10.1007/BF02757029}{{\em Lett. Nuovo Cim.} {\bfseries
  4} (1972) 737--740}.

\bibitem{Hawking:1974sw}
S.~W. Hawking, ``{Particle Creation by Black Holes},''
  \href{http://dx.doi.org/10.1007/BF02345020}{{\em Commun. Math. Phys.}
  {\bfseries 43} (1975) 199--220}.
[erratum,ibid,167(1975)].

\bibitem{Jacobson:1995ab}
T.~Jacobson, ``{Thermodynamics of space-time: The Einstein equation of
  state},'' \href{http://dx.doi.org/10.1103/PhysRevLett.75.1260}{{\em
  Phys.Rev.Lett.} {\bfseries 75} (1995) 1260--1263},
\href{http://arxiv.org/abs/gr-qc/9504004}{{\ttfamily arXiv:gr-qc/9504004
  [gr-qc]}}.

\bibitem{Padmanabhan:2003gd}
T.~Padmanabhan, ``{Gravity and the thermodynamics of horizons},''
  \href{http://dx.doi.org/10.1016/j.physrep.2004.10.003}{{\em Phys.Rept.}
  {\bfseries 406} (2005) 49--125},
\href{http://arxiv.org/abs/gr-qc/0311036}{{\ttfamily arXiv:gr-qc/0311036
  [gr-qc]}}.

\bibitem{Padmanabhan:2009vy}
T.~Padmanabhan, ``{Thermodynamical Aspects of Gravity: New insights},''
  \href{http://dx.doi.org/10.1088/0034-4885/73/4/046901}{{\em Rept. Prog.
  Phys.} {\bfseries 73} (2010) 046901},
\href{http://arxiv.org/abs/0911.5004}{{\ttfamily arXiv:0911.5004 [gr-qc]}}.

\bibitem{Padmanabhan:2013nxa}
T.~Padmanabhan, ``{General Relativity from a Thermodynamic Perspective},'' {\em
  Gen.Rel.Grav.} {\bfseries 46} (2014) 1673,
\href{http://arxiv.org/abs/1312.3253}{{\ttfamily arXiv:1312.3253 [gr-qc]}}.

\bibitem{Chakraborty:2015hna}
S.~Chakraborty and T.~Padmanabhan, ``{Thermodynamical interpretation of the
  geometrical variables associated with null surfaces},''
  \href{http://dx.doi.org/10.1103/PhysRevD.92.104011}{{\em Phys. Rev.}
  {\bfseries D92} no.~10, (2015) 104011},
\href{http://arxiv.org/abs/1508.04060}{{\ttfamily arXiv:1508.04060 [gr-qc]}}.

\bibitem{Chakraborty:2015aja}
S.~Chakraborty, K.~Parattu, and T.~Padmanabhan, ``{Gravitational field
  equations near an arbitrary null surface expressed as a thermodynamic
  identity},'' \href{http://dx.doi.org/10.1007/JHEP10(2015)097}{{\em JHEP}
  {\bfseries 10} (2015) 097},
\href{http://arxiv.org/abs/1505.05297}{{\ttfamily arXiv:1505.05297 [gr-qc]}}.

\bibitem{Chakraborty:2014rga}
S.~Chakraborty and T.~Padmanabhan, ``{Evolution of Spacetime arises due to the
  departure from Holographic Equipartition in all Lanczos-Lovelock Theories of
  Gravity},'' \href{http://dx.doi.org/10.1103/PhysRevD.90.124017}{{\em Phys.
  Rev.} {\bfseries D90} no.~12, (2014) 124017},
\href{http://arxiv.org/abs/1408.4679}{{\ttfamily arXiv:1408.4679 [gr-qc]}}.

\bibitem{Chakraborty:2015wma}
S.~Chakraborty, ``{Lanczos-Lovelock gravity from a thermodynamic
  perspective},'' \href{http://dx.doi.org/10.1007/JHEP08(2015)029}{{\em JHEP}
  {\bfseries 08} (2015) 029},
\href{http://arxiv.org/abs/1505.07272}{{\ttfamily arXiv:1505.07272 [gr-qc]}}.

\bibitem{Padmanabhan:2016eld}
T.~Padmanabhan, ``{The Atoms Of Space, Gravity and the Cosmological
  Constant},'' \href{http://dx.doi.org/10.1142/S0218271816300202}{{\em Int. J.
  Mod. Phys.} {\bfseries D25} no.~07, (2016) 1630020},
\href{http://arxiv.org/abs/1603.08658}{{\ttfamily arXiv:1603.08658 [gr-qc]}}.

\bibitem{Padmanabhan:2015zmr}
T.~Padmanabhan, ``{Gravity and/is Thermodynamics},''
  \href{http://dx.doi.org/10.18520/v109/i12/2236-2242}{{\em Curr. Sci.}
  {\bfseries 109} (2015) 2236--2242},
\href{http://arxiv.org/abs/1512.06546}{{\ttfamily arXiv:1512.06546 [gr-qc]}}.

\bibitem{Padmanabhan:2015pza}
T.~Padmanabhan, ``{Distribution function of the Atoms of Spacetime and the
  Nature of Gravity},'' \href{http://dx.doi.org/10.3390/e17117420}{{\em
  Entropy} {\bfseries 17} (2015) 7420--7452},
\href{http://arxiv.org/abs/1508.06286}{{\ttfamily arXiv:1508.06286 [gr-qc]}}.

\bibitem{Chakraborty:2016dwb}
S.~Chakraborty, S.~Bhattacharya, and T.~Padmanabhan, ``{Entropy of a generic
  null surface from its associated Virasoro algebra},''
  \href{http://dx.doi.org/10.1016/j.physletb.2016.10.059}{{\em Phys. Lett.}
  {\bfseries B763} (2016) 347--351},
\href{http://arxiv.org/abs/1605.06988}{{\ttfamily arXiv:1605.06988 [gr-qc]}}.

\bibitem{Mathur:2009hf}
S.~D. Mathur, ``{The Information paradox: A Pedagogical introduction},''
  \href{http://dx.doi.org/10.1088/0264-9381/26/22/224001}{{\em Class. Quant.
  Grav.} {\bfseries 26} (2009) 224001},
\href{http://arxiv.org/abs/0909.1038}{{\ttfamily arXiv:0909.1038 [hep-th]}}.

\bibitem{Harlow:2014yka}
D.~Harlow, ``{Jerusalem Lectures on Black Holes and Quantum Information},''
  \href{http://dx.doi.org/10.1103/RevModPhys.88.015002}{{\em Rev. Mod. Phys.}
  {\bfseries 88} (2016) 15002},
  \href{http://arxiv.org/abs/1409.1231}{{\ttfamily arXiv:1409.1231 [hep-th]}}.
[Rev. Mod. Phys.88,15002(2016)].

\bibitem{Polchinski:2016hrw}
J.~Polchinski, \href{http://dx.doi.org/10.1142/9789813149441_0006}{``{The Black
  Hole Information Problem},''} in {\em {Proceedings, Theoretical Advanced
  Study Institute in Elementary Particle Physics: New Frontiers in Fields and
  Strings (TASI 2015): Boulder, CO, USA, June 1-26, 2015}}, pp.~353--397.
\newblock 2017.
\newblock \href{http://arxiv.org/abs/1609.04036}{{\ttfamily arXiv:1609.04036
  [hep-th]}}.
\newblock
\url{https://inspirehep.net/record/1486509/files/arXiv:1609.04036.pdf}.
\newblock

\bibitem{Hawking:1974rv}
S.~Hawking, ``{Black hole explosions},''
\href{http://dx.doi.org/10.1038/248030a0}{{\em Nature} {\bfseries 248} (1974)
  30--31}.

\bibitem{Gibbons:1976ue}
G.~Gibbons and S.~Hawking, ``{Action Integrals and Partition Functions in
  Quantum Gravity},''
\href{http://dx.doi.org/10.1103/PhysRevD.15.2752}{{\em Phys.Rev.} {\bfseries
  D15} (1977) 2752--2756}.

\bibitem{Mann:2014yxa}
R.~B. Mann, ``{The Firewall Phenomenon},''
\href{http://dx.doi.org/10.1007/978-3-319-10852-0_3}{{\em Fundam. Theor. Phys.}
  {\bfseries 178} (2015) 71--113}.

\bibitem{Kiefer:2001wn}
C.~Kiefer, ``{Hawking radiation from decoherence},''
  \href{http://dx.doi.org/10.1088/0264-9381/18/22/101}{{\em Class. Quant.
  Grav.} {\bfseries 18} (2001) L151},
\href{http://arxiv.org/abs/gr-qc/0110070}{{\ttfamily arXiv:gr-qc/0110070
  [gr-qc]}}.

\bibitem{Demers:1995tr}
J.-G. Demers and C.~Kiefer, ``{Decoherence of black holes by Hawking
  radiation},'' \href{http://dx.doi.org/10.1103/PhysRevD.53.7050}{{\em Phys.
  Rev.} {\bfseries D53} (1996) 7050--7061},
\href{http://arxiv.org/abs/hep-th/9511147}{{\ttfamily arXiv:hep-th/9511147
  [hep-th]}}.

\bibitem{Maldacena:2013xja}
J.~Maldacena and L.~Susskind, ``{Cool horizons for entangled black holes},''
  \href{http://dx.doi.org/10.1002/prop.201300020}{{\em Fortsch. Phys.}
  {\bfseries 61} (2013) 781--811},
\href{http://arxiv.org/abs/1306.0533}{{\ttfamily arXiv:1306.0533 [hep-th]}}.

\bibitem{Visser:2014ypa}
M.~Visser, ``{Thermality of the Hawking flux},''
  \href{http://dx.doi.org/10.1007/JHEP07(2015)009}{{\em JHEP} {\bfseries 07}
  (2015) 009},
\href{http://arxiv.org/abs/1409.7754}{{\ttfamily arXiv:1409.7754 [gr-qc]}}.

\bibitem{Alonso-Serrano:2015trn}
A.~Alonso-Serrano and M.~Visser, ``{On burning a lump of coal},''
  \href{http://dx.doi.org/10.1016/j.physletb.2016.04.023}{{\em Phys. Lett.}
  {\bfseries B757} (2016) 383--386},
\href{http://arxiv.org/abs/1511.01162}{{\ttfamily arXiv:1511.01162 [gr-qc]}}.

\bibitem{Alonso-Serrano:2015bcr}
A.~Alonso-Serrano and M.~Visser, ``{Entropy/information flux in Hawking
  radiation},''
\href{http://arxiv.org/abs/1512.01890}{{\ttfamily arXiv:1512.01890 [gr-qc]}}.

\bibitem{Mathur:2008kg}
S.~D. Mathur, ``{Tunneling into fuzzball states},''
  \href{http://dx.doi.org/10.1007/s10714-009-0837-3}{{\em Gen. Rel. Grav.}
  {\bfseries 42} (2010) 113--118},
\href{http://arxiv.org/abs/0805.3716}{{\ttfamily arXiv:0805.3716 [hep-th]}}.

\bibitem{Papadodimas:2013wnh}
K.~Papadodimas and S.~Raju, ``{Black Hole Interior in the Holographic
  Correspondence and the Information Paradox},''
  \href{http://dx.doi.org/10.1103/PhysRevLett.112.051301}{{\em Phys. Rev.
  Lett.} {\bfseries 112} no.~5, (2014) 051301},
\href{http://arxiv.org/abs/1310.6334}{{\ttfamily arXiv:1310.6334 [hep-th]}}.

\bibitem{Modak:2014vya}
S.~K. Modak, L.~Ortíz, I.~Peña, and D.~Sudarsky, ``{Non-Paradoxical Loss of
  Information in Black Hole Evaporation in a Quantum Collapse Model},''
  \href{http://dx.doi.org/10.1103/PhysRevD.91.124009}{{\em Phys. Rev.}
  {\bfseries D91} no.~12, (2015) 124009},
\href{http://arxiv.org/abs/1408.3062}{{\ttfamily arXiv:1408.3062 [gr-qc]}}.

\bibitem{Chakraborty:2015nwa}
S.~Chakraborty, S.~Singh, and T.~Padmanabhan, ``{A quantum peek inside the
  black hole event horizon},''
  \href{http://dx.doi.org/10.1007/JHEP06(2015)192}{{\em JHEP} {\bfseries 06}
  (2015) 192},
\href{http://arxiv.org/abs/1503.01774}{{\ttfamily arXiv:1503.01774 [gr-qc]}}.

\bibitem{Singh:2014paa}
S.~Singh and S.~Chakraborty, ``{Black hole kinematics: The “in”-vacuum
  energy density and flux for different observers},''
  \href{http://dx.doi.org/10.1103/PhysRevD.90.024011}{{\em Phys. Rev.}
  {\bfseries D90} no.~2, (2014) 024011},
\href{http://arxiv.org/abs/1404.0684}{{\ttfamily arXiv:1404.0684 [gr-qc]}}.

\bibitem{Khlopov:2008qy}
M.~{\relax Yu}. Khlopov, ``{Primordial Black Holes},''
  \href{http://dx.doi.org/10.1088/1674-4527/10/6/001}{{\em Res. Astron.
  Astrophys.} {\bfseries 10} (2010) 495--528},
\href{http://arxiv.org/abs/0801.0116}{{\ttfamily arXiv:0801.0116 [astro-ph]}}.

\bibitem{Khlopov:2004tn}
M.~{\relax Yu}. Khlopov, A.~Barrau, and J.~Grain, ``{Gravitino production by
  primordial black hole evaporation and constraints on the inhomogeneity of the
  early universe},'' \href{http://dx.doi.org/10.1088/0264-9381/23/6/004}{{\em
  Class. Quant. Grav.} {\bfseries 23} (2006) 1875--1882},
\href{http://arxiv.org/abs/astro-ph/0406621}{{\ttfamily arXiv:astro-ph/0406621
  [astro-ph]}}.

\bibitem{Khlopov:1985jw}
M.~Khlopov, B.~A. Malomed, and I.~B. Zeldovich, ``{Gravitational instability of
  scalar fields and formation of primordial black holes},''
{\em Mon. Not. Roy. Astron. Soc.} {\bfseries 215} (1985) 575--589.

\bibitem{Gray:2015pma}
F.~Gray, S.~Schuster, A.~Van–Brunt, and M.~Visser, ``{The Hawking cascade
  from a black hole is extremely sparse},''
  \href{http://dx.doi.org/10.1088/0264-9381/33/11/115003}{{\em Class. Quant.
  Grav.} {\bfseries 33} no.~11, (2016) 115003},
\href{http://arxiv.org/abs/1506.03975}{{\ttfamily arXiv:1506.03975 [gr-qc]}}.

\bibitem{Hod:2000kb}
S.~Hod, ``{Discrete black hole radiation and the information loss paradox},''
  \href{http://dx.doi.org/10.1016/S0375-9601(02)00013-0}{{\em Phys. Lett.}
  {\bfseries A299} (2002) 144--148},
\href{http://arxiv.org/abs/gr-qc/0012076}{{\ttfamily arXiv:gr-qc/0012076
  [gr-qc]}}.

\bibitem{Hod:2000it}
S.~Hod, ``{Gravitation, the quantum, and Bohr's correspondence principle},''
  \href{http://dx.doi.org/10.1023/A:1026753914838}{{\em Gen. Rel. Grav.}
  {\bfseries 31} (1999) 1639},
\href{http://arxiv.org/abs/gr-qc/0002002}{{\ttfamily arXiv:gr-qc/0002002
  [gr-qc]}}.

\bibitem{Chakraborty:2016fye}
S.~Chakraborty and K.~Lochan, ``{Quantum leaps of black holes: Magnifying
  glasses of quantum gravity},''
  \href{http://dx.doi.org/10.1142/S0218271816440247}{{\em Int. J. Mod. Phys.}
  {\bfseries D25} (2016) 1644024},
\href{http://arxiv.org/abs/1606.04348}{{\ttfamily arXiv:1606.04348 [gr-qc]}}.

\bibitem{Lochan:2015bha}
K.~Lochan and S.~Chakraborty, ``{Discrete quantum spectrum of black holes},''
  \href{http://dx.doi.org/10.1016/j.physletb.2016.01.060}{{\em Phys. Lett.}
  {\bfseries B755} (2016) 37--42},
\href{http://arxiv.org/abs/1509.09010}{{\ttfamily arXiv:1509.09010 [gr-qc]}}.

\bibitem{Saini:2015dea}
A.~Saini and D.~Stojkovic, ``{Radiation from a collapsing object is manifestly
  unitary},'' \href{http://dx.doi.org/10.1103/PhysRevLett.114.111301}{{\em
  Phys. Rev. Lett.} {\bfseries 114} no.~11, (2015) 111301},
\href{http://arxiv.org/abs/1503.01487}{{\ttfamily arXiv:1503.01487 [gr-qc]}}.

\bibitem{Page:1993wv}
D.~N. Page, ``{Information in black hole radiation},''
  \href{http://dx.doi.org/10.1103/PhysRevLett.71.3743}{{\em Phys. Rev. Lett.}
  {\bfseries 71} (1993) 3743--3746},
\href{http://arxiv.org/abs/hep-th/9306083}{{\ttfamily arXiv:hep-th/9306083
  [hep-th]}}.

\bibitem{Braunstein:2006sj}
S.~L. Braunstein and A.~K. Pati, ``{Quantum information cannot be completely
  hidden in correlations: Implications for the black-hole information
  paradox},'' \href{http://dx.doi.org/10.1103/PhysRevLett.98.080502}{{\em Phys.
  Rev. Lett.} {\bfseries 98} (2007) 080502},
\href{http://arxiv.org/abs/gr-qc/0603046}{{\ttfamily arXiv:gr-qc/0603046
  [gr-qc]}}.

\bibitem{Braunstein:2005zz}
S.~L. Braunstein and P.~van Loock, ``{Quantum information with continuous
  variables},''
\href{http://dx.doi.org/10.1103/RevModPhys.77.513}{{\em Rev. Mod. Phys.}
  {\bfseries 77} (2005) 513--577}.

\bibitem{Braunstein:2009my}
S.~L. Braunstein, S.~Pirandola, and K.~Życzkowski, ``{Better Late than Never:
  Information Retrieval from Black Holes},''
  \href{http://dx.doi.org/10.1103/PhysRevLett.110.101301}{{\em Phys. Rev.
  Lett.} {\bfseries 110} no.~10, (2013) 101301},
\href{http://arxiv.org/abs/0907.1190}{{\ttfamily arXiv:0907.1190 [quant-ph]}}.

\bibitem{Bose:1996pi}
S.~Bose, L.~Parker, and Y.~Peleg, ``{Predictability and semiclassical
  approximation at the onset of black hole formation},''
  \href{http://dx.doi.org/10.1103/PhysRevD.54.7490}{{\em Phys. Rev.} {\bfseries
  D54} (1996) 7490--7505},
\href{http://arxiv.org/abs/hep-th/9606152}{{\ttfamily arXiv:hep-th/9606152
  [hep-th]}}.

\bibitem{Susskind:1993if}
L.~Susskind, L.~Thorlacius, and J.~Uglum, ``{The Stretched horizon and black
  hole complementarity},''
  \href{http://dx.doi.org/10.1103/PhysRevD.48.3743}{{\em Phys. Rev.} {\bfseries
  D48} (1993) 3743--3761},
\href{http://arxiv.org/abs/hep-th/9306069}{{\ttfamily arXiv:hep-th/9306069
  [hep-th]}}.

\bibitem{Almheiri:2012rt}
A.~Almheiri, D.~Marolf, J.~Polchinski, and J.~Sully, ``{Black Holes:
  Complementarity or Firewalls?},''
  \href{http://dx.doi.org/10.1007/JHEP02(2013)062}{{\em JHEP} {\bfseries 02}
  (2013) 062},
\href{http://arxiv.org/abs/1207.3123}{{\ttfamily arXiv:1207.3123 [hep-th]}}.

\bibitem{Price:1986yy}
R.~H. Price and K.~S. Thorne, ``{Membrane Viewpoint on Black Holes: Properties
  and Evolution of the Stretched Horizon},''
\href{http://dx.doi.org/10.1103/PhysRevD.33.915}{{\em Phys. Rev.} {\bfseries
  D33} (1986) 915--941}.

\bibitem{Parikh:1998mg}
M.~K. Parikh, {\em {Membrane horizons: The Black hole's new clothes}}.
\newblock PhD thesis, Princeton U., 1998.
\newblock \href{http://arxiv.org/abs/hep-th/9907002}{{\ttfamily
  arXiv:hep-th/9907002 [hep-th]}}.
\newblock
\url{http://wwwlib.umi.com/dissertations/fullcit?p9920459-MC}.
\newblock

\bibitem{Bousso:2012as}
R.~Bousso, ``{Complementarity Is Not Enough},''
  \href{http://dx.doi.org/10.1103/PhysRevD.87.124023}{{\em Phys. Rev.}
  {\bfseries D87} no.~12, (2013) 124023},
\href{http://arxiv.org/abs/1207.5192}{{\ttfamily arXiv:1207.5192 [hep-th]}}.

\bibitem{Avery:2012tf}
S.~G. Avery, B.~D. Chowdhury, and A.~Puhm, ``{Unitarity and fuzzball
  complementarity: 'Alice fuzzes but may not even know it!'},''
  \href{http://dx.doi.org/10.1007/JHEP09(2013)012}{{\em JHEP} {\bfseries 09}
  (2013) 012},
\href{http://arxiv.org/abs/1210.6996}{{\ttfamily arXiv:1210.6996 [hep-th]}}.

\bibitem{Chowdhury:2012vd}
B.~D. Chowdhury and A.~Puhm, ``{Is Alice burning or fuzzing?},''
  \href{http://dx.doi.org/10.1103/PhysRevD.88.063509}{{\em Phys. Rev.}
  {\bfseries D88} (2013) 063509},
\href{http://arxiv.org/abs/1208.2026}{{\ttfamily arXiv:1208.2026 [hep-th]}}.

\bibitem{Mathur:2013gua}
S.~D. Mathur and D.~Turton, ``{The flaw in the firewall argument},''
  \href{http://dx.doi.org/10.1016/j.nuclphysb.2014.05.012}{{\em Nucl. Phys.}
  {\bfseries B884} (2014) 566--611},
\href{http://arxiv.org/abs/1306.5488}{{\ttfamily arXiv:1306.5488 [hep-th]}}.

\bibitem{Almheiri:2013hfa}
A.~Almheiri, D.~Marolf, J.~Polchinski, D.~Stanford, and J.~Sully, ``{An
  Apologia for Firewalls},''
  \href{http://dx.doi.org/10.1007/JHEP09(2013)018}{{\em JHEP} {\bfseries 09}
  (2013) 018},
\href{http://arxiv.org/abs/1304.6483}{{\ttfamily arXiv:1304.6483 [hep-th]}}.

\bibitem{Lee:2013vga}
B.-H. Lee and D.-h. Yeom, ``{Status report: black hole complementarity
  controversy},''
  \href{http://dx.doi.org/10.1016/j.nuclphysbps.2013.10.082}{{\em Nucl. Phys.
  Proc. Suppl.} {\bfseries 246-247} (2014) 178--182},
\href{http://arxiv.org/abs/1302.6006}{{\ttfamily arXiv:1302.6006 [gr-qc]}}.

\bibitem{Harlow:2013tf}
D.~Harlow and P.~Hayden, ``{Quantum Computation vs. Firewalls},''
  \href{http://dx.doi.org/10.1007/JHEP06(2013)085}{{\em JHEP} {\bfseries 06}
  (2013) 085},
\href{http://arxiv.org/abs/1301.4504}{{\ttfamily arXiv:1301.4504 [hep-th]}}.

\bibitem{Bousso:2013wia}
R.~Bousso, ``{Firewalls from double purity},''
  \href{http://dx.doi.org/10.1103/PhysRevD.88.084035}{{\em Phys. Rev.}
  {\bfseries D88} no.~8, (2013) 084035},
\href{http://arxiv.org/abs/1308.2665}{{\ttfamily arXiv:1308.2665 [hep-th]}}.

\bibitem{Papadodimas:2012aq}
K.~Papadodimas and S.~Raju, ``{An Infalling Observer in AdS/CFT},''
  \href{http://dx.doi.org/10.1007/JHEP10(2013)212}{{\em JHEP} {\bfseries 10}
  (2013) 212},
\href{http://arxiv.org/abs/1211.6767}{{\ttfamily arXiv:1211.6767 [hep-th]}}.

\bibitem{Nomura:2012cx}
Y.~Nomura, J.~Varela, and S.~J. Weinberg, ``{Black Holes, Information, and
  Hilbert Space for Quantum Gravity},''
  \href{http://dx.doi.org/10.1103/PhysRevD.87.084050}{{\em Phys. Rev.}
  {\bfseries D87} (2013) 084050},
\href{http://arxiv.org/abs/1210.6348}{{\ttfamily arXiv:1210.6348 [hep-th]}}.

\bibitem{Susskind:2012uw}
L.~Susskind, ``{The Transfer of Entanglement: The Case for Firewalls},''
\href{http://arxiv.org/abs/1210.2098}{{\ttfamily arXiv:1210.2098 [hep-th]}}.

\bibitem{Brustein:2012jn}
R.~Brustein, ``{Origin of the blackhole information paradox},''
  \href{http://dx.doi.org/10.1002/prop.201300037}{{\em Fortsch. Phys.}
  {\bfseries 62} (2014) 255--265},
\href{http://arxiv.org/abs/1209.2686}{{\ttfamily arXiv:1209.2686 [hep-th]}}.

\bibitem{Giveon:2012kp}
A.~Giveon and N.~Itzhaki, ``{String Theory Versus Black Hole
  Complementarity},'' \href{http://dx.doi.org/10.1007/JHEP12(2012)094}{{\em
  JHEP} {\bfseries 12} (2012) 094},
\href{http://arxiv.org/abs/1208.3930}{{\ttfamily arXiv:1208.3930 [hep-th]}}.

\bibitem{Shenker:2013pqa}
S.~H. Shenker and D.~Stanford, ``{Black holes and the butterfly effect},''
  \href{http://dx.doi.org/10.1007/JHEP03(2014)067}{{\em JHEP} {\bfseries 03}
  (2014) 067},
\href{http://arxiv.org/abs/1306.0622}{{\ttfamily arXiv:1306.0622 [hep-th]}}.

\bibitem{Hooft:2016vug}
G.~'t~Hooft, ``{The firewall transformation for black holes and some of its
  implications},''
\href{http://arxiv.org/abs/1612.08640}{{\ttfamily arXiv:1612.08640 [gr-qc]}}.

\bibitem{Ong:2014maa}
Y.~C. Ong, B.~McInnes, and P.~Chen, ``{Cold black holes in the Harlow–Hayden
  approach to firewalls},''
  \href{http://dx.doi.org/10.1016/j.nuclphysb.2014.12.024}{{\em Nucl. Phys.}
  {\bfseries B891} (2015) 627--654},
\href{http://arxiv.org/abs/1403.4886}{{\ttfamily arXiv:1403.4886 [hep-th]}}.

\bibitem{Hutchinson:2013kka}
J.~Hutchinson and D.~Stojkovic, ``{Icezones instead of firewalls: extended
  entanglement beyond the event horizon and unitary evaporation of a black
  hole},'' \href{http://dx.doi.org/10.1088/0264-9381/33/13/135006}{{\em Class.
  Quant. Grav.} {\bfseries 33} no.~13, (2016) 135006},
\href{http://arxiv.org/abs/1307.5861}{{\ttfamily arXiv:1307.5861 [hep-th]}}.

\bibitem{Mathur:2016ffb}
S.~D. Mathur, ``{What prevents gravitational collapse in string theory?},''
  \href{http://dx.doi.org/10.1142/S0218271816440181}{{\em Int. J. Mod. Phys.}
  {\bfseries D25} no.~12, (2016) 1644018},
\href{http://arxiv.org/abs/1609.05222}{{\ttfamily arXiv:1609.05222 [hep-th]}}.

\bibitem{Mathur:2014roa}
S.~D. Mathur, ``{Remnants, Fuzzballs or Wormholes?},''
  \href{http://dx.doi.org/10.1142/S0218271814420243}{{\em Int. J. Mod. Phys.}
  {\bfseries D23} no.~12, (2014) 1442024},
\href{http://arxiv.org/abs/1406.0807}{{\ttfamily arXiv:1406.0807 [hep-th]}}.

\bibitem{Mathur:2008nj}
S.~D. Mathur, ``{Fuzzballs and the information paradox: A Summary and
  conjectures},''
\href{http://arxiv.org/abs/0810.4525}{{\ttfamily arXiv:0810.4525 [hep-th]}}.

\bibitem{Mathur:2005zp}
S.~D. Mathur, ``{The Fuzzball proposal for black holes: An Elementary
  review},'' \href{http://dx.doi.org/10.1002/prop.200410203}{{\em Fortsch.
  Phys.} {\bfseries 53} (2005) 793--827},
\href{http://arxiv.org/abs/hep-th/0502050}{{\ttfamily arXiv:hep-th/0502050
  [hep-th]}}.

\bibitem{Strominger:1996sh}
A.~Strominger and C.~Vafa, ``{Microscopic origin of the Bekenstein-Hawking
  entropy},'' \href{http://dx.doi.org/10.1016/0370-2693(96)00345-0}{{\em Phys.
  Lett.} {\bfseries B379} (1996) 99--104},
\href{http://arxiv.org/abs/hep-th/9601029}{{\ttfamily arXiv:hep-th/9601029
  [hep-th]}}.

\bibitem{Mathur:1997wb}
S.~D. Mathur, ``{Emission rates, the correspondence principle and the
  information paradox},''
  \href{http://dx.doi.org/10.1016/S0550-3213(98)00336-8}{{\em Nucl. Phys.}
  {\bfseries B529} (1998) 295--320},
\href{http://arxiv.org/abs/hep-th/9706151}{{\ttfamily arXiv:hep-th/9706151
  [hep-th]}}.

\bibitem{Lunin:2002qf}
O.~Lunin and S.~D. Mathur, ``{Statistical interpretation of Bekenstein entropy
  for systems with a stretched horizon},''
  \href{http://dx.doi.org/10.1103/PhysRevLett.88.211303}{{\em Phys. Rev. Lett.}
  {\bfseries 88} (2002) 211303},
\href{http://arxiv.org/abs/hep-th/0202072}{{\ttfamily arXiv:hep-th/0202072
  [hep-th]}}.

\bibitem{Kraus:2015zda}
P.~Kraus and S.~D. Mathur, ``{Nature abhors a horizon},''
  \href{http://dx.doi.org/10.1142/S0218271815430038}{{\em Int. J. Mod. Phys.}
  {\bfseries D24} no.~12, (2015) 1543003},
\href{http://arxiv.org/abs/1505.05078}{{\ttfamily arXiv:1505.05078 [hep-th]}}.

\bibitem{Gibbons:2013tqa}
G.~W. Gibbons and N.~P. Warner, ``{Global structure of five-dimensional
  fuzzballs},'' \href{http://dx.doi.org/10.1088/0264-9381/31/2/025016}{{\em
  Class. Quant. Grav.} {\bfseries 31} (2014) 025016},
\href{http://arxiv.org/abs/1305.0957}{{\ttfamily arXiv:1305.0957 [hep-th]}}.

\bibitem{Chen:2015gux}
P.~Chen, Y.~C. Ong, D.~N. Page, M.~Sasaki, and D.-h. Yeom, ``{Naked Black Hole
  Firewalls},'' \href{http://dx.doi.org/10.1103/PhysRevLett.116.161304}{{\em
  Phys. Rev. Lett.} {\bfseries 116} no.~16, (2016) 161304},
\href{http://arxiv.org/abs/1511.05695}{{\ttfamily arXiv:1511.05695 [hep-th]}}.

\bibitem{Mathur:2011wg}
S.~D. Mathur and C.~J. Plumberg, ``{Correlations in Hawking radiation and the
  infall problem},'' \href{http://dx.doi.org/10.1007/JHEP09(2011)093}{{\em
  JHEP} {\bfseries 09} (2011) 093},
\href{http://arxiv.org/abs/1101.4899}{{\ttfamily arXiv:1101.4899 [hep-th]}}.

\bibitem{Mathur:2012zp}
S.~D. Mathur, ``{Black Holes and Beyond},''
  \href{http://dx.doi.org/10.1016/j.aop.2012.05.001}{{\em Annals Phys.}
  {\bfseries 327} (2012) 2760--2793},
\href{http://arxiv.org/abs/1205.0776}{{\ttfamily arXiv:1205.0776 [hep-th]}}.

\bibitem{Chowdhury:2007jx}
B.~D. Chowdhury and S.~D. Mathur, ``{Radiation from the non-extremal
  fuzzball},'' \href{http://dx.doi.org/10.1088/0264-9381/25/13/135005}{{\em
  Class. Quant. Grav.} {\bfseries 25} (2008) 135005},
\href{http://arxiv.org/abs/0711.4817}{{\ttfamily arXiv:0711.4817 [hep-th]}}.

\bibitem{Giddings:1992hh}
S.~B. Giddings, ``{Black holes and massive remnants},''
  \href{http://dx.doi.org/10.1103/PhysRevD.46.1347}{{\em Phys. Rev.} {\bfseries
  D46} (1992) 1347--1352},
\href{http://arxiv.org/abs/hep-th/9203059}{{\ttfamily arXiv:hep-th/9203059
  [hep-th]}}.

\bibitem{Chen:2014jwq}
P.~Chen, Y.~C. Ong, and D.-h. Yeom, ``{Black Hole Remnants and the Information
  Loss Paradox},'' \href{http://dx.doi.org/10.1016/j.physrep.2015.10.007}{{\em
  Phys. Rept.} {\bfseries 603} (2015) 1--45},
\href{http://arxiv.org/abs/1412.8366}{{\ttfamily arXiv:1412.8366 [gr-qc]}}.

\bibitem{Aharonov:1987tp}
Y.~Aharonov, A.~Casher, and S.~Nussinov, ``{The Unitarity Puzzle and Planck
  Mass Stable Particles},''
\href{http://dx.doi.org/10.1016/0370-2693(87)91320-7}{{\em Phys. Lett.}
  {\bfseries B191} (1987) 51}.

\bibitem{Nikolic:2015vga}
H.~Nikolic, ``{Gravitational crystal inside the black hole},''
  \href{http://dx.doi.org/10.1142/S0217732315502016}{{\em Mod. Phys. Lett.}
  {\bfseries A30} no.~37, (2015) 1550201},
\href{http://arxiv.org/abs/1505.04088}{{\ttfamily arXiv:1505.04088 [hep-th]}}.

\bibitem{Vaz:2012zq}
C.~Vaz and K.~Lochan, ``{Tunneling during quantum collapse in AdS spacetime},''
  \href{http://dx.doi.org/10.1103/PhysRevD.87.024045}{{\em Phys. Rev.}
  {\bfseries D87} no.~2, (2013) 024045},
\href{http://arxiv.org/abs/1204.5466}{{\ttfamily arXiv:1204.5466 [gr-qc]}}.

\bibitem{Saini:2014qpa}
A.~Saini and D.~Stojkovic, ``{Nonlocal (but also nonsingular) physics at the
  last stages of gravitational collapse},''
  \href{http://dx.doi.org/10.1103/PhysRevD.89.044003}{{\em Phys. Rev.}
  {\bfseries D89} no.~4, (2014) 044003},
\href{http://arxiv.org/abs/1401.6182}{{\ttfamily arXiv:1401.6182 [gr-qc]}}.

\bibitem{Greenwood:2008ht}
E.~Greenwood and D.~Stojkovic, ``{Quantum gravitational collapse:
  Non-singularity and non-locality},''
  \href{http://dx.doi.org/10.1088/1126-6708/2008/06/042}{{\em JHEP} {\bfseries
  06} (2008) 042},
\href{http://arxiv.org/abs/0802.4087}{{\ttfamily arXiv:0802.4087 [gr-qc]}}.

\bibitem{Vaz:2014rya}
C.~Vaz, ``{Black holes as Gravitational Atoms},''
  \href{http://dx.doi.org/10.1142/S0218271814410028}{{\em Int. J. Mod. Phys.}
  {\bfseries D23} no.~12, (2014) 1441002},
\href{http://arxiv.org/abs/1405.4898}{{\ttfamily arXiv:1405.4898 [gr-qc]}}.

\bibitem{Vaz:2014era}
C.~Vaz, ``{Quantum gravitational dust collapse does not result in a black
  hole},'' \href{http://dx.doi.org/10.1016/j.nuclphysb.2014.12.021}{{\em Nucl.
  Phys.} {\bfseries B891} (2015) 558--569},
\href{http://arxiv.org/abs/1407.3823}{{\ttfamily arXiv:1407.3823 [gr-qc]}}.

\bibitem{Sarkar:2016kqt}
S.~Sarkar, C.~Vaz, and L.~C.~R. Wijewardhana, ``{Quantum dust collapse in 2+1
  dimension},'' \href{http://dx.doi.org/10.1103/PhysRevD.93.043017}{{\em Phys.
  Rev.} {\bfseries D93} no.~4, (2016) 043017},
\href{http://arxiv.org/abs/1602.01141}{{\ttfamily arXiv:1602.01141 [gr-qc]}}.

\bibitem{Sahu:2015dea}
S.~Sahu, K.~Lochan, and D.~Narasimha, ``{Gravitational lensing by self-dual
  black holes in loop quantum gravity},''
  \href{http://dx.doi.org/10.1103/PhysRevD.91.063001}{{\em Phys. Rev.}
  {\bfseries D91} (2015) 063001},
\href{http://arxiv.org/abs/1502.05619}{{\ttfamily arXiv:1502.05619 [gr-qc]}}.

\bibitem{Chakraborty:2016lxo}
S.~Chakraborty and S.~SenGupta, ``{Strong gravitational lensing --- A probe for
  extra dimensions and Kalb-Ramond field},''
\href{http://arxiv.org/abs/1611.06936}{{\ttfamily arXiv:1611.06936 [gr-qc]}}.

\bibitem{Ramallo:2013bua}
A.~V. Ramallo, ``{Introduction to the AdS/CFT correspondence},''
  \href{http://dx.doi.org/10.1007/978-3-319-12238-0_10}{{\em Springer Proc.
  Phys.} {\bfseries 161} (2015) 411--474},
\href{http://arxiv.org/abs/1310.4319}{{\ttfamily arXiv:1310.4319 [hep-th]}}.

\bibitem{Maldacena:1997re}
J.~M. Maldacena, ``{The Large N limit of superconformal field theories and
  supergravity},'' \href{http://dx.doi.org/10.1023/A:1026654312961}{{\em Int.
  J. Theor. Phys.} {\bfseries 38} (1999) 1113--1133},
  \href{http://arxiv.org/abs/hep-th/9711200}{{\ttfamily arXiv:hep-th/9711200
  [hep-th]}}.
[Adv. Theor. Math. Phys.2,231(1998)].

\bibitem{Witten:1998qj}
E.~Witten, ``{Anti-de Sitter space and holography},'' {\em Adv. Theor. Math.
  Phys.} {\bfseries 2} (1998) 253--291,
\href{http://arxiv.org/abs/hep-th/9802150}{{\ttfamily arXiv:hep-th/9802150
  [hep-th]}}.

\bibitem{Aharony:1999ti}
O.~Aharony, S.~S. Gubser, J.~M. Maldacena, H.~Ooguri, and Y.~Oz, ``{Large N
  field theories, string theory and gravity},''
  \href{http://dx.doi.org/10.1016/S0370-1573(99)00083-6}{{\em Phys. Rept.}
  {\bfseries 323} (2000) 183--386},
\href{http://arxiv.org/abs/hep-th/9905111}{{\ttfamily arXiv:hep-th/9905111
  [hep-th]}}.

\bibitem{Hartnoll:2009sz}
S.~A. Hartnoll, ``{Lectures on holographic methods for condensed matter
  physics},'' \href{http://dx.doi.org/10.1088/0264-9381/26/22/224002}{{\em
  Class. Quant. Grav.} {\bfseries 26} (2009) 224002},
\href{http://arxiv.org/abs/0903.3246}{{\ttfamily arXiv:0903.3246 [hep-th]}}.

\bibitem{DHoker:2002nbb}
E.~D'Hoker and D.~Z. Freedman, ``{Supersymmetric gauge theories and the AdS /
  CFT correspondence},'' in {\em {Strings, Branes and Extra Dimensions: TASI
  2001: Proceedings}}, pp.~3--158.
\newblock 2002.
\newblock
\href{http://arxiv.org/abs/hep-th/0201253}{{\ttfamily arXiv:hep-th/0201253
  [hep-th]}}.
\newblock

\bibitem{Sachdev:2010ch}
S.~Sachdev, ``{Condensed Matter and AdS/CFT},''
  \href{http://arxiv.org/abs/1002.2947}{{\ttfamily arXiv:1002.2947 [hep-th]}}.
[Lect. Notes Phys.828,273(2011)].

\bibitem{Nishioka:2009un}
T.~Nishioka, S.~Ryu, and T.~Takayanagi, ``{Holographic Entanglement Entropy: An
  Overview},'' \href{http://dx.doi.org/10.1088/1751-8113/42/50/504008}{{\em J.
  Phys.} {\bfseries A42} (2009) 504008},
\href{http://arxiv.org/abs/0905.0932}{{\ttfamily arXiv:0905.0932 [hep-th]}}.

\bibitem{Einstein:1935rr}
A.~Einstein, B.~Podolsky, and N.~Rosen, ``{Can quantum mechanical description
  of physical reality be considered complete?},''
\href{http://dx.doi.org/10.1103/PhysRev.47.777}{{\em Phys. Rev.} {\bfseries 47}
  (1935) 777--780}.

\bibitem{Einstein:1935tc}
A.~Einstein and N.~Rosen, ``{The Particle Problem in the General Theory of
  Relativity},''
\href{http://dx.doi.org/10.1103/PhysRev.48.73}{{\em Phys. Rev.} {\bfseries 48}
  (1935) 73--77}.

\bibitem{Fuller:1962zza}
R.~W. Fuller and J.~A. Wheeler, ``{Causality and Multiply Connected
  Space-Time},''
\href{http://dx.doi.org/10.1103/PhysRev.128.919}{{\em Phys. Rev.} {\bfseries
  128} (1962) 919--929}.

\bibitem{Visser:1995cc}
M.~Visser, {\em {Lorentzian wormholes: From Einstein to Hawking}}.
\newblock
1995.
\newblock

\bibitem{Marolf:2012xe}
D.~Marolf and A.~C. Wall, ``{Eternal Black Holes and Superselection in
  AdS/CFT},'' \href{http://dx.doi.org/10.1088/0264-9381/30/2/025001}{{\em
  Class. Quant. Grav.} {\bfseries 30} (2013) 025001},
\href{http://arxiv.org/abs/1210.3590}{{\ttfamily arXiv:1210.3590 [hep-th]}}.

\bibitem{Israel:1976ur}
W.~Israel, ``{Thermo field dynamics of black holes},''
\href{http://dx.doi.org/10.1016/0375-9601(76)90178-X}{{\em Phys. Lett.}
  {\bfseries A57} (1976) 107--110}.

\bibitem{Maldacena:2001kr}
J.~M. Maldacena, ``{Eternal black holes in anti-de Sitter},''
  \href{http://dx.doi.org/10.1088/1126-6708/2003/04/021}{{\em JHEP} {\bfseries
  04} (2003) 021},
\href{http://arxiv.org/abs/hep-th/0106112}{{\ttfamily arXiv:hep-th/0106112
  [hep-th]}}.

\bibitem{Bryan:2016wzx}
K.~L.~H. Bryan and A.~J.~M. Medved, ``{Black holes and information: A new take
  on an old paradox},''
\href{http://arxiv.org/abs/1603.07569}{{\ttfamily arXiv:1603.07569 [hep-th]}}.

\bibitem{Papadodimas:2013jku}
K.~Papadodimas and S.~Raju, ``{State-Dependent Bulk-Boundary Maps and Black
  Hole Complementarity},''
  \href{http://dx.doi.org/10.1103/PhysRevD.89.086010}{{\em Phys. Rev.}
  {\bfseries D89} no.~8, (2014) 086010},
\href{http://arxiv.org/abs/1310.6335}{{\ttfamily arXiv:1310.6335 [hep-th]}}.

\bibitem{Papadodimas:2015jra}
K.~Papadodimas and S.~Raju, ``{Remarks on the necessity and implications of
  state-dependence in the black hole interior},''
  \href{http://dx.doi.org/10.1103/PhysRevD.93.084049}{{\em Phys. Rev.}
  {\bfseries D93} no.~8, (2016) 084049},
\href{http://arxiv.org/abs/1503.08825}{{\ttfamily arXiv:1503.08825 [hep-th]}}.

\bibitem{Papadodimas:2015xma}
K.~Papadodimas and S.~Raju, ``{Local Operators in the Eternal Black Hole},''
  \href{http://dx.doi.org/10.1103/PhysRevLett.115.211601}{{\em Phys. Rev.
  Lett.} {\bfseries 115} no.~21, (2015) 211601},
\href{http://arxiv.org/abs/1502.06692}{{\ttfamily arXiv:1502.06692 [hep-th]}}.

\bibitem{Banerjee:2016mhh}
S.~Banerjee, J.-W. Bryan, K.~Papadodimas, and S.~Raju, ``{A toy model of black
  hole complementarity},''
  \href{http://dx.doi.org/10.1007/JHEP05(2016)004}{{\em JHEP} {\bfseries 05}
  (2016) 004},
\href{http://arxiv.org/abs/1603.02812}{{\ttfamily arXiv:1603.02812 [hep-th]}}.

\bibitem{Ghosh:2016fvm}
S.~Ghosh and S.~Raju, ``{The Breakdown of String Perturbation Theory for Many
  External Particles},''
\href{http://arxiv.org/abs/1611.08003}{{\ttfamily arXiv:1611.08003 [hep-th]}}.

\bibitem{schlieder1965}
S.~Schlieder, ``Some remarks about the localization of states in a quantum
  field theory,'' {\em Comm. Math. Phys.} {\bfseries 1} no.~4, (1965) 265--280.
  \url{http://projecteuclid.org/euclid.cmp/1103758945}.

\bibitem{GellMann:1991ck}
M.~Gell-Mann and J.~B. Hartle, ``{Time symmetry and asymmetry in quantum
  mechanics and quantum cosmology},'' in {\em {4th International Conference on
  Ion Sources (ICIS 1991) Bensheim, Germany, September 30-October 4, 1991}},
  pp.~1151--1174.
\newblock 1991.
\newblock \href{http://arxiv.org/abs/gr-qc/9304023}{{\ttfamily
  arXiv:gr-qc/9304023 [gr-qc]}}.
\newblock
  \url{http://tuvalu.santafe.edu/~mgm/Site/Publications_files/MGM%20106.pdf}.
\newblock
[,0311(1991)].

\bibitem{Bennett:1992tv}
C.~H. Bennett, G.~Brassard, C.~Crepeau, R.~Jozsa, A.~Peres, and W.~K. Wootters,
  ``{Teleporting an unknown quantum state via dual classical and
  Einstein-Podolsky-Rosen channels},''
\href{http://dx.doi.org/10.1103/PhysRevLett.70.1895}{{\em Phys. Rev. Lett.}
  {\bfseries 70} (1993) 1895--1899}.

\bibitem{Horowitz:2003he}
G.~T. Horowitz and J.~M. Maldacena, ``{The Black hole final state},''
  \href{http://dx.doi.org/10.1088/1126-6708/2004/02/008}{{\em JHEP} {\bfseries
  02} (2004) 008},
\href{http://arxiv.org/abs/hep-th/0310281}{{\ttfamily arXiv:hep-th/0310281
  [hep-th]}}.

\bibitem{Gottesman:2003up}
D.~Gottesman and J.~Preskill, ``{Comment on `The Black hole final state'},''
  \href{http://dx.doi.org/10.1088/1126-6708/2004/03/026}{{\em JHEP} {\bfseries
  03} (2004) 026},
\href{http://arxiv.org/abs/hep-th/0311269}{{\ttfamily arXiv:hep-th/0311269
  [hep-th]}}.

\bibitem{Unruh:1995gn}
W.~G. Unruh and R.~M. Wald, ``{On evolution laws taking pure states to mixed
  states in quantum field theory},''
  \href{http://dx.doi.org/10.1103/PhysRevD.52.2176}{{\em Phys. Rev.} {\bfseries
  D52} (1995) 2176--2182},
\href{http://arxiv.org/abs/hep-th/9503024}{{\ttfamily arXiv:hep-th/9503024
  [hep-th]}}.

\bibitem{BOHM:1966zz}
D.~Bohm and J.~Bub, ``{A Proposed Solution of the Measurement Problem in
  Quantum Mechanics by a Hidden Variable Theory},''
\href{http://dx.doi.org/10.1103/RevModPhys.38.453}{{\em Rev. Mod. Phys.}
  {\bfseries 38} (1966) 453--469}.

\bibitem{Weinberg:1989us}
S.~Weinberg, ``{Testing Quantum Mechanics},''
\href{http://dx.doi.org/10.1016/0003-4916(89)90276-5}{{\em Annals Phys.}
  {\bfseries 194} (1989) 336}.

\bibitem{Weinberg:2011jg}
S.~Weinberg, ``{Collapse of the State Vector},''
  \href{http://dx.doi.org/10.1103/PhysRevA.85.062116}{{\em Phys. Rev.}
  {\bfseries A85} (2012) 062116},
\href{http://arxiv.org/abs/1109.6462}{{\ttfamily arXiv:1109.6462 [quant-ph]}}.

\bibitem{Zurek:2003zz}
W.~H. Zurek, ``{Decoherence, einselection, and the quantum origins of the
  classical},''
\href{http://dx.doi.org/10.1103/RevModPhys.75.715}{{\em Rev. Mod. Phys.}
  {\bfseries 75} (2003) 715--775}.

\bibitem{Zeh:1970zz}
H.~D. Zeh, ``{On the interpretation of measurement in quantum theory},''
\href{http://dx.doi.org/10.1007/BF00708656}{{\em Found. Phys.} {\bfseries 1}
  (1970) 69--76}.

\bibitem{Perez:2005gh}
A.~Perez, H.~Sahlmann, and D.~Sudarsky, ``{On the quantum origin of the seeds
  of cosmic structure},''
  \href{http://dx.doi.org/10.1088/0264-9381/23/7/008}{{\em Class. Quant. Grav.}
  {\bfseries 23} (2006) 2317--2354},
\href{http://arxiv.org/abs/gr-qc/0508100}{{\ttfamily arXiv:gr-qc/0508100
  [gr-qc]}}.

\bibitem{Sudarsky:2009za}
D.~Sudarsky, ``{Shortcomings in the Understanding of Why Cosmological
  Perturbations Look Classical},''
  \href{http://dx.doi.org/10.1142/S0218271811018937}{{\em Int. J. Mod. Phys.}
  {\bfseries D20} (2011) 509--552},
\href{http://arxiv.org/abs/0906.0315}{{\ttfamily arXiv:0906.0315 [gr-qc]}}.

\bibitem{Canate:2012ua}
P.~Cañate, P.~Pearle, and D.~Sudarsky, ``{Continuous spontaneous localization
  wave function collapse model as a mechanism for the emergence of cosmological
  asymmetries in inflation},''
  \href{http://dx.doi.org/10.1103/PhysRevD.87.104024}{{\em Phys. Rev.}
  {\bfseries D87} no.~10, (2013) 104024},
\href{http://arxiv.org/abs/1211.3463}{{\ttfamily arXiv:1211.3463 [gr-qc]}}.

\bibitem{Martin:2012pea}
J.~Martin, V.~Vennin, and P.~Peter, ``{Cosmological Inflation and the Quantum
  Measurement Problem},''
  \href{http://dx.doi.org/10.1103/PhysRevD.86.103524}{{\em Phys. Rev.}
  {\bfseries D86} (2012) 103524},
\href{http://arxiv.org/abs/1207.2086}{{\ttfamily arXiv:1207.2086 [hep-th]}}.

\bibitem{Das:2013qwa}
S.~Das, K.~Lochan, S.~Sahu, and T.~P. Singh, ``{Quantum to classical transition
  of inflationary perturbations: Continuous spontaneous localization as a
  possible mechanism},'' \href{http://dx.doi.org/10.1103/PhysRevD.89.109902,
  10.1103/PhysRevD.88.085020}{{\em Phys. Rev.} {\bfseries D88} no.~8, (2013)
  085020}, \href{http://arxiv.org/abs/1304.5094}{{\ttfamily arXiv:1304.5094
  [astro-ph.CO]}}.
[Erratum: Phys. Rev.D89,no.10,109902(2014)].

\bibitem{Okon:2013lsa}
E.~Okon and D.~Sudarsky, ``{Benefits of Objective Collapse Models for Cosmology
  and Quantum Gravity},''
  \href{http://dx.doi.org/10.1007/s10701-014-9772-6}{{\em Found. Phys.}
  {\bfseries 44} (2014) 114--143},
\href{http://arxiv.org/abs/1309.1730}{{\ttfamily arXiv:1309.1730 [gr-qc]}}.

\bibitem{Lochan:2014dca}
K.~Lochan, K.~Parattu, and T.~Padmanabhan, ``{Quantum Evolution Leading to
  Classicality: A Concrete Example},''
  \href{http://dx.doi.org/10.1007/s10714-014-1841-9}{{\em Gen. Rel. Grav.}
  {\bfseries 47} no.~1, (2015) 1841},
\href{http://arxiv.org/abs/1404.2605}{{\ttfamily arXiv:1404.2605 [gr-qc]}}.

\bibitem{Tumulka:2005ki}
R.~Tumulka, ``{On spontaneous wave function collapse and quantum field
  theory},'' \href{http://dx.doi.org/10.1098/rspa.2005.1636}{{\em Proc. Roy.
  Soc. Lond.} {\bfseries A462} (2006) 1897--1908},
\href{http://arxiv.org/abs/quant-ph/0508230}{{\ttfamily arXiv:quant-ph/0508230
  [quant-ph]}}.

\bibitem{Bedingham:2010hz}
D.~J. Bedingham, ``{Relativistic state reduction dynamics},''
  \href{http://dx.doi.org/10.1007/s10701-010-9510-7}{{\em Found. Phys.}
  {\bfseries 41} (2011) 686--704},
\href{http://arxiv.org/abs/1003.2774}{{\ttfamily arXiv:1003.2774 [quant-ph]}}.

\bibitem{Pearle:2014tda}
P.~Pearle, ``{Relativistic dynamical collapse model},''
  \href{http://dx.doi.org/10.1103/PhysRevD.91.105012}{{\em Phys. Rev.}
  {\bfseries D91} no.~10, (2015) 105012},
\href{http://arxiv.org/abs/1412.6723}{{\ttfamily arXiv:1412.6723 [quant-ph]}}.

\bibitem{Callan:1992rs}
C.~G. Callan, Jr., S.~B. Giddings, J.~A. Harvey, and A.~Strominger,
  ``{Evanescent black holes},''
  \href{http://dx.doi.org/10.1103/PhysRevD.45.R1005}{{\em Phys. Rev.}
  {\bfseries D45} no.~4, (1992) R1005},
\href{http://arxiv.org/abs/hep-th/9111056}{{\ttfamily arXiv:hep-th/9111056
  [hep-th]}}.

\bibitem{Giddings:1992ff}
S.~B. Giddings and W.~M. Nelson, ``{Quantum emission from two-dimensional black
  holes},'' \href{http://dx.doi.org/10.1103/PhysRevD.46.2486}{{\em Phys. Rev.}
  {\bfseries D46} (1992) 2486--2496},
\href{http://arxiv.org/abs/hep-th/9204072}{{\ttfamily arXiv:hep-th/9204072
  [hep-th]}}.

\bibitem{Lochan:2016cxt}
K.~Lochan, S.~Chakraborty, and T.~Padmanabhan, ``{Dynamic realization of the
  Unruh effect for a geodesic observer},''
\href{http://arxiv.org/abs/1603.01964}{{\ttfamily arXiv:1603.01964 [gr-qc]}}.

\bibitem{Ashtekar:2008jd}
A.~Ashtekar, V.~Taveras, and M.~Varadarajan, ``{Information is Not Lost in the
  Evaporation of 2-dimensional Black Holes},''
  \href{http://dx.doi.org/10.1103/PhysRevLett.100.211302}{{\em Phys. Rev.
  Lett.} {\bfseries 100} (2008) 211302},
\href{http://arxiv.org/abs/0801.1811}{{\ttfamily arXiv:0801.1811 [gr-qc]}}.

\bibitem{Modak:2016uwr}
S.~K. Modak and D.~Sudarsky, ``{Modelling non-paradoxical loss of information
  in black hole evaporation},''
\href{http://arxiv.org/abs/1607.05410}{{\ttfamily arXiv:1607.05410 [gr-qc]}}.

\bibitem{Okon:2016qlh}
E.~Okon and D.~Sudarsky, ``{Black Holes, Information Loss and the Measurement
  Problem},''
\href{http://arxiv.org/abs/1607.01255}{{\ttfamily arXiv:1607.01255 [gr-qc]}}.

\bibitem{Bedingham:2016aus}
D.~Bedingham, S.~K. Modak, and D.~Sudarsky, ``{Relativistic collapse dynamics
  and black hole information loss},''
  \href{http://dx.doi.org/10.1103/PhysRevD.94.045009}{{\em Phys. Rev.}
  {\bfseries D94} no.~4, (2016) 045009},
\href{http://arxiv.org/abs/1604.06537}{{\ttfamily arXiv:1604.06537 [gr-qc]}}.

\bibitem{Okon:2014dpa}
E.~Okon and D.~Sudarsky, ``{The Black Hole Information Paradox and the Collapse
  of the Wave Function},''
  \href{http://dx.doi.org/10.1007/s10701-015-9877-6}{{\em Found. Phys.}
  {\bfseries 45} no.~4, (2015) 461--470},
\href{http://arxiv.org/abs/1406.2011}{{\ttfamily arXiv:1406.2011 [gr-qc]}}.

\bibitem{Modak:2014qja}
S.~K. Modak, L.~Ortíz, I.~Peña, and D.~Sudarsky, ``{Black hole evaporation:
  information loss but no paradox},''
  \href{http://dx.doi.org/10.1007/s10714-015-1960-y}{{\em Gen. Rel. Grav.}
  {\bfseries 47} no.~10, (2015) 120},
\href{http://arxiv.org/abs/1406.4898}{{\ttfamily arXiv:1406.4898 [gr-qc]}}.

\bibitem{Hawking:2016msc}
S.~W. Hawking, M.~J. Perry, and A.~Strominger, ``{Soft Hair on Black Holes},''
  \href{http://dx.doi.org/10.1103/PhysRevLett.116.231301}{{\em Phys. Rev.
  Lett.} {\bfseries 116} no.~23, (2016) 231301},
\href{http://arxiv.org/abs/1601.00921}{{\ttfamily arXiv:1601.00921 [hep-th]}}.

\bibitem{Hawking:2016sgy}
S.~W. Hawking, M.~J. Perry, and A.~Strominger, ``{Superrotation Charge and
  Supertranslation Hair on Black Holes},''
\href{http://arxiv.org/abs/1611.09175}{{\ttfamily arXiv:1611.09175 [hep-th]}}.

\bibitem{Bondi:1962px}
H.~Bondi, M.~G.~J. van~der Burg, and A.~W.~K. Metzner, ``{Gravitational waves
  in general relativity. 7. Waves from axisymmetric isolated systems},''
\href{http://dx.doi.org/10.1098/rspa.1962.0161}{{\em Proc. Roy. Soc. Lond.}
  {\bfseries A269} (1962) 21--52}.

\bibitem{Sachs:1962wk}
R.~K. Sachs, ``{Gravitational waves in general relativity. 8. Waves in
  asymptotically flat space-times},''
\href{http://dx.doi.org/10.1098/rspa.1962.0206}{{\em Proc. Roy. Soc. Lond.}
  {\bfseries A270} (1962) 103--126}.

\bibitem{Strominger:2013jfa}
A.~Strominger, ``{On BMS Invariance of Gravitational Scattering},''
  \href{http://dx.doi.org/10.1007/JHEP07(2014)152}{{\em JHEP} {\bfseries 07}
  (2014) 152},
\href{http://arxiv.org/abs/1312.2229}{{\ttfamily arXiv:1312.2229 [hep-th]}}.

\bibitem{Weinberg:1965nx}
S.~Weinberg, ``{Infrared photons and gravitons},''
\href{http://dx.doi.org/10.1103/PhysRev.140.B516}{{\em Phys. Rev.} {\bfseries
  140} (1965) B516--B524}.

\bibitem{Cachazo:2014fwa}
F.~Cachazo and A.~Strominger, ``{Evidence for a New Soft Graviton Theorem},''
\href{http://arxiv.org/abs/1404.4091}{{\ttfamily arXiv:1404.4091 [hep-th]}}.

\bibitem{Ashtekar:1978zz}
A.~Ashtekar and R.~O. Hansen, ``{A unified treatment of null and spatial
  infinity in general relativity. I - Universal structure, asymptotic
  symmetries, and conserved quantities at spatial infinity},''
\href{http://dx.doi.org/10.1063/1.523863}{{\em J. Math. Phys.} {\bfseries 19}
  (1978) 1542--1566}.

\bibitem{Ashtekar:1981sf}
A.~Ashtekar, ``{Asymptotic Quantization of the Gravitational Field},''
\href{http://dx.doi.org/10.1103/PhysRevLett.46.573}{{\em Phys. Rev. Lett.}
  {\bfseries 46} (1981) 573--576}.

\bibitem{Kulish:1970ut}
P.~P. Kulish and L.~D. Faddeev, ``{Asymptotic conditions and infrared
  divergences in quantum electrodynamics},''
  \href{http://dx.doi.org/10.1007/BF01066485}{{\em Theor. Math. Phys.}
  {\bfseries 4} (1970) 745}.
[Teor. Mat. Fiz.4,153(1970)].

\bibitem{Ware:2013zja}
J.~Ware, R.~Saotome, and R.~Akhoury, ``{Construction of an asymptotic S matrix
  for perturbative quantum gravity},''
  \href{http://dx.doi.org/10.1007/JHEP10(2013)159}{{\em JHEP} {\bfseries 10}
  (2013) 159},
\href{http://arxiv.org/abs/1308.6285}{{\ttfamily arXiv:1308.6285 [hep-th]}}.

\bibitem{Barnich:2009se}
G.~Barnich and C.~Troessaert, ``{Symmetries of asymptotically flat 4
  dimensional spacetimes at null infinity revisited},''
  \href{http://dx.doi.org/10.1103/PhysRevLett.105.111103}{{\em Phys. Rev.
  Lett.} {\bfseries 105} (2010) 111103},
\href{http://arxiv.org/abs/0909.2617}{{\ttfamily arXiv:0909.2617 [gr-qc]}}.

\bibitem{Barnich:2011ct}
G.~Barnich and C.~Troessaert, ``{Supertranslations call for superrotations},''
  {\em PoS} (2010) 010, \href{http://arxiv.org/abs/1102.4632}{{\ttfamily
  arXiv:1102.4632 [gr-qc]}}.
[Ann. U. Craiova Phys.21,S11(2011)].

\bibitem{Barnich:2011mi}
G.~Barnich and C.~Troessaert, ``{BMS charge algebra},''
  \href{http://dx.doi.org/10.1007/JHEP12(2011)105}{{\em JHEP} {\bfseries 12}
  (2011) 105},
\href{http://arxiv.org/abs/1106.0213}{{\ttfamily arXiv:1106.0213 [hep-th]}}.

\bibitem{Kapec:2015vwa}
D.~Kapec, V.~Lysov, S.~Pasterski, and A.~Strominger, ``{Higher-Dimensional
  Supertranslations and Weinberg's Soft Graviton Theorem},''
\href{http://arxiv.org/abs/1502.07644}{{\ttfamily arXiv:1502.07644 [gr-qc]}}.

\bibitem{Hollands:2016oma}
S.~Hollands, A.~Ishibashi, and R.~M. Wald, ``{BMS Supertranslations and Memory
  in Four and Higher Dimensions},''
\href{http://arxiv.org/abs/1612.03290}{{\ttfamily arXiv:1612.03290 [gr-qc]}}.

\bibitem{deBoer:2003vf}
J.~de~Boer and S.~N. Solodukhin, ``{A Holographic reduction of Minkowski
  space-time},'' \href{http://dx.doi.org/10.1016/S0550-3213(03)00494-2}{{\em
  Nucl. Phys.} {\bfseries B665} (2003) 545--593},
\href{http://arxiv.org/abs/hep-th/0303006}{{\ttfamily arXiv:hep-th/0303006
  [hep-th]}}.

\bibitem{Banks:2003vp}
T.~Banks, ``{A Critique of pure string theory: Heterodox opinions of diverse
  dimensions},''
\href{http://arxiv.org/abs/hep-th/0306074}{{\ttfamily arXiv:hep-th/0306074
  [hep-th]}}.

\bibitem{Hogan:1993xj}
P.~A. Hogan, ``{A Spherical impulse gravity wave},''
\href{http://dx.doi.org/10.1103/PhysRevLett.70.117}{{\em Phys. Rev. Lett.}
  {\bfseries 70} (1993) 117--118}.

\bibitem{Nutku:1977wp}
Y.~Nutku and M.~Halil, ``{Colliding Impulsive Gravitational Waves},''
\href{http://dx.doi.org/10.1103/PhysRevLett.39.1379}{{\em Phys. Rev. Lett.}
  {\bfseries 39} (1977) 1379--1382}.

\bibitem{Penrose:1972xrn}
R.~Penrose, ``{The geometry of impulsive gravitational waves},'' in {\em
  General relativity: Papers in honour of J.L. Synge}, L.~O'Raifeartaigh, ed.,
  pp.~101--115.
\newblock
1972.
\newblock

\bibitem{Strominger:2016wns}
A.~Strominger and A.~Zhiboedov, ``{Superrotations and Black Hole Pair
  Creation},''
\href{http://arxiv.org/abs/1610.00639}{{\ttfamily arXiv:1610.00639 [hep-th]}}.

\bibitem{He:2014laa}
T.~He, V.~Lysov, P.~Mitra, and A.~Strominger, ``{BMS supertranslations and
  Weinberg’s soft graviton theorem},''
  \href{http://dx.doi.org/10.1007/JHEP05(2015)151}{{\em JHEP} {\bfseries 05}
  (2015) 151},
\href{http://arxiv.org/abs/1401.7026}{{\ttfamily arXiv:1401.7026 [hep-th]}}.

\bibitem{Parattu:2015gga}
K.~Parattu, S.~Chakraborty, B.~R. Majhi, and T.~Padmanabhan, ``{A Boundary Term
  for the Gravitational Action with Null Boundaries},''
  \href{http://dx.doi.org/10.1007/s10714-016-2093-7}{{\em Gen. Rel. Grav.}
  {\bfseries 48} no.~7, (2016) 94},
\href{http://arxiv.org/abs/1501.01053}{{\ttfamily arXiv:1501.01053 [gr-qc]}}.

\bibitem{Donnay:2015abr}
L.~Donnay, G.~Giribet, H.~A. Gonzalez, and M.~Pino, ``{Supertranslations and
  Superrotations at the Black Hole Horizon},''
  \href{http://dx.doi.org/10.1103/PhysRevLett.116.091101}{{\em Phys. Rev.
  Lett.} {\bfseries 116} no.~9, (2016) 091101},
\href{http://arxiv.org/abs/1511.08687}{{\ttfamily arXiv:1511.08687 [hep-th]}}.

\bibitem{Compere:2016hzt}
G.~Compère and J.~Long, ``{Classical static final state of collapse with
  supertranslation memory},''
  \href{http://dx.doi.org/10.1088/0264-9381/33/19/195001}{{\em Class. Quant.
  Grav.} {\bfseries 33} no.~19, (2016) 195001},
\href{http://arxiv.org/abs/1602.05197}{{\ttfamily arXiv:1602.05197 [gr-qc]}}.

\bibitem{Eling:2016xlx}
C.~Eling and Y.~Oz, ``{On the Membrane Paradigm and Spontaneous Breaking of
  Horizon BMS Symmetries},''
  \href{http://dx.doi.org/10.1007/JHEP07(2016)065}{{\em JHEP} {\bfseries 07}
  (2016) 065},
\href{http://arxiv.org/abs/1605.00183}{{\ttfamily arXiv:1605.00183 [hep-th]}}.

\bibitem{Lochan:2015oba}
K.~Lochan and T.~Padmanabhan, ``{Extracting information about the initial state
  from the black hole radiation},''
  \href{http://dx.doi.org/10.1103/PhysRevLett.116.051301}{{\em Phys. Rev.
  Lett.} {\bfseries 116} no.~5, (2016) 051301},
\href{http://arxiv.org/abs/1507.06402}{{\ttfamily arXiv:1507.06402 [gr-qc]}}.

\bibitem{Lochan:2016nbs}
K.~Lochan, S.~Chakraborty, and T.~Padmanabhan, ``{Information retrieval from
  black holes},'' \href{http://dx.doi.org/10.1103/PhysRevD.94.044056}{{\em
  Phys. Rev.} {\bfseries D94} no.~4, (2016) 044056},
\href{http://arxiv.org/abs/1604.04987}{{\ttfamily arXiv:1604.04987 [gr-qc]}}.

\bibitem{Lochan:2014xja}
K.~Lochan and T.~Padmanabhan, ``{Inertial nonvacuum states viewed from the
  Rindler frame},'' \href{http://dx.doi.org/10.1103/PhysRevD.91.044002}{{\em
  Phys. Rev.} {\bfseries D91} no.~4, (2015) 044002},
\href{http://arxiv.org/abs/1411.7019}{{\ttfamily arXiv:1411.7019 [gr-qc]}}.

\bibitem{Chatwin-Davies:2015hna}
A.~Chatwin-Davies, A.~S. Jermyn, and S.~M. Carroll, ``{How to Recover a Qubit
  That Has Fallen Into a Black Hole},''
  \href{http://dx.doi.org/10.1103/PhysRevLett.115.261302}{{\em Phys. Rev.
  Lett.} {\bfseries 115} no.~26, (2015) 261302},
\href{http://arxiv.org/abs/1507.03592}{{\ttfamily arXiv:1507.03592 [hep-th]}}.

\bibitem{Dupuis}
F.~{Dupuis}, ``{The decoupling approach to quantum information theory},'' {\em
  PhD Thesis} (2010) , \href{http://arxiv.org/abs/1004.1641}{{\ttfamily
  1004.1641}}.

\bibitem{tHooftTS}
G.~'t~Hooft, ``{Dimensional reduction in quantum gravity},'' {\em
  Salamfestschrift: A Collection of Talks} {\bfseries 4} (1993) .

\bibitem{Hayden:2007cs}
P.~Hayden and J.~Preskill, ``{Black holes as mirrors: Quantum information in
  random subsystems},''
  \href{http://dx.doi.org/10.1088/1126-6708/2007/09/120}{{\em JHEP} {\bfseries
  09} (2007) 120},
\href{http://arxiv.org/abs/0708.4025}{{\ttfamily arXiv:0708.4025 [hep-th]}}.

\bibitem{Hooft:2015jea}
G.~'t~Hooft, ``{Diagonalizing the Black Hole Information Retrieval Process},''
\href{http://arxiv.org/abs/1509.01695}{{\ttfamily arXiv:1509.01695 [gr-qc]}}.

\bibitem{tHooft:1984kcu}
G.~'t~Hooft, ``{On the Quantum Structure of a Black Hole},''
\href{http://dx.doi.org/10.1016/0550-3213(85)90418-3}{{\em Nucl. Phys.}
  {\bfseries B256} (1985) 727--745}.

\bibitem{tHooft:1996rdg}
G.~'t~Hooft, ``{The Scattering matrix approach for the quantum black hole: An
  Overview},'' \href{http://dx.doi.org/10.1142/S0217751X96002145}{{\em Int. J.
  Mod. Phys.} {\bfseries A11} (1996) 4623--4688},
\href{http://arxiv.org/abs/gr-qc/9607022}{{\ttfamily arXiv:gr-qc/9607022
  [gr-qc]}}.

\bibitem{tHooft:1986cfy}
G.~'t~Hooft, ``{Strings From Gravity},''
\href{http://dx.doi.org/10.1088/0031-8949/1987/T15/019}{{\em Phys. Scripta}
  {\bfseries T15} (1987) 143}.

\bibitem{Aichelburg:1970dh}
P.~C. Aichelburg and R.~U. Sexl, ``{On the Gravitational field of a massless
  particle},''
\href{http://dx.doi.org/10.1007/BF00758149}{{\em Gen. Rel. Grav.} {\bfseries 2}
  (1971) 303--312}.

\bibitem{Dray:1984ha}
T.~Dray and G.~'t~Hooft, ``{The Gravitational Shock Wave of a Massless
  Particle},''
\href{http://dx.doi.org/10.1016/0550-3213(85)90525-5}{{\em Nucl. Phys.}
  {\bfseries B253} (1985) 173--188}.

\end{thebibliography}\endgroup

\bibliographystyle{./utphys1}

\end{document}